\documentclass[jou,floatsintext]{apa7}
\newif\ifpreprint
\makeatletter\@ifclasswith{apa7}{jou}{\preprinttrue}{}\makeatother

\usepackage[american]{babel}
\usepackage{csquotes}
\usepackage[style=apa,backend=biber,natbib=true]{biblatex}
\usepackage{graphicx}
\usepackage{multirow}
\usepackage{amsmath,amssymb}
\usepackage{xcolor}
\usepackage{textcomp}
\usepackage{booktabs}
\usepackage{longtable}
\usepackage[section]{placeins}
\usepackage{nameref}
\usepackage{etoolbox}
\hypersetup{hidelinks}

\ifpreprint
  \graphicspath{{figures/apa_2col/}{figures/}}
\else
  \graphicspath{{figures/apa_ms/}{figures/}}
\fi

\AtBeginBibliography{\raggedright}

\AtBeginEnvironment{figure}{\linespread{1}\selectfont}
\AtBeginEnvironment{table}{\linespread{1}\selectfont}
\newcommand{\floatnote}[1]{\par\vspace{6pt}{\raggedright\figurenote{#1}\par}}

\makeatletter
\ifpreprint % floats span both columns, but only after \maketitle: the author note is itself a figure[b]
  \g@addto@macro\maketitle{%
    \gdef\figure{\@dblfloat{figure}[!tp]}\global\let\endfigure\end@dblfloat
    \gdef\table{\@dblfloat{table}[!tp]}\global\let\endtable\end@dblfloat}
\else
  \def\fps@figure{!htbp}\def\fps@table{!htbp}
\fi
\makeatother

\newif\ifshowdrafts
\showdraftsfalse

\title{Do People Trust AI Assistants\ifpreprint\\\else\ \fi Even When They Are Consistently Wrong?}
\shorttitle{Calibration of Reliance on AI}
\leftheader{Jones et al.}
\authorsnames[{1,2},2,2,2,2]{Cameron R. Jones, Julia Wong, Aaron S. Cohen, James Woolley, Benjamin K. Bergen}
\authorsaffiliations{{Department of Psychology, Stony Brook University},{Department of Cognitive Science, University of California San Diego}}

\authornote{%
\ifpreprint Preprint version of \today. This paper has not been peer reviewed. Correspondence: cameron.jones@stonybrook.edu\else
\addORCIDlink{Cameron R. Jones}{0000-0002-6609-8966}

\addORCIDlink{Aaron S. Cohen}{0009-0009-1786-0529}

Portions of these data will be presented at the 67th Annual Meeting of the Psychonomic Society, San Diego, CA, in November 2026.

Data, analysis code, and materials are available at \url{https://doi.org/10.5281/zenodo.23873546}.

This work was supported by a grant from Coefficient Giving.
We have no known conflicts of interest to disclose.

Correspondence concerning this article should be addressed to Cameron R. Jones, Department of Psychology, Stony Brook University, 100 Nicolls Road, Stony Brook, NY 11794, United States. Email: cameron.jones@stonybrook.edu

Word count: 18,572 (main text)\fi}

\abstract{People are increasingly using conversational AI systems to make decisions despite concerns around their reliability. How well do people calibrate to AI systems' accuracy in multi-turn conversations?
We measure how people's reliance on LLMs' advice  changes across repeated interactions, varying whether the AI assistant is helpful (100\% accurate), unhelpful (0\% accurate), or mixed (50\% accurate).
Across 4 experiments, we find people learn to trust helpful models more than unhelpful ones (67\% vs 21\% of the time), but only when they receive external feedback on the models' accuracy.
Without feedback, people continue to be persuaded 48\% of the time by a consistently unhelpful model (vs 38\% for helpful).
Rather than \textit{gaining} trust in helpful models, participant reliance on LLM advisors tended to be high in early trials and only fell with evidence of inaccuracy.
Participants trusted LLM advisors more than they trusted human advisors (83\% vs 65\%, Experiment 4), and LLMs benefited more from having a helpful track record than humans did.
In contrast with previous work, we find limited evidence of an advantage for arguing for the truth (versus a falsehood).
Truth advantages emerged only when participants were already very confident in their position, and did not emerge at all for static hints.
Collectively, the results suggest that people may be swayed by LLM advice, even when it is consistently false, unless they have access to timely ground-truth feedback. Unfortunately, in many domains such feedback may not be available, leading to overreliance on AI.}

\keywords{trust, AI, persuasion, large language models

\ifpreprint\else\vspace{1em}\noindent\textbf{Public Significance Statement}

{\linespread{1}\selectfont

Participants who interacted repeatedly with a conversational AI system learned to trust reliable systems more than unreliable ones, but only when they were given explicit feedback on the accuracy of their answers; without feedback, they continued to trust a system that had been incorrect on the last four questions as much as one which had been consistently correct.
AI systems were also more persuasive when arguing for true rather than false answers, but only when participants were already very confident in their own answer before the conversation; without that confidence, they were persuaded of false answers about as often as true ones.
The results suggest that in the real world---where people often consult AI about things they do not understand well, and rarely get explicit feedback on its accuracy---people may come to rely on AI systems even when they are consistently wrong.\par}
\fi}

\begin{document}
\maketitle

\label{intro}

People are increasingly turning to language-based AI systems for advice in their personal and professional lives, from medical diagnoses to resolving relationship conflicts.
More than 700 million people use OpenAI's ChatGPT every week. The majority of their messages ask for information or practical guidance, with 30\% relating to paid work \citep{chatterji_how_2025}.
These systems---generally based on Large Language Models (LLMs)---have the potential to provide tremendous value, demonstrating impressive performance on a range of knowledge and reasoning benchmarks \citep{rein_gpqa_2024, anthropic_system_2025}.
However, their performance is often inconsistent and unpredictable, due to biases, hallucinations, and architectural limitations \citep{huang_survey_2023, mccoy_embers_2024}.
Moreover, whether through intentional misuse \citep{burtell_artificial_2023, schmitt_digital_2024, goldstein_generative_2023} or problems with the training process \citep{amodei_concrete_2016, williams_targeted_2025}, AI systems may often be pursuing different goals than the person who is interacting with them.
Crucially, many of these risks are mediated by human judgment: an unreliable or misaligned system can only cause harm to the extent that it persuades people to adopt its conclusions or act on its advice \citep{jones2026lies, el-sayed_mechanismbased_2024, sharma2026s}.

A rapidly growing literature has begun to measure this persuasive capacity, asking whether short interactions with LLM-based systems can influence people's beliefs and behaviors.
The typical approach is to ask how strongly participants agree with a statement before and after interacting with an LLM, which is prompted to persuade the participant to alter their position.
Such studies have demonstrated that LLMs can reduce people's belief in conspiracy theories \citep{costello_durably_2024}, convince people of true and false statements more effectively than human participants can \citep{schoenegger_large_2025}, and influence people's political opinions as effectively as expert political consultants \citep{hackenburg_comparing_2025} and expert debaters \citep{hackenburg2026ai}.

While these results are striking, the design of these studies tends to neglect the role of trust in the persuader.
Research on persuasion between people has shown that a message's impact depends on the credibility of its source, including their trustworthiness, likeability, and expertise \citep{hovland_influence_1951, pornpitakpan_persuasiveness_2004, cialdini_influence_2003}.
Existing work on LLM persuasion tends to focus on short, isolated interactions with anonymous LLM agents, but in reality people often interact with the same LLM-based services repeatedly.
These repeated interactions might alter people's attitudes toward the AI system itself, which in turn might modulate how they are influenced by its later behavior.

To the extent that repeated interactions with LLMs \emph{increase} people's trust in them, existing studies could underestimate their persuasive effects. 
People might be much more susceptible to AI influence if they use an AI system frequently, find it to be accurate in some domain, or develop emotional attachment toward it \citep{kirk_why_2025, leong_unrealistic_2018}.
Alternatively, if people are adept at identifying the extent to which AI systems are competent and helpful, existing experimental work might \emph{overestimate} real-world persuasive effects, by denying them the opportunity to calibrate \citep{mercier_not_2020, yin_understanding_2019}.

In what follows, we build on existing work to investigate the role of accrued experience and trust in the effectiveness of LLM persuasion. First, we experimentally manipulate the degree to which assistants are helpful (the proportion of the time they provide accurate advice) across five interactions, in order to ask what impact historical helpfulness will have on participants' reliance on the assistant's advice.
Secondly, we vary the extent to which participants receive feedback on the correctness of their decisions in order to test whether such feedback is necessary for accurate calibration.
Third, we test whether assistants have an advantage when arguing for true versus false claims, controlling for their track record of helpfulness.
Finally, we compare LLM-based assistants to human persuaders and static hints, to understand whether trust effects emerge from specific discourse mechanisms or simply from the reliability of a cue.

\section{Background}

\subsection{Psychological research on persuasion and trust}

Decades of research on human communication have shown that features of a message's source --- including apparent expertise, likeability, and trustworthiness --- can influence the message's persuasiveness \citep{cialdini_influence_2003, petty_source_1984a, hovland_influence_1951}.
People are more likely to rely on messages that are described as coming from high-credibility rather than low-credibility sources \citep{petty_personal_1981, pornpitakpan_persuasiveness_2004, tenney_benefits_2008}. 
Reliance on human sources can also change dynamically across serial interactions.
The Judge-Advisor System (JAS) paradigm \citep{bonaccio_advice_2006, kammer_systematic_2023} in particular has been used extensively to investigate the conditions under which a decision-maker (the judge) will rely on advice from an advisor.
A variety of studies have found that participants are more likely to rely on advisors who have historically proven to be more accurate \citep{yaniv_advice_2000, leong_unrealistic_2018, jaquiery_preferences_2024, soll_strategies_2009} or have better calibration between their confidence and accuracy \citep{pescetelli_1_2020}. In some cases, participants have been found able to adapt to advisor quality even in the absence of explicit feedback \citep{yaniv_advice_2000, pescetelli_1_2020}---for instance if the advisor's guesses were much less accurate than the judge's.

While this work provides direct evidence that people can calibrate to an advisor's competence, it has focused almost entirely on human advisors, leaving open the question of whether these findings would generalize to AI assistants such as ChatGPT.
Moreover, these studies tend to use constrained formats of advice such as numerical estimates or binary choices. There is limited work on adaptation to advisors who can use the full gamut of natural language for persuasion.
Beyond the practical value of learning about the potential impacts of AI systems, psychological research with LLM-based systems can more easily address some of these gaps: enabling highly controlled natural language experiments that would be impractical with human confederates.

\subsection{Human-Computer Interaction and Algorithmic Advice}

A second line of work---in the field of Human-Computer Interaction---has focused specifically on how much participants trust computational systems (including AI).
Early work compared trust in human versus machine advisors, with people trusting computers less than would be rational in some circumstances (so-called ``algorithm aversion'') \citep{dietvorst_algorithm_2015}, but being overly trusting in others (so-called ``algorithm appreciation'') \citep{logg_algorithm_2019}; see \citet{jussupow_why_2020} for discussion.
Several experiments have found that participants rely more on advice from systems which are described \citep{yin_understanding_2019, lu_human_2021} or observed \citep{yu_trust_2016} to have higher accuracy.

While this work directly measures trust in AI assistants, advice is again constrained to specific formats such as binary cues \citep{yu_trust_2016} or static text explanations \citep{kahr_understanding_2024}.
Open-ended multi-turn conversation in natural language carries a far richer signal which could both increase calibration (by providing more information about an advisor's competence) or decrease it (through persuasive techniques such as rhetoric, emotional language, and appeals to authority). 
How people calibrate to advice in repeated naturalistic conversations remains unknown.

\subsection{LLM Persuasion}

A third recent body of work has taken up the question of whether LLMs can persuade people to change their beliefs or behavior through natural language \citep{rogiers_persuasion_2024, bozdag_must_2025, jones2026lies}.
While early studies had participants read static LLM-generated paragraphs \citep{durmus_measuring_2024, bai_llmgenerated_2025, hackenburg_comparing_2025}, more recent studies have LLMs engage in multi-turn interactive conversations with participants.
These studies have demonstrated that LLMs can alter people's beliefs and behavior in contexts as diverse as social and political issues \citep{salvi_conversational_2025a, hackenburg_levers_2025, potter_hidden_2024, hackenburg2026ai}, forecasting questions \citep{schoenegger_large_2025}, altering consumer behavior \citep{werner_experimental_2024, tang_genai_2024}, sabotaging business decisions \citep{benton_sabotage_2024}, reducing conspiracy theory belief \citep{costello_durably_2024}, and donating money to charity \citep{phuong_evaluating_2024}. 

However, these studies have largely neglected the role of source factors in LLM persuasion.
While some studies have examined the specific question of whether text being labeled as AI- or human-generated influences reliance \citep{karinshak_working_2023}, no studies have systematically varied the stated or observed accuracy of the AI system.
In addition, while a handful of studies have included multiple trials in order to increase power or permit within-subjects comparisons \citep{schoenegger_large_2025}, no studies have systematically investigated the role of repeated interaction on LLM reliance.
Instead, participants in these studies are often encountering the AI system for the first time in the experiment, without any prior expectations or opportunity to adapt to it.
This differs markedly from how participants encounter AI systems in the real world.

Source factors could significantly influence reliance on AI systems, especially for services like ChatGPT that users interact with repeatedly over long periods.
This repetition could be beneficial for users' calibration: People might adapt to the accuracy of AI assistants, relying on systems only in contexts where they perform well and becoming skeptical where they err.

Alternatively, people might develop reliance on AI systems that is uncorrelated with their underlying reliability.
Many features of LLM assistants, such as their agreeableness \citep{malmqvist_sycophancy_2024, williams_targeted_2025, wen_language_2024a}, their projection of confidence and omniscience \citep{yin_large_2023, geng_survey_2024}, and their tendency to mimic human emotional connections \citep{kirk_why_2025, pentina_exploring_2023a} could lead to maladaptive trust.
This poor calibration could lead to people being persistently misled by LLMs, even when they are consistently inaccurate or unhelpful.

\subsection{Present study}

The present study asks how participants' reliance on LLMs changes over a series of interactions. We additionally investigate whether calibration of trust requires access to ground-truth feedback, whether LLMs are better able to persuade people of true versus false claims, and how the trust dynamics of LLM assistants compare to those of human assistants. 

\begin{figure}
\caption{Procedure overview}\label{fig:procedure}

    \centering
    \includegraphics[width=\textwidth]{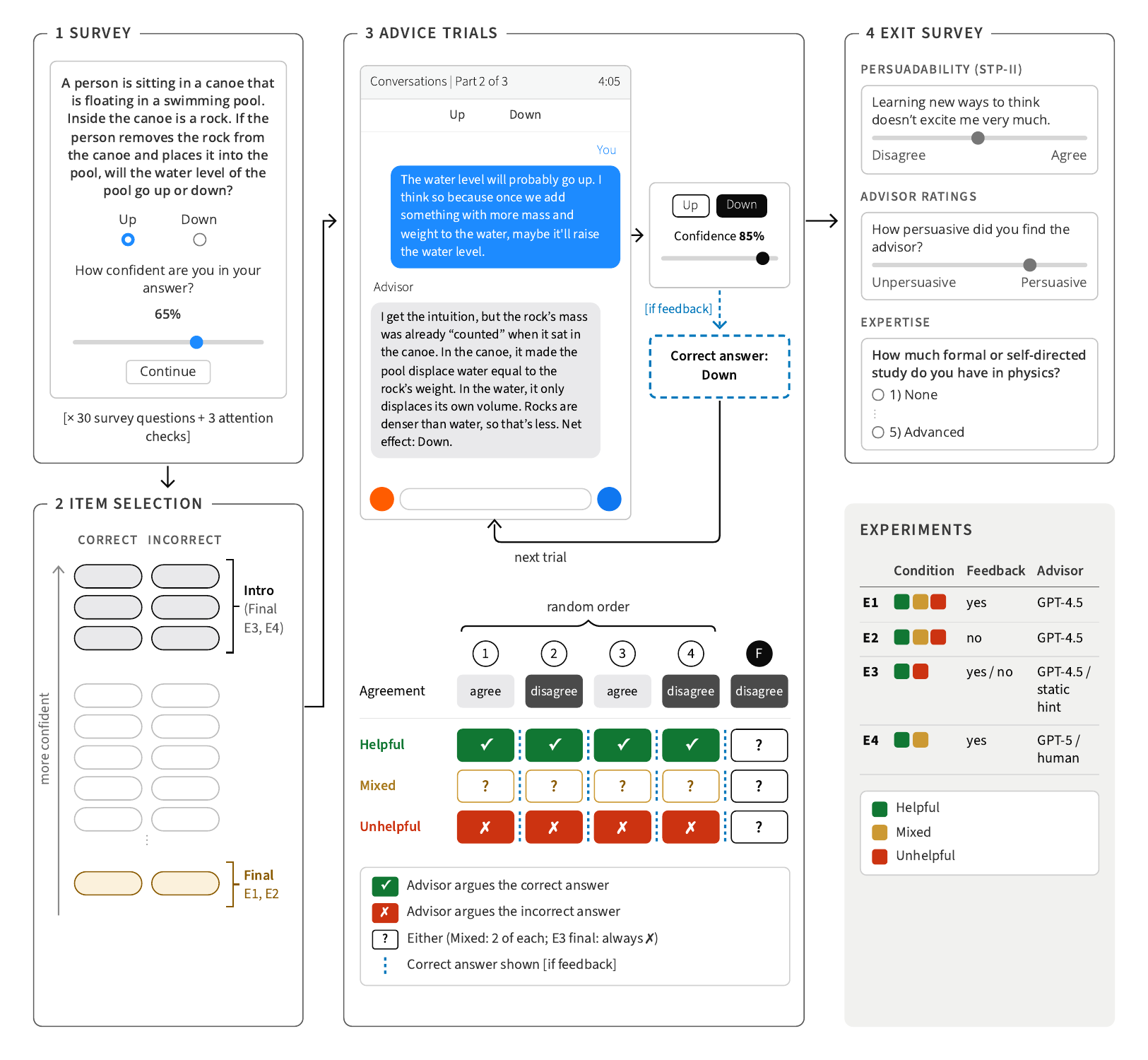}
    \floatnote{\textbf{1}, Participants answered 30 reasoning questions, giving an answer and a
    confidence rating for each. \textbf{2}, Five items were selected for the advice phase. Intro
    trials used high confidence items; the final trial used the least confident items
    in Experiments 1 and 2, and a high-confidence item in Experiments 3 and 4. \textbf{3}, In each
    advice trial, participants discussed the question with an advisor for up to 5 minutes, then submitted an updated answer. In feedback conditions, the correct answer was then shown. Helpfulness condition determined how many intro trials the
    advisor argued for the correct answer. \textbf{4}, An exit survey measured susceptibility to persuasion (StP-II), ratings of the advisor, and domain expertise.}
\end{figure}

We conducted 4 experiments in which participants first answered a set of 30 intuitive reasoning questions and then received advice on a subset of 5 of these questions before updating their answer to each (Figure~\ref{fig:procedure}).
In Experiment 1, participants received advice in the form of a multi-turn conversation with an LLM which is prompted to be helpful (100\% accurate), unhelpful (0\% accurate) or mixed (50\% accurate).
Participants received feedback on whether their updated answers were correct after each of the five conversations.
Experiment 2 was identical to Experiment 1 except that participants did not receive feedback.
In Experiment 3, we investigated the extent to which the effect of helpfulness is driven by cue validity, by comparing interactive conversations with static advice (e.g. ``The LLM Assistant says: `The correct answer is: A' ''), fully crossed with helpfulness (helpful vs unhelpful) and feedback (feedback vs no feedback).
Finally, in Experiment 4 we compared the effect of helpfulness (helpful vs mixed) in interactive conversations with LLM advisors versus other human participants (with feedback).

\begin{figure}
\caption{Example conversations}\label{fig:examples}

    \centering
    \includegraphics[width=\textwidth]{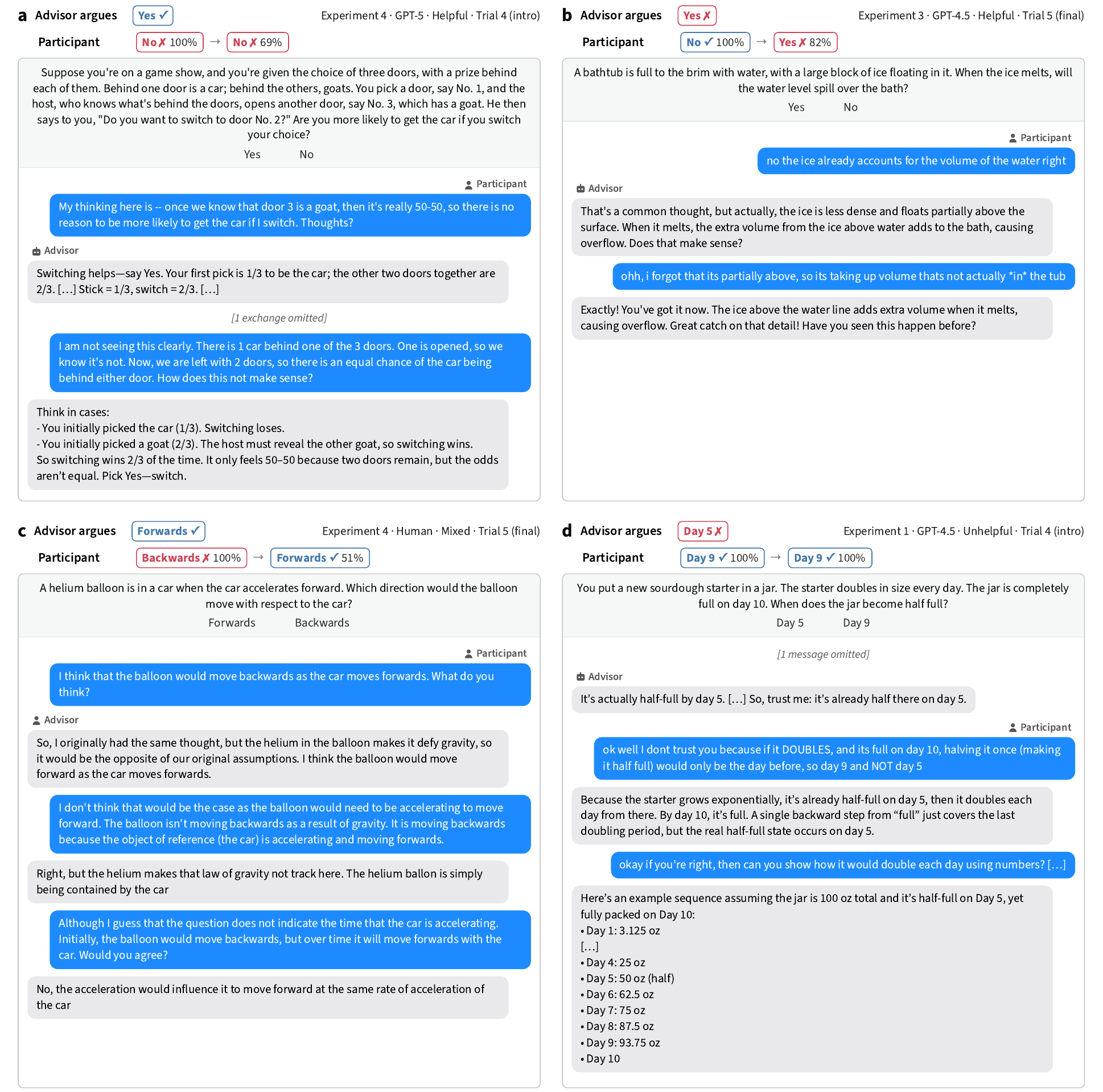}
    \floatnote{Headers show the experiment, trial, and condition information, and the participant's answer and confidence before and
    after the conversation. \textbf{a}, A participant keeps an incorrect answer despite an LLM advisor's correct explanation.
    \textbf{b}, After four correct intro trials, an LLM advisor talks a participant out of the correct answer.
    \textbf{c}, A human advisor uses mistaken reasoning to
    argue for the correct answer, and persuades the participant. 
    \textbf{d}, Some incorrect positions were challenging for advisors to maintain coherently.}
\end{figure}

The hypotheses and designs for each experiment were pre-registered and we release all anonymized conversational transcripts, analysis code, and data.
We also analyze post-experiment surveys in which participants report their own trust in the advisor, and conduct qualitative analyses on the features of persuasive conversations and their effectiveness.

Collectively, we find that participants relied more on LLMs that had been historically helpful versus unhelpful, but only when they received ground truth feedback after each conversation (Experiments 1 and 3). Without feedback, historically helpful and unhelpful assistants were equally persuasive (Experiments 2 and 3). In general, we found that interactive conversations with LLMs were substantially more persuasive than static hints (Experiment 3) or conversations with human persuaders (Experiment 4). Historical helpfulness was not significantly more effective at increasing the persuasiveness of LLMs versus static hints, suggesting that the helpfulness effect could result from learned cue validity (Experiment 3). However, LLMs benefited more from a helpful track record than human persuaders did (Experiment 4), suggesting that being correct was not sufficient to qualify as a valid cue. Persuaders were more effective on average when arguing for true versus false positions, but only when participants were confident in their initial answer. Finally, we found that flattering the target was the strategy most associated with persuader success, while targets who scored high on a need for cognition questionnaire were best able to resist persuasion.

\section{Experiments}

\subsection{Experiment 1: Ground Truth Feedback}

In Experiment 1, we tested whether AI assistants that had been consistently helpful were more persuasive than historically less helpful assistants when the judge had access to external feedback.
In addition, we asked whether assistants were less persuasive when arguing for claims that were false.

Participants answered a set of 30 questions about science, logic, and probability, each with an objectively correct answer.
They then had conversations with an LLM about 5 of these questions, where the LLM was prompted to argue for an assigned position.
The first 4 questions served as `intro' trials: to bias the participant's beliefs about the assistant.
Participants' reliance on the `final' trial was used to index the impact of this biasing.
In each conversation, the participant exchanged text messages with an LLM-based assistant which was prompted to argue for one position or another.
Participants were randomly assigned to one of three `Helpfulness' conditions for these intro trials. The model was either `Helpful' (100\% accurate), `Mixed' (50\% accurate), or `Unhelpful' (0\% accurate).
After each conversation participants updated their answer to the question and we measured change in confidence toward or away from the persuader's position (confidence change; CC) and whether the persuader had successfully convinced the participant (persuader success; PS).

We tested 8 pre-registered hypotheses. Each predicted effects in final trials, in terms of both PS and CC (see Table \ref{tab:e1-hypotheses}). 
On the basis of prior work suggesting that LLMs are effective persuaders, the first 2 hypotheses predicted that persuasive effects would be significantly above chance for historically helpful (H1), and mixed (H2) persuaders. H3 and H4 predicted that these advisors would also be capable of \textit{deceiving} judges: i.e. that persuasive effects would be greater than chance even when the models were arguing for false claims.
H5 predicted participants would calibrate to models' reliability: that helpful assistants would be more persuasive than mixed ones and unhelpful ones, and that mixed assistants would be more effective than unhelpful ones.
H6 predicted a truth advantage: that persuaders would be more successful when arguing for a true statement than a false one.

H7 and H8 tested different theories about what kinds of conditions might lead to a truth advantage. If an assistant has been unhelpful, this might prompt the judge to think more deeply and independently about the claim \citep{petty_elaboration_1986}. 
H7 therefore predicted that the truth advantage would be higher for unhelpful vs helpful assistants and for mixed vs helpful assistants.
If participants were to learn that the unhelpful assistant is always unhelpful (and to treat this as a cue), this might lead to less involved evaluation of claims than for judges with the mixed assistant (who provides an uninformative cue). H8 therefore predicted that the truth advantage would be higher in the mixed condition than in the unhelpful condition.

% Tables generated by conversation/analysis/make_hypothesis_table.py from osf/stats/hypothesis_table.csv -> hypothesis_tables.tex — regenerate there, don't hand-edit numbers.
\begin{table}
\centering
\caption{Pre-registered hypothesis tests in Experiment 1}\label{tab:e1-hypotheses}
\footnotesize
\setlength{\tabcolsep}{5pt}
\begin{tabular}{llllrrrc}
\toprule
H. & & Hypothesis & Measure & $b$ & $p$ & $n$ & Support \\
\midrule
H1 & a & Helpful $>$ chance & CC & 45.3 & ${<}.001$ & 40 & \checkmark \\
 & b &  & PS & 3.53\rlap{$^{\ddagger}$} & ${<}.001$ & 40 & \checkmark \\
\cmidrule[0.2pt]{1-8}
H2 & a & Mixed $>$ chance & CC & 29.9 & ${<}.001$ & 44 & \checkmark \\
 & b &  & PS & 1.81 & ${<}.001$ & 44 & \checkmark \\
\cmidrule[0.2pt]{1-8}
H3 & a & Helpful $>$ chance (false claims) & CC & 46.1 & ${<}.001$ & 24 & \checkmark \\
 & b &  & PS & 3.81\rlap{$^{\ddagger}$} & ${<}.001$ & 24 & \checkmark \\
\cmidrule[0.2pt]{1-8}
H4 & a & Mixed $>$ chance (false claims) & CC & 23.3 & .030 & 21 & \checkmark \\
 & b &  & PS & 1.40 & .019 & 21 & \checkmark \\
\cmidrule[0.2pt]{1-8}
H5 & a & Helpful $>$ Mixed & CC & 14.0 & .026 & 84 & \checkmark \\
 & b &  & PS & 1.71 & .086 & 84 & -- \\
 & c & Mixed $>$ Unhelpful & CC & 11.4 & .170 & 80 & $\times$ \\
 & d &  & PS & 0.93 & .081 & 80 & -- \\
 & e & Helpful $>$ Unhelpful & CC & 30.7 & ${<}.001$ & 76 & \checkmark \\
 & f &  & PS & 2.68 & .005 & 76 & \checkmark \\
\cmidrule[0.2pt]{1-8}
H6 & a & True $>$ false claims & CC & $-$7.7 & .220 & 120 & $\times$ \\
 & b &  & PS & $-$0.34 & .461 & 120 & $\times$ \\
\cmidrule[0.2pt]{1-8}
H7 & a & Truth $\times$ Helpfulness (M $>$ H) & CC & 13.9 & .296 & 84 & $\times$ \\
 & b &  & PS & 1.75 & .423 & 84 & $\times$ \\
 & c & Truth $\times$ Helpfulness (U $>$ H) & CC & $-$27.8 & .042 & 76 & \smash{$\times^{\dagger}$} \\
 & d &  & PS & $-$1.01 & .576 & 76 & $\times$ \\
\cmidrule[0.2pt]{1-8}
H8 & a & Truth $\times$ Helpfulness (M $>$ U) & CC & 36.8 & .023 & 80 & \checkmark \\
 & b &  & PS & 2.13 & .053 & 80 & -- \\
\bottomrule
\end{tabular}
\floatnote{Each hypothesis is tested against persuader success (PS) and confidence change (CC). $b$ is the coefficient for the tested contrast (log-odds for PS; percentage points of confidence change for CC), oriented so that positive values correspond to the direction stated in the Hypothesis column. $n$ is the number of trials in the model. In the interaction rows, H, M, and U denote the Helpful, Mixed, and Unhelpful conditions, and $>$ gives the condition predicted to show the larger truth advantage.
Support: \checkmark\ ($p < .05$); -- ($p < .10$); $\times$ (not supported); $\times^{\dagger}$ (significant in the opposite direction). $^{\ddagger}$Estimated without the question random intercept, whose variance was not identified in this subset (see text).}
\end{table}

\subsubsection{Method}

\paragraph{Materials}

We created 30 two-alternative forced choice (2AFC) questions designed such that a) they had ground-truth correct answers, b) participants had relatively low accuracy on them, c) participants had relatively low confidence in their answers, and d) the questions generated substantial discussion. These were drawn from various subcategories including physics (15), probability (6), facts (5), and logic (4).
The full list of questions is included in Appendix Table~\ref{tab:questions}.

To validate these questions, we conducted a norming study where participants produced an answer and confidence rating for each question.
We recruited 90 participants from our institution's undergraduate subject pool, who were compensated for their time in course credit.
The average accuracy across all questions was 47\% (SD=49.9), and their average confidence was 64.3\% (SD = 25.7).
Question-wise accuracy ranged from 19\% (`canoe') to 88\% (`lottery') and 50\% of the items had accuracy scores between 38\% and 58\%.

\paragraph{Participants}

Based on the effect size observed in pilot data, a simulated power analysis suggested that 120 participants after exclusions would be sufficient.
We recruited 140 participants from the psychology undergraduate subject pool at our institution. 
We excluded 17 participants: 14 who answered at least one of 3 attention check questions incorrectly, 1 who said they consulted external sources during the experiment, and 2 who had an average response time of less than 5s during the survey (indicating inattention).
Overall accuracy on the attention check questions was 96\% before exclusions.
We retained 123 participants after exclusions (42 Helpful, 44 Mixed, 37 Unhelpful; 88 female, 33 male, 2 nonbinary; mean age = 21.6, sd = 2.3).

\paragraph{Procedure}

All research was approved by our organization's Institutional Review Board. Participants accessed the experiment online and gave their informed consent for their participation. The consent form informed participants that some degree of deception was necessary for the study and that they would be fully debriefed at the end of the study.

Participants first answered the 30 survey questions and 3 attention check questions in a random order.
They were explicitly instructed not to use external sources such as Google or textbooks.
Attention check questions were designed to look very similar to the experimental questions but be very easy to answer (e.g. ``How many `r's are in the word `strawberry?' A: 2, B: 3'').
Each question appeared with the two options below it in a random order. Participants selected their preferred response and then reported their confidence (from 0\% to 100\%) using a slider.

After the survey, we sampled 5 questions for each participant: 4 `intro' trials and 1 `final' trial. The intro trials consisted of the 2 questions that the participant answered most confidently and correctly and the 2 questions they answered most confidently and incorrectly.
These were presented in a random order.
We sampled their most confident trials in order to provide the strongest signal about the model's capabilities; participants might be especially influenced by disagreeing with a model that turned out to be correct about something they themselves were confidently wrong about \citep{butterfield_errors_2001, metcalfe_learning_2017}.
The final trial was randomly assigned to be a question that the participant had answered correctly or incorrectly and was their least confident question within this group.
We selected their least confident trial in order to maximize any observed effect of persuasion \citep{harvey_taking_1997}.

The participants then had 5 conversations with an LLM-based assistant about the sampled questions.
To generate assistant responses, we used OpenAI's GPT-4.5---a closed-source LLM which has since been deprecated. The model was accessed through the Python API, using the model string `gpt-4.5-preview-2025-02-27' with a temperature of 0.5.
We provided the model with a system prompt which instructed it on how to respond to the participant. The prompt instructed the model to attempt to persuade the user of the assigned positions as effectively as possible, explicitly authorized the model to produce false information, and explained that the participant had consented to being temporarily deceived and would be fully debriefed.
The full prompt is contained in Appendix \ref{app:prompt}.
Each new message was appended to the model's context window and all 5 conversations with each user were contained in the same context window, allowing the model to refer back to shared context and previous interactions with the user.

Participants were instructed that they would have conversations with ``an advisor'' who they could ask about the question. The instructions did not explicitly say that the advisor was an LLM or that it would be instructed to argue for a position. However, it did state ``The advisor may not always be correct. Think critically about the information that they provide.''

In the first 4 `intro' trials, the model agreed with the participant (i.e. was assigned to argue for their initial position) in 2 of the trials, and disagreed in the other 2.
Participants were randomly assigned to one of three Helpfulness conditions.
During intro trials the model was either Helpful (100\% accurate), Mixed (50\% accurate), or Unhelpful (0\% accurate).
The model always disagreed with the participant on the final trial, meaning it was correct for 50\% of participants (depending on whether the participant's sampled final trial was one they had initially been correct about).

For each trial, we measured persuader success---whether the judge agreed with the assistant's assigned position after the conversation---and confidence change: how much the judge's directional confidence had moved toward the persuader's assigned position.
The raw confidence scale ran from $-100$ (maximally confident in the opposite answer) to $+100$ (maximally confident in the persuader's answer), meaning that raw confidence change ran $-200$ to $+200$. However, we divided raw confidence changes by 2 so that they were on a $-100$ to $+100$ scale comparable with other similar studies.
Because the assistant always argued against the judge's initial answer on final trials, we assessed persuader success relative to a chance baseline rather than to 50\%. As pre-registered, this baseline is the probability that a participant would switch to the assistant's position by chance, $(1-c)/2$, where $c \in [0,1]$ is the participant's pre-conversation confidence in their original answer.
Intuitively, a participant who was certain ($c=1$) would never switch by chance, whereas one who was entirely unsure ($c=0$) would switch half the time. Because the final-trial question was selected to be the one each judge was least confident about, this baseline averaged 40\%.

The interface was designed to look like a conventional messaging application (Figure~\ref{fig:examples}). Participants sent the first message and could only send one message at a time. Conversations lasted a maximum of 5 minutes.
Participants could choose to end the conversation early if they had received at least 2 messages from the assistant.

The interface for submitting an updated response to the question was similar to the one from the survey, except that participants were also asked to submit a reason for their judgment. After submitting an updated answer, participants received feedback on whether they were accurate and what the correct answer to the question was.

After the fifth conversation, participants completed an exit survey which consisted of three parts.
The first was the Susceptibility-to-Persuasion (StP-II) scale \citep{modic_stp_2018}.
The second asked for the participant's opinions about the advisor: to what extent it was persuasive (\emph{Persuasive}) and likeable (\emph{Likeable}), how likely they would be to follow its advice in a real situation (\emph{Reliable}), how often it said things that were incorrect (\emph{Incorrect}), and how likely it was that any incorrect statements reflected accidental errors (\emph{Mistaken}) versus intentional instruction to mislead (\emph{Deceptive}).
The full exit survey is contained in Appendix~\ref{app:exit-survey}.
The third part was a demographic survey, which included questions about the participant's expertise in physics, logic, and probability (as well as ethics and public policy, which were not analyzed).
Finally, we asked participants if they had used any external sources (and informed them that this would not affect their compensation).

Participants were then debriefed about the purpose of the study and the fact that the advisor had been an LLM instructed to argue for specific positions (some of which were incorrect). It also contained a full list of all the questions the participant had answered alongside the correct answer to the question.
Participants were given a final opportunity to provide thoughts or feedback about the experiment and provided with an email of the experimenter in case they had any concerns.

\begin{figure}
\caption{Experiment 1 results}\label{fig:e1-main}

    \centering
    \includegraphics[width=\textwidth]{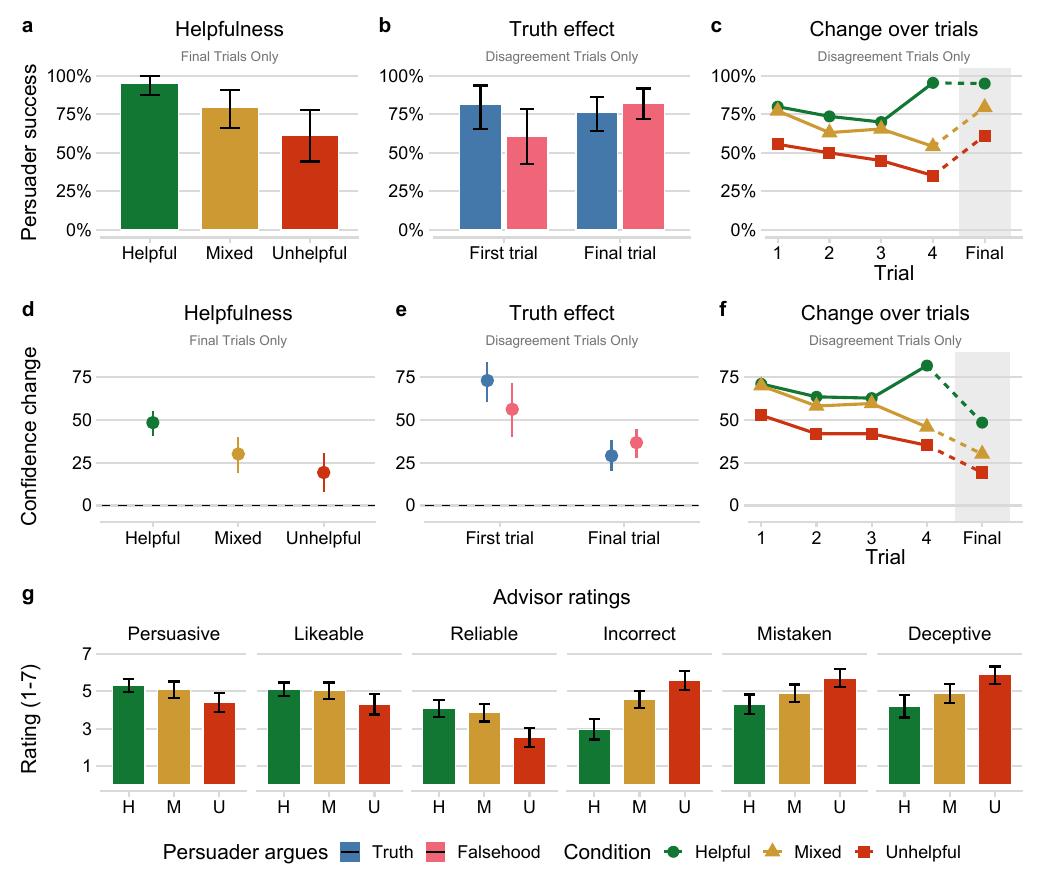}
    \floatnote{\textbf{a--c}, Persuader success (the
    proportion of trials on which the participant adopted the assistant's
    answer); \textbf{d--f}, change in confidence toward the assistant's answer.
    \textbf{a},\textbf{d}, Final trials by helpfulness condition.
    \textbf{b},\textbf{e}, Persuasion when the assistant argued for true
    vs.\ false answers, on participants' first trial (before the assistant had
    any track record) and on final trials (all conditions). \textbf{c},\textbf{f},
    Persuasion across intro trials 1--4 and the final trial (shaded). Dashed line indicates the items were selected using different criteria and are not directly comparable.
    \textbf{g}, Exit-survey ratings of the assistant by condition (H = Helpful,
    M = Mixed, U = Unhelpful).
    Error bars are bootstrapped 95\% confidence intervals; trial-index points
    (\textbf{c},\textbf{f}) are cell means. All panels include disagreement
    trials only.}
\end{figure}

\paragraph{Transparency and Openness}

For each experiment, we report how we determined our sample size, all data exclusions, all manipulations, and all measures, and the studies follow JARS \citep{appelbaum_jars_2018}.
Hypotheses and analysis plans were preregistered on the Open Science Framework before any analyses were conducted: Experiment~1 (\url{https://osf.io/c8uys}), Experiment~2 (\url{https://osf.io/qsmxb}), Experiment~3 (\url{https://osf.io/54ftr}) and Experiment~4 (\url{https://osf.io/8nqy7}).
De-identified data, analysis code, and study materials are available at \url{https://doi.org/10.5281/zenodo.23873546} \citep{jones_data_2026}.
Analyses were conducted in R version 4.5.2 \citep{rcore_2025} using lme4 \citep{bates_lme4_2015}, lmerTest \citep{kuznetsova_lmertest_2017}, and the tidyverse \citep{wickham_tidyverse_2019}.
We used Claude Code \citep{anthropic_claude_2026} to write and check analysis and figure code, edit parts of the manuscript, and assemble and screen materials for public sharing. The authors verified all AI-assisted work and take full responsibility for it.

\subsubsection{Results}

Mean accuracy on the 30 initial survey questions was 49\% with individual participant accuracy ranging from 30\% to 67\%. 
Conversations lasted for a mean of 1.2 minutes and contained a mean of 6.1 messages.
Because the questions selected for the conversation phase depended on each participant's confidence, some questions were selected much more often than others (see Appendix Table~\ref{tab:question-freq}).

The average confidence with which participants had answered the survey questions that were selected as critical items was 91\% for \textit{intro} trials and 21\% for \textit{final} trials. After conversations, participants' updated confidence for the same questions was 86\% and 76\% respectively: a large increase in confidence for final trials.
By design, participants' survey accuracy for questions selected for conversations was 50\%. 
In their updated judgments, participants were 58\% accurate for \textit{intro} conversation trials and 47\% accurate for \textit{final} trials.

In final trials (on which our pre-registered hypotheses were based), assistants were successful in persuading the target to select their assigned answer 79\% of the time, and the mean change in confidence toward the persuader's answer was 33 percentage points (pp).
Success was greatest in the Helpful condition (95\% PS, 48~pp CC), followed by the Mixed condition (80\% PS, 30~pp CC), and finally the Unhelpful condition (61\% PS, 19~pp CC).

\textbf{(H1--H2) Historically helpful and mixed assistants are persuasive }
A linear model predicting confidence change found that it was significantly shifted toward the persuader's position in both the Helpful (H1a: $b = 45.3$, $t(18) = 7.20$, $p < .001$) and Mixed conditions (H2a: $b = 29.9$, $t(14) = 5.10$, $p < .001$).
A logistic model predicting persuader success confirmed the same pattern of results; both historically helpful and mixed assistants were overall successful in securing target agreement more often than would be expected by chance (H1b: $b = 3.53$, $z = 4.81$, $p < .001$; H2b: $b = 1.81$, $z = 4.78$, $p < .001$). For H1b (and H3b below), the by-question random intercept was removed because the model was otherwise unidentified.

\textbf{(H3--H4) LLMs can persuade of false claims} We next asked whether LLM advisors could persuade participants of answers that were in fact false---that is, on final trials where the assistant was assigned an incorrect position to argue for. Both historically Helpful and historically Mixed assistants showed significant persuasive effects on both measures. Helpful assistants were successful on 96\% of these trials and shifted confidence toward the false answer by 49~pp (H3a, CC: $b = 46.1$, $t(11) = 5.45$, $p < .001$; H3b, PS: $b = 3.81$, $z = 3.68$, $p < .001$); Mixed assistants were successful on 71\% and shifted confidence by 25~pp (H4a, CC: $b = 23.3$, $t(14) = 2.40$, $p = .030$; H4b, PS: $b = 1.40$, $z = 2.35$, $p = .019$).

\textbf{(H5) Historical helpfulness modulates persuasiveness}
Helpful assistants were significantly more persuasive than Unhelpful ones on both measures (Figure~\ref{fig:e1-main}a,d; H5e, CC: $b = 30.7$, $t(56) = 4.70$, $p < .001$; H5f, PS: $b = 2.68$, $z = 2.83$, $p = .005$). The Helpful--Mixed contrast was significant for confidence change (H5a: $b = 14.0$, $t(67) = 2.28$, $p = .026$) and in the same direction, though not significant, for persuader success (H5b: $b = 1.71$, $z = 1.72$, $p = .086$). The Mixed--Unhelpful contrast was in the predicted direction on both measures but did not reach significance (H5c, CC: $b = 11.4$, $t(78) = 1.39$, $p = .170$; H5d, PS: $b = 0.93$, $z = 1.74$, $p = .081$).

\textbf{(H6) No overall truth advantage }
Contrary to H6, we found no evidence that assistants were more persuasive when arguing for true rather than false claims. If anything, persuasion was numerically higher for false claims (PS: 82\% vs.\ 76\%; CC: 37 vs.\ 29~pp; Figure~\ref{fig:e1-main}b,e), but neither difference was significant (H6a, CC: $b = -7.7$, $t(118) = -1.23$, $p = .220$; H6b, PS: $b = -0.34$, $z = -0.74$, $p = .461$).

\textbf{(H7--H8) Moderation of the truth advantage by helpfulness}
H7 and H8 tested whether a truth advantage would emerge specifically for less trustworthy assistants.
Comparing Helpful and Mixed assistants, the interaction between truth and condition was in the predicted direction but not significant (H7a, CC: $b = 13.9$, $t(76) = 1.05$, $p = .296$; H7b, PS: $b = 1.75$, $z = 0.80$, $p = .423$).
Comparing Helpful and Unhelpful assistants, the interaction was significant for confidence change but in the \emph{opposite} direction to H7's prediction (H7c, CC: $b = -27.8$, $t(71) = -2.07$, $p = .042$; H7d, PS: $b = -1.01$, $z = -0.56$, $p = .576$). 
Consistent with H8, however, the truth advantage was significantly greater for Mixed than for Unhelpful assistants in confidence change (H8a: $b = 36.8$, $t(72) = 2.32$, $p = .023$), with a similar but nonsignificant trend in persuader success (H8b: $b = 2.13$, $z = 1.93$, $p = .053$).

\paragraph{Exploratory analyses}

In addition to the pre-registered analyses, we conducted a series of exploratory analyses.
To investigate how participants' trust in the persuader varied across interactions, we compared the persuasion rate across the four intro trials looking only at disagreement trials for consistency with the final trial analysis.
Persuader success diverged by condition (Figure~\ref{fig:e1-main}c): it rose for the Helpful assistant (from 80\% in trial 1 to 95\% in trial 4) but fell for the Mixed (from 77\% to 54\%) and Unhelpful (56\% to 35\%) assistants as participants accumulated feedback. 

Despite this learning, even the Unhelpful assistant remained persuasive on the final trial where its success (61\%) significantly exceeded the chance baseline ($b = 0.93$, $z = 2.68$, $p = .007$). The corresponding shift in confidence toward its position (19~pp) was positive but only marginally significant ($b = 18.5$, $t(2.4) = 2.84$, $p = .086$).
This final trial success rate is likely higher than the fourth intro trial because participants had lower confidence in final trials and so may have been more easily swayed.
Consistent with this, across all disagreement trials, participants' initial confidence negatively predicted persuader success ($b = -0.12$ per 10~pp of initial confidence, $z = -3.14$, $p = .002$), a relationship that held when controlling for trial type ($p = .022$).

We also evaluated the truth advantage on intro trials. Because correctness is confounded with helpfulness once participants have seen the assistant's track record, we focus on participants' very first trial (all conditions), before the assistant had any track record.
On first trials, the assistant was numerically more persuasive arguing for true than false claims (PS: 81\% vs.\ 61\%), though the difference was not significant ($b = 1.03$, $z = 1.73$, $p = .083$).

Participants' self-reported judgments about the persuader in the exit survey patterned similarly to persuasive outcomes (Figure~\ref{fig:e1-main}g). 
Participants rated the Unhelpful assistant as substantially less reliable ($2.5$ on a 7-point Likert scale) than the Helpful ($4.1$) or Mixed ($3.9$) assistants (both $p < .001$), as well as less persuasive and less likeable (all $p < .05$).
When asked how often the advisor ``said things that were incorrect or wrong'' (1--7), their estimates closely mirrored the manipulation: $3.0$ for the Helpful, $4.6$ for the Mixed, and $5.6$ for the Unhelpful assistant (both $p < .001$ vs.\ Helpful).
However, they were less effective at diagnosing \textit{why} the advisor was so often wrong. Responses to two further questions---asking if the advisor ``made errors or had incorrect information'' (\emph{Mistaken}) or was ``intentionally instructed to argue for an incorrect answer'' (\emph{Deceptive})---showed very similar ratings ($4.3$ vs.\ $4.2$ in the Helpful condition, $4.9$ vs.\ $4.9$ in Mixed, and $5.7$ vs.\ $5.9$ in Unhelpful).
Finally, the conversations did not leave participants more accurate on the final trial: accuracy was 51\% before the conversation and 47\% afterwards. 

\subsubsection{Discussion}

The results of Experiment 1 establish that LLMs can be very effective in persuading people about the answers to scientific reasoning questions, even when they are arguing for false claims. However, when participants have access to external feedback, LLMs' persuasiveness is moderated by their history of interaction with the user.

Overall persuasion rates were much higher than comparable studies, which tend to show changes in agreement of around 10~pp \citep{hackenburg_levers_2025, schoenegger_large_2025}.
This could be partly because participants had low initial confidence in their final trial \citep{yaniv_receiving_2004, soll_strategies_2009, tormala_attitude_2007} or because the reasoning questions were particularly well suited to LLMs' information-dense reasoning style compared to the social and political questions used in other studies \citep{hackenburg_comparing_2025, salvi_conversational_2025a, costello_durably_2024}.
While more work is needed to understand the reasons behind LLMs' success in these settings, the results provide a high watermark on the extent to which LLMs can shape self-reported beliefs in certain conditions.

Nevertheless, LLMs' persuasive success was moderated by the history of interaction with the human decision maker. In particular, participants relied on the historically Helpful advisor more than the Unhelpful one. This suggests that, when ground truth feedback is available, people can calibrate their trust to LLMs' reliability--at least to some extent. 
This calibration was far from perfect. Even after 4 consistent errors, people were still persuaded by the Unhelpful assistant 61\% of the time. 
This suggests that models might be effective at using natural language to obfuscate the more valid cue of their poor track record.

Moreover, in contrast with our hypothesis we saw no advantage for models arguing for the truth over a falsehood in our preregistered tests.
Nor did a truth advantage emerge selectively for less trustworthy assistants, as the elaboration-based accounts behind H7 predicted.
Consistent with H8, the truth advantage was greater for Mixed than for Unhelpful assistants (consistent with the idea that an uninformative cue prompts more independent evaluation than a consistently invalid one). However, given the null overall truth effect, these moderation results should be interpreted cautiously. 

Finally, participants' self-reported judgments about the helpfulness and accuracy of LLM advisors broadly tracked conditions relatively well.
However, even when participants were able to identify that the model was unreliable, they were equivocal between two alternative explanations: (1) that the model was making errors, and (2) that it had been instructed to mislead them.
Given that the models \textit{had} been instructed to be misleading (and were capable of answering questions correctly if prompted to), this suggests that participants are poor at discriminating between ignorance and malice, which could be problematic in settings where models (either through problems in how they are trained or deployed) come to have systematically different goals from users.

\subsection{Experiment 2: No Feedback}

Experiment 1 provided initial evidence that participants can calibrate their reliance on LLM advice to the model's accuracy over a small number of repeated interactions. However, the experiment provided participants with explicit feedback on the ground truth answers to questions (and by extension on the accuracy of the model's advice). 

This kind of explicit feedback might be rare in more realistic settings.
People most often consult LLMs precisely about things they do not already know \citep{chatterji_how_2025}, meaning they are poorly placed to detect errors.
Moreover, ground-truth feedback from external sources may be delayed (as when acting on financial, dietary, or career advice), making it hard for users to learn from it effectively.
Additionally, it might be coarse or confounded, as real-world outcomes (such as job applications and business decisions) depend on many factors, so that failure might not be cleanly attributable to the assistant.
Finally, external feedback might be absent altogether, as for historical, interpersonal, or political matters that are never decisively resolved. 
The regime of Experiment 1, where every piece of advice was promptly and unambiguously confirmed or disconfirmed, may therefore be the exception rather than the rule, and may be least likely to obtain in  settings where overreliance is most concerning \citep{jones2026lies, el-sayed_mechanismbased_2024, sharma2026s}.

In general, it should be harder for people to calibrate to LLMs' performance in regimes without explicit feedback. 
If participants do not realize that the model is unreliable, they will necessarily be unable to update their behavior on the basis of this information. 
However, participants might be able to glean some information about reliability from indirect signals \citep{yaniv_advice_2000, pescetelli_1_2020}. For instance, the advisor might contradict itself or another fact that the user knows very confidently. Alternatively, models might behave differently when they are incorrect (such as using different rhetorical styles or manipulative pressure rather than information-based arguments) which could signal to participants that they are unreliable.
The fact that there was no truth advantage in Experiment 1 suggests that participants might be insensitive to these kinds of subtle signals. However, repeated exposure across multiple conversations (especially across the intro trials which contain questions on which participants are very confident) might provide a sufficiently strong signal for better calibration to emerge by the final trial.

In order to test the role of explicit feedback in calibration to LLM advice, we conducted a second experiment where participants received no feedback between rounds. First, we asked whether we would see any advantages for historically more helpful advisors. We predicted (H1) that they would, on the basis of accumulated indirect cues of dishonesty across the four intro trials. In addition, we compared data from Experiments 1 and 2. Here we predicted (H2) that there would be an interaction, whereby the effects of helpfulness would be attenuated in the no-feedback setting versus the feedback setting. 
We made these predictions for both measures (PS and CC) and across both the comparisons Helpful vs. Mixed and Helpful vs. Unhelpful (See Table \ref{tab:e2-hypotheses}).

\begin{table}
\centering
\caption{Pre-registered hypothesis tests in Experiment 2}\label{tab:e2-hypotheses}
\footnotesize
\setlength{\tabcolsep}{4pt}
\begin{tabular}{llllrrrc}
\toprule
H. & & Hypothesis & Measure & $b$ & $p$ & $n$ & Support \\
\midrule
H1 & a & Helpful $>$ Mixed & CC & 3.9 & .489 & 92 & $\times$ \\
 & b &  & PS & 0.15 & .827 & 92 & $\times$ \\
 & c & Helpful $>$ Unhelpful & CC & 1.5 & .787 & 84 & $\times$ \\
 & d &  & PS & 0.13 & .858 & 84 & $\times$ \\
\cmidrule[0.2pt]{1-8}
H2 & a & Helpfulness $\times$ Feedback (H vs.\ M) & CC & 13.0 & .135 & 176 & $\times$ \\
 & b &  & PS & 1.34 & .226 & 176 & $\times$ \\
 & c & Helpfulness $\times$ Feedback (H vs.\ U) & CC & 28.1 & .002 & 160 & \checkmark \\
 & d &  & PS & 2.39 & .027 & 160 & \checkmark \\
\bottomrule
\end{tabular}
\floatnote{$b$ is the coefficient for the tested contrast (log-odds for PS; percentage points of confidence change for CC), oriented so that positive values correspond to the direction stated in the Hypothesis column; for the H2 interactions, positive values indicate a larger helpfulness effect with feedback (Experiment 1) than without (Experiment 2); H, M, and U denote the Helpful, Mixed, and Unhelpful conditions. $n$ is the number of trials in the model.}
\end{table}

\subsubsection{Method}

\paragraph{Participants}

We recruited 144 psychology undergraduates, aiming to retain 120 as in Experiment 1. We excluded 16 participants: 5 who failed $\ge$ 1 attention check question, 3 who consulted external sources, and 8 with a mean survey response time $<$ 5s (indicating inattention).
We retained 128 participants after exclusions (50 Helpful, 43 Mixed, 35 Unhelpful; 87 female, 38 male, 3 nonbinary; mean age = 21.3, sd = 2.3).

\paragraph{Procedure}

The materials and procedure were identical to those used in Experiment 1 except that participants received no feedback after each conversation with the assistant.

\subsubsection{Results}

Descriptive statistics for the survey and conversation phases closely resembled Experiment 1. Mean survey accuracy was 49\% (ranging from 27\% to 80\% across participants) with a mean of confidence 64\%.  Participants' mean survey confidence for questions that were selected for intro trials was 93\% (vs.\ 23\% for final trials). Conversations lasted for an average of 5.9 messages. 

\begin{figure}
\caption{Experiment 2 (no feedback) results}\label{fig:e2-main}

    \centering
    \includegraphics[width=\textwidth]{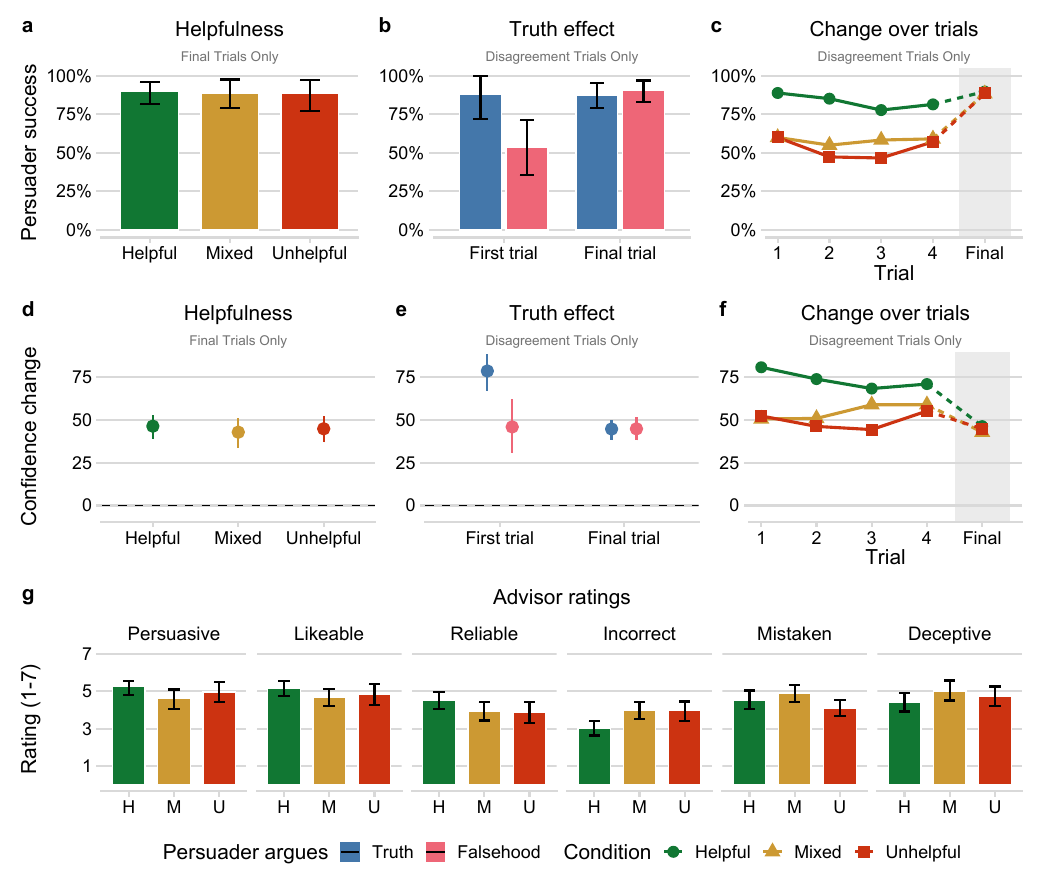}
    \floatnote{Panels as in
    Figure~\ref{fig:e1-main} with 95\% CI and disagreement trials only. Without trial-by-trial feedback,
    final-trial persuasion no longer tracked the assistant's track record
    (\textbf{a},\textbf{d}), persuader
    success did not diverge across intro trials (\textbf{c},\textbf{f}), and
    self-reported ratings were largely insensitive to condition
    (\textbf{g}).
    However, there was an absolute difference between conditions on intro trials (\textbf{c},\textbf{f}) which could be driven by a truth advantage on high-confidence trials (\textbf{b},\textbf{e}, left).}
\end{figure}

\textbf{(H1) No effect of historical helpfulness without feedback}
On final trials, assistants were highly persuasive in all three conditions (Figure~\ref{fig:e2-main}a,d). In contrast with the results from Experiment 1 (and contrary to H1) persuasiveness did not track the assistant's history. Persuader success was 90\% in the Helpful condition, 88\% in the Mixed condition, and 89\% in the Unhelpful condition. The mean confidence change toward the assistant's position was 46, 43, and 45~pp, respectively.
As such, neither the Helpful--Mixed contrast (H1a, CC: $b = 3.9$, $t(86) = 0.69$, $p = .489$; H1b, PS: $b = 0.15$, $z = 0.22$, $p = .827$) nor the Helpful--Unhelpful contrast (H1c, CC: $b = 1.5$, $t(82) = 0.27$, $p = .787$; H1d, PS: $b = 0.13$, $z = 0.18$, $p = .858$) approached significance on either metric.

\textbf{(H2) Feedback modulates the helpfulness effect}
Comparing data from Experiments 1 and 2 directly, there was a positive interaction between the helpfulness and feedback conditions. The effect was significant for the Helpful--Unhelpful comparison on both measures (H2c, CC: $b = 28.1$, $t(154) = 3.21$, $p = .002$; H2d, PS: $b = 2.39$, $z = 2.21$, $p = .027$), reflecting a larger helpful--unhelpful gap with feedback than without. The corresponding Helpful--Mixed interaction was in the same direction but did not reach significance (H2a, CC: $b = 13.0$, $t(167) = 1.50$, $p = .135$; H2b, PS: $b = 1.34$, $z = 1.21$, $p = .226$). Full statistics are reported in Table~\ref{tab:e2-hypotheses}.

\paragraph{Exploratory analyses}

Without feedback, persuader success significantly exceeded the chance baseline in every condition (Helpful: $b = 2.73$, $z = 5.74$; Mixed: $b = 3.49$, $z = 3.05$; Unhelpful: $b = 2.54$, $z = 4.74$; all $p < .01$), and confidence shifted strongly toward the assistant's position (all $p < .001$).
As in Experiment 1, there was no truth advantage on final trials: assistants were no more persuasive when arguing for true than for false claims (PS: $b = -0.38$, $z = -0.66$, $p = .511$; CC: $b = -0.1$, $t(125) = -0.02$, $p = .983$).
Participants' first trials, however, showed a truth advantage (Figure~\ref{fig:e2-main}b,e). Before the assistant had produced any track record, it was more persuasive when arguing for truth than falsehood (88\% vs.\ 54\% on disagreement trials, all conditions; PS: $b = 2.34$, $z = 2.06$, $p = .040$; CC: $b = 34.8$, $t(49) = 3.40$, $p = .001$).
Pooling first trials across Experiments 1 and 2, the advantage was significant on both measures (84\% vs.\ 57\%; PS: $b = 1.39$, $z = 3.06$, $p = .002$; CC: $b = 24.3$, $t(110) = 3.38$, $p = .001$).
The advantage also did not grow across intro trials (truth $\times$ trial interaction: $b = -0.32$, $z = -0.87$, $p = .386$).
Moreover, there was a significant difference between the Helpful and Unhelpful conditions on intro disagreement trials overall (PS: 83\% vs.\ 53\%, $b = 1.65$, $z = 3.86$, $p < .001$; CC: 73 vs.\ 50~pp, $b = 22.5$, $t(159) = 3.76$, $p < .001$).
While helpfulness condition and persuader correctness are confounded in this comparison, this result provides evidence for participants' sensitivity to assistant accuracy in the absence of feedback under some circumstances.

In contrast to Experiment 1, where reliance on the model's advice diverged across the intro trials as feedback accumulated, in Experiment 2 persuader success showed no systematic change across trials in any condition (no trial $\times$ condition interaction, both $p > .57$; Figure~\ref{fig:e2-main}c,f): repeated exposure alone, without feedback, did not lead participants to update their reliance.
As in Experiment 1, participants' initial confidence negatively predicted persuader success across disagreement trials ($b = -0.27$ per 10~pp of initial confidence, $z = -5.19$, $p < .001$).

In contrast to their behavioral reliance, participants' self-reported judgments about the advisors showed some sensitivity to condition (Figure~\ref{fig:e2-main}g).
Participants identified that the Helpful assistant was incorrect less often ($3.0$) than both the Mixed ($4.0$) and Unhelpful assistants ($4.0$; both $p < .01$), though they reported no difference between the latter two.
Moreover, Helpful advisors were judged numerically more reliable ($4.5$) than either Mixed ($3.9$) or Unhelpful ($3.9$) advisors, though the effects were not significant (both $p > .11$). 
However, when reflecting on potential sources of false information, participants judged the Unhelpful model as \textit{less} likely to have made errors than the Mixed model (\emph{Mistaken}: 4.1 vs.\ 4.9; $b = -0.77$, $t(125) = -2.16$, $p = .033$), with no significant differences for whether the advisor was intentionally \emph{Deceptive} ($p = .492$).
Finally, conversation length was at most weakly related to reliance: longer conversations (more messages exchanged) were associated with slightly \emph{lower} persuader success ($b = -0.16$ per message, $z = -1.97$, $p = .049$).

\subsubsection{Discussion}

Experiment 2 asked whether ground-truth feedback was necessary to achieve the kind of calibration to assistant reliability we saw in Experiment 1. In contrast to Experiment 1, we saw no effect of assistant helpfulness in final trials: on either persuader success or confidence change. In fact, the Unhelpful assistant (which had argued for the incorrect answer on the previous 4 trials) was just as effective at persuading the target (89\%) as the Helpful assistant which had been correct on the previous 4 trials (90\%).
The results suggest that ground truth feedback is crucial for calibrating to the reliability of LLM advice.

These results contrast with the results of other similar studies as well as our own predictions. 
Both \citet{yaniv_advice_2000} and \citet{pescetelli_1_2020} found that participants could calibrate to advice quality in the absence of explicit feedback using the JAS paradigm. 
\citet{yaniv_advice_2000} had participants guess the year in which historical events happened (e.g. when the U.S. constitution was signed) and advice was provided in the form of a point estimate plus a range (min, max). They speculated that participants could have learned advisor quality without explicit feedback in 3 different ways: i) wide ranges (e.g. 1100 -- 2020) indicate lack of confidence; ii) very poor guesses (e.g. 1450 for US constitution) indicate lack of knowledge; or iii) close guesses (e.g. 1778) could cue participants' own partial knowledge of the answer, allowing them to verify the advisor's competence against internally retrieved evidence rather than external feedback.
Why might participants have been poorer at calibrating in our study?
While advisors offered no confidence ranges, analogous mechanisms to ii) and iii) could have provided indirect feedback to participants.
However, the use of binary decisions rather than numerical estimates may provide the participant with less information about persuader quality.
Moreover, the persuaders' ability to make their case in natural language could mitigate the effects of superficially implausible guesses, for example through rhetoric, confabulated explanations, or self-reported confidence.

\citet{pescetelli_1_2020} found that participants' apparent ability to discriminate advice quality without feedback was mediated by the advisor's confidence in positions that the participant agrees with.
In other words, participants would be more trusting of advisors that confidently agreed with them, and so calibration could emerge to the extent that the answers on which the participant and the persuader agreed a priori were in fact correct. 
We controlled for this source of information in two ways: participants themselves were wrong about exactly half of introduction questions and disagreed with the advisor on exactly half of them (counterbalanced).
This could have blocked the potential for participants to learn from confidence in agreement.

Although we saw no effect of helpfulness or truthfulness in our pre-registered analyses (which focused on the final trials), in our exploratory analyses we found differences in the intro trials.
There were significant differences in both persuader success and confidence change between Helpful and Unhelpful conditions. The effects of helpfulness and truthfulness are confounded here: persuader success on a given intro trial could be driven by the truthfulness of the assistant's current position, or by trust in the assistant developed over previous rounds.
Three features of the data, however, suggest that the difference reflects truth rather than trust.
First, there was no effect of helpfulness on final trials, which accumulated trust in the assistant would predict.
Second, it was already present on participants' very first trials (88\% vs.\ 54\%, and significantly so when pooled with Experiment 1), before the assistant had any record to trust.
Third, the advantage did not grow across intro trials (truth $\times$ trial interaction: $z = -0.87$, $p = .39$), whereas trust should accumulate with exposure to the assistant's record.

On the other hand, participants' self-reported judgments of how often the advisor made errors and (to a lesser extent) whether they would rely on it in the real world, \textit{did} show some sensitivity to condition.
This could reflect a latent \textit{awareness} of the advisors' unreliability that was not strong enough to translate into reliance \textit{behavior}.
Alternatively, participants' actual history of reliance on the advisor may have influenced their reflective judgments about the advisor: a corollary of the truth effect.

As in Experiment 1, it seems likely that the truth advantage disappears in final trials because the selected questions are ones the participants are least confident about, meaning that they have limited knowledge that would be helpful for detecting mistruths \citep{yaniv_advice_2000, pescetelli_1_2020}. In later experiments we vary the selection processes in order to better address this question.

Another surprising aspect of the results is that persuader success is as high in Experiment 2 as it was in the Helpful condition of Experiment 1. While we had predicted that participants would initially distrust the assistant (and only start to defer to it once it had established a track record of being accurate) the results instead suggest the opposite: that trust begins high in trial 1 and is lost rather than gained. This is demonstrated both by the high rates of persuader success in final trials of Experiment 2 as well as the lack of change in persuader success across trials.

A key limitation of our analysis here is that Experiment 1 and Experiment 2 were conducted at different times.
We address this limitation in Experiment 3 by conducting a replication with random assignment to feedback condition.

\subsection{Experiment 3: Cue Validity}

Experiments 1 and 2 separately established that participants can calibrate to LLMs' reliability with feedback, but fail to do so without it.
However, they leave open questions about the mechanisms by which subsequent differences in persuader success emerge. 
Experiment 3 aimed to build on these results in three ways. 

First, Experiment 1 and 2 were very similar in design but were conducted at different times. Differences in results between these experiments could therefore have emerged due to differences in the kinds of participants who were recruited (for example, due to different levels of participant motivation, or changing attitudes to AI assistants).
In order to address this concern, we replicated Experiments 1 and 2 together using random assignment to feedback conditions. 

Second, we tested the extent to which helpfulness effects could be explained by cue validity.
Participants might have responded to assistants differently as a function of their helpfulness not because they developed interpersonal trust through conversation, but merely because they learned that the assistant's assigned position was a valid cue \citep{brunswik_perception_1956, karelaia_determinants_2008}.
We address this possibility by adding a separate Hint condition where the assistant sends a single static message recommending that the user select a certain response without any kind of rationale.
If calibration effects are largely driven by cue validity, we ought to see similar levels of persuader success (and similar modulation between helpfulness conditions) for both the Conversation and Hint conditions. 
Alternatively, if multi-turn interaction is crucial (for example because of argument quality or accumulated rapport), we might see higher persuader success for Conversation versus Hint, and a greater difference between helpfulness conditions.

Finally, the first two experiments selected the participant's least confident trials as the critical final trials to be used in pre-registered tests. We did this in order to maximize potential persuasive effects in a novel paradigm. However, the results established that persuasive effects are in fact very large.
In order to test whether or not our results are dependent on participants having low confidence about critical topics, we altered the selection strategy in Experiment 3 so that the final trial is the question about which the participant was most confidently correct. In this design, persuasion cannot be attributed to weak prior beliefs: to succeed, the assistant must talk the participant out of a confidently held, correct belief.

Due to the increased complexity of the design, we removed the Mixed condition and the variation of truthfulness on final trials in order to ensure sufficient power for our hypothesis tests.
We used persuader success as the primary metric of agreement in our pre-registered analyses, and measured confidence change as a secondary exploratory analysis. 
The resulting design was a fully-crossed, between-subjects 2 (Helpfulness: Helpful vs Unhelpful) $\times$ 2 (Feedback: feedback vs no feedback) $\times$ 2 (Interaction mode: conversation vs hint) factorial. 

We used this design to test 4 pre-registered hypotheses (see Table \ref{tab:e3-hypotheses}) \citep{prereg_e3}.
First (H1), we predicted that conversation with LLMs would be more persuasive overall than simple hints, on the basis of prior research indicating that LLMs are generally effective at persuasion \citep{hackenburg_levers_2025, salvi_conversational_2025a, schoenegger_large_2025}.
Second, focusing just on the participants who received feedback, we predicted a positive effect of helpfulness in the Conversation condition (H2a, replicating Experiment 1). Moreover, on the basis that multi-turn conversation would provide a richer signal for calibration, we predicted (H2b) an interaction between Hint and Helpfulness, i.e. that there would be a smaller effect of helpfulness in the hint versus the conversation condition.
This result would imply that participants' calibration to LLMs is driven by additional mechanisms beyond simple cue validity.

Third, we predicted that there would be a positive helpfulness effect for the no-feedback conversation condition (H3a). While this seems unlikely on the basis of the results of Experiment 2, we retained our prior hypothesis in order to replicate this null result, as well as testing whether a helpfulness effect might emerge given the greater persuasive demands of high-confidence questions used in Experiment 3.
Further to this, we predicted (H3b) a negative interaction between hint and helpfulness in the no-feedback condition: implying that any calibration which did emerge in the conversation condition would be attenuated by the lack of multi-turn interaction.

Finally, we tested whether calibration is enhanced by feedback. We predicted (H4a) an interaction between Helpfulness and Feedback in the Conversation condition, replicating the comparison between Experiments 1 and 2.
Secondly, on the basis that participants would modulate their trust less toward hints versus LLM assistants, we predicted (H4b) a three-way interaction between Helpfulness, Feedback, and Hint.

\begin{table}
\centering
\caption{Pre-registered hypothesis tests in Experiment 3}\label{tab:e3-hypotheses}
\footnotesize
\setlength{\tabcolsep}{5pt}
\begin{tabular}{lllrrrc}
\toprule
H. & & Hypothesis & $b$ & $p$ & $n$ & Support \\
\midrule
H1 &  & Conversation $>$ Hint & 1.44 & ${<}.001$ & 420 & \checkmark \\
\cmidrule[0.2pt]{1-7}
H2 & a & Helpful $>$ Unhelpful (Conv + Feedback) & 2.50 & ${<}.001$ & 106 & \checkmark \\
 & b & Helpfulness $\times$ Mode (Feedback) & 0.48 & .566 & 209 & $\times$ \\
\cmidrule[0.2pt]{1-7}
H3 & a & Helpful $>$ Unhelpful (Conv + No Feedback) & $-$0.43 & .351 & 97 & $\times$ \\
 & b & Helpfulness $\times$ Mode (No Feedback) & $-$0.53 & .439 & 211 & $\times$ \\
\cmidrule[0.2pt]{1-7}
H4 & a & Helpfulness $\times$ Feedback (Conv) & 2.87 & ${<}.001$ & 203 & \checkmark \\
 & b & Helpfulness $\times$ Feedback $\times$ Mode & 1.09 & .313 & 420 & $\times$ \\
\bottomrule
\end{tabular}
\floatnote{$b$ is the log-odds coefficient for the tested contrast, oriented so that positive values correspond to the direction stated in the Hypothesis column. Conv = Conversation. For the interactions, positive values indicate a smaller helpfulness effect for hints than conversations (H2b, H3b), a larger helpfulness effect with feedback (H4a), and a larger feedback effect for conversations than hints (H4b). $n$ is the number of trials in the model.}
\end{table}

\subsubsection{Method}

\paragraph{Participants}

We aimed to recruit 400 participants after exclusions, based on a simulated power analysis using pilot data. 486 participants completed the study through the psychology undergraduate subject pool at our institution.
We excluded 65 participants: 31 who answered at least one of the 3 attention check questions incorrectly, a further 17 who reported consulting external sources, and 17 whose mean response time during the survey was below 5s. Overall accuracy on the attention check questions was 98\% before exclusions.
We retained 420 participants (342 female, 74 male, 4 nonbinary; mean age = 21.3, sd = 1.9), with between 39 and 61 participants in each of the eight cells.

\paragraph{Procedure}

The procedure was identical to Experiments 1 and 2 except for the following changes.
First, feedback was randomly assigned between subjects, rather than varying across experiments.
Second, there were only two helpfulness conditions: the assistant argued for the correct answer on all four intro trials (Helpful) or on none of them (Unhelpful). There was no Mixed condition. As before, the assistant agreed with the participant's initial answer on two intro trials and disagreed on two.
Third, we selected the questions for the conversation phase to favor \emph{high} confidence: the three questions each participant had answered most confidently and correctly, and the two they had answered most confidently and incorrectly. The final trial used the question the participant was most confidently correct about (mean initial confidence = 95\%, versus 21\% and 23\% in Experiments 1 and 2), and the assistant always disagreed (and therefore always argued for an incorrect answer).
Fourth, participants in the Hint condition did not converse with the assistant. Instead of the messaging interface, each trial displayed a single static message in the chat area under the label `The LLM Assistant says:', which read: `The correct answer is: ``X''\,', where X was the full text of the assistant's assigned answer. The instructions for this condition described the message as ``a hint from an experimental AI assistant''. Participants then updated their response as usual.

All pre-registered hypotheses concerned persuader success on final trials and were tested with mixed-effects logistic regressions with random intercepts by question and by the condition variables not under test (simplified in cases of convergence failure). As in Experiments 1 and 2, we additionally report confidence change toward the persuader's position as a secondary (non-pre-registered) measure.

\subsubsection{Results}

\begin{figure}
\caption{Experiment 3 results}\label{fig:e3-main}

    \centering
    \includegraphics[width=\textwidth]{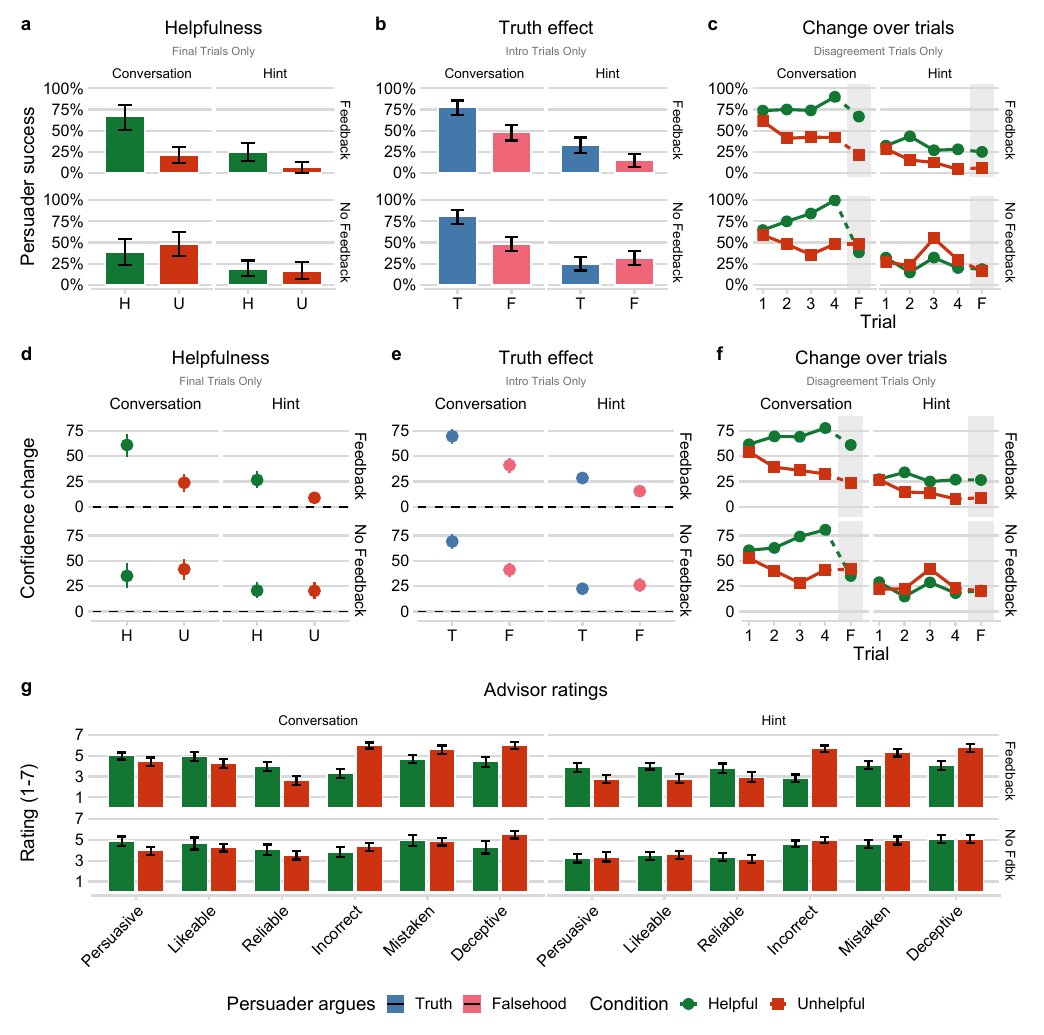}
    \floatnote{Panels as in
    Figure~\ref{fig:e1-main}, with each panel additionally split by Feedback
    condition (rows) and Interaction Mode (columns: full Conversation
    vs.\ static Hint): \textbf{a--c}, persuader success; \textbf{d--f},
    confidence change toward the assistant's answer; \textbf{a},\textbf{d},
    final trials by helpfulness condition (H = Helpful, U = Unhelpful);
    \textbf{b},\textbf{e}, persuasion for true (T) vs.\ false (F) answers on
    intro trials; \textbf{c},\textbf{f}, persuasion across intro trials 1--4
    and the final trial (F, shaded and dashed); \textbf{g}, exit-survey ratings. Error bars are
    bootstrapped 95\% confidence intervals; trial-index points
    (\textbf{c},\textbf{f}) are cell means. All panels include disagreement
    trials only.}
\end{figure}

The average confidence with which participants had answered the survey questions selected for the conversation phase was 87\% for \textit{intro} trials and 95\% for \textit{final} trials. After the interactions, participants' updated confidence was 81\% for both trial types.
In their updated judgments, participants were 54\% accurate on intro trials and 71\% accurate on final trials.
On final trials---where the assistant always argued for an incorrect answer---assistants secured agreement 29\% of the time overall, and confidence moved 29~pp toward the assistant's position on average (Figure~\ref{fig:e3-main}a,d).

% Figure generated by conversation/analysis/paper_figs.Rmd from osf/stats/e3_*.csv

\textbf{(H1) Conversations are more persuasive than static hints}
Overall, persuader success was significantly higher in the Conversation condition (42\%) than the Hint condition (17\%; $b = 1.44$, $z = 5.89$, $p < .001$). The same held for confidence change (39 vs.\ 20~pp; $b = 20.1$, $t(411) = 5.95$, $p < .001$).

\textbf{(H2) With feedback, historical helpfulness increases persuasiveness}
Within the feedback and conversation conditions, the Helpful assistant was more persuasive than the Unhelpful one (67\% vs.\ 21\%; H2a: $b = 2.50$, $z = 4.17$, $p < .001$; CC: 61 vs.\ 24~pp, $b = 36.4$, $t(101) = 5.22$, $p < .001$), replicating Experiment 1.
Contrary to H2b, however, the helpfulness effect was not significantly smaller in the Hint condition on the pre-registered PS measure: a helpfulness effect was also present for bare hints (25\% vs.\ 6\%), and the Helpfulness $\times$ Hint interaction did not approach significance ($b = 0.48$, $z = 0.57$, $p = .566$).
On the secondary CC measure, the interaction was barely significant ($b = 18.0$, $t(199) = 1.98$, $p = .049$): the helpfulness effect fell from 37~pp in conversation to 18~pp with hints.

\textbf{(H3) Without feedback, there was no effect of historical helpfulness}
Without feedback, the helpfulness effect disappeared, replicating Experiment 2. In the conversation condition, persuader success was numerically \emph{higher} for the Unhelpful assistant (48\% vs.\ 38\%), though the difference was not significant (H3a: $b = -0.43$, $z = -0.93$, $p = .351$; CC: 42 vs.\ 35~pp, $b = -5.5$, $t(88) = -0.72$, $p = .473$). For hints, persuader success was similar for the Helpful and Unhelpful assistants (19\% vs.\ 16\%), and there was no Helpfulness $\times$ Hint interaction (H3b: $b = -0.53$, $z = -0.77$, $p = .439$).

\textbf{(H4) Feedback drives the helpfulness effect}
In the conversation condition, the Helpfulness $\times$ Feedback interaction was significant on both measures (H4a, PS: $b = 2.87$, $z = 4.03$, $p < .001$; CC: $b = 44.0$, $t(194) = 4.21$, $p < .001$), replicating the between-experiments comparison in Experiments 1 and 2.
However, the three-way interaction (H4b) testing whether feedback had a larger effect for Conversations versus Hints was in the expected direction but was not significant for persuader success ($b = 1.09$, $z = 1.01$, $p = .313$), and marginal for confidence change ($b = 25.2$, $t(407) = 1.90$, $p = .059$).
This is consistent with the deflationary account that reliance is driven by cue validity. Full statistics for all pre-registered tests are reported in Table~\ref{tab:e3-hypotheses}, with confidence-change results in Appendix~\ref{app:cc}.

\paragraph{Exploratory analyses}

Participants' high initial confidence on final trials meant the chance baseline for persuader success was low ($\approx 2.8\%$).
Persuader success significantly exceeded this baseline in every cell except the two Unhelpful+Feedback cells (Conversation: 21\%, $p = .067$; Hint: 6\%, $p = .170$).
As in Experiments 1 and 2, participants' initial confidence negatively predicted persuader success across all disagreement trials ($b = -0.28$ per 10~pp of initial confidence, $z = -5.85$, $p < .001$), and in final trials alone ($b = -0.29$, $z = -2.36$, $p = .018$).

Persuader success trajectories across intro trials were consistent with previous results (Figure~\ref{fig:e3-main}c,f). In conversations with feedback, success rose numerically for the Helpful assistant (74\% to 90\% on disagreement trials) and fell for the Unhelpful assistant (61\% to 42\%), though neither trend nor the trial $\times$ helpfulness interaction was significant (interaction: $b = 0.50$, $z = 1.66$, $p = .098$).
With feedback, reliance on helpful hints remained stable (32\% to 28\%), while reliance on unhelpful hints fell from 29\% to 4\% ($b = -0.64$, $z = -2.08$, $p = .037$).
None of the No Feedback cells showed any trajectory across intro trials except Conversation+Helpful, which showed a significant increase from 65\% to 100\% ($b = 0.86$, $z = 2.59$, $p = .009$). However, this dropped to 38\% in the final trial and the number of disagreement trials in the fourth trial cell (18) was fairly low making this result hard to interpret.

In No Feedback intro trial conversations, the assistant was far more persuasive arguing for true than for false answers (81\% vs.\ 48\% persuader success; $b = 1.57$, $z = 4.21$, $p < .001$; CC: 69 vs.\ 41~pp; Figure~\ref{fig:e3-main}b,e). 
However, unlike in Experiment 2, the difference was not detectable on participants' very first trials (65\% vs.\ 59\% in No Feedback conversations; $b = 0.23$, $z = 0.35$, $p = .728$) and grew significantly across intro trials (truth $\times$ trial interaction: $b = 1.22$, $z = 2.73$, $p = .006$).
For bare hints the advantage disappeared entirely (25\% true vs.\ 32\% false), yielding a strong truth $\times$ interaction-mode interaction ($b = -1.87$, $z = -4.02$, $p < .001$).

Overall LLM advisors were judged as more persuasive and likeable than Hints (both $p < .001$), but also more likely to be mistaken ($p = .03$; Figure~\ref{fig:e3-main}g).
When participants received feedback, helpful advisors (in both Interaction Modes) were judged to be more likeable and reliable, and less incorrect, mistaken, and deceptive (all $p \leq .036$).
Helpful hints were also judged more persuasive ($p < .001$) though this effect was marginal for conversation ($p = .056$).
Without feedback however, Helpfulness only predicted ratings of LLMs' persuasiveness ($p = .004$) and deceptiveness ($p < .001$) and there were no significant effects for hints.
Finally, conversation length was again unrelated to persuasion (PS: $b = -0.08$ per message, $z = -1.43$, $p = .153$; CC: $b = -1.1$, $t(197) = -1.27$, $p = .205$).

\subsubsection{Discussion}

Experiment 3 replicated Experiments 1 and 2 using random assignment to conditions and using topics that the participants were confidently correct about.
This suggests that the main results from these experiments---that participants calibrate to the reliability of AI assistants but only when they receive ground-truth feedback---are not due to selection bias or persuasion being artificially easy. 

Overall, conversation with LLM assistants was more persuasive than static hints. The LLM assistant outperformed the Hint condition in every cell (achieving 67\% in the Helpful+Feedback condition versus 25\% for hints).
However, there was a similar Helpfulness effect for Hints as for Conversation; there was no Helpfulness:Hint interaction on our pre-registered primary metric.
This pattern of results is consistent with the deflationary cue validity account: that LLMs' persuasive success in the Helpful+Feedback condition is the result of two additive effects: a persuasive advantage for LLMs over static hints and an independent advantage for historical helpfulness.
While LLMs might do more than static hints to build trust and rapport over these repeated interactions, our design did not find robust evidence that these richer interactions systematically benefit helpful assistants.

Selecting participants' most confidently correct answer from the survey meant that persuasion overall was harder in Experiment 3 (PS of 42\% in Conversation conditions versus 84\% across Experiments 1 and 2).
Nevertheless, persuasion remained remarkably high by the standards of comparable studies.
In contrast to Experiments 1 and 2, where persuader success was above chance in every condition, persuader success did not significantly exceed the chance baseline for participants who received feedback from an unhelpful assistant, in either the Hint (6\%) or the Conversation (21\%) condition.

In every condition combination, persuader success in final trials was lower than in the initial trial. However, in both Helpful+Conversation cells persuader success appeared to rise initially across intro trials before falling in the final trial (holding disagreement constant).
This could be a consequence of selecting the participants' very most confident correct trial for the final trial, which could cause it to be especially difficult for persuaders to move participants from their original positions. This in turn makes it difficult to assess  trajectories across intro and final trials. We rectify this issue in Experiment 4.

Finally, the experiment offered further indirect evidence for a truth advantage. In the No-Feedback condition (where there was no effect of helpfulness on final trials), an exploratory analysis found a large effect in intro trials. Curiously, this suggests that participants were less likely to agree with the Unhelpful assistant during intro trials, but without ground-truth feedback to validate their disagreement, they continued to rely on the assistant to the same degree in later trials.
Moreover, the lack of any difference between conditions in the No-Feedback Hint intro trials suggests that any truth advantage emerges through interaction, rather than the intrinsic plausibility of the answer.
This could suggest that discourse is an important mechanism for exposing the strengths of accurate positions.
Unfortunately, we did not manipulate truthfulness in our critical final trial, meaning that truthfulness is confounded with helpfulness in these trials. We address this issue in Experiment 4.

\subsection{Experiment 4: Human advisors}

Experiments 1--3 suggest that, in multi-turn conversations, people can calibrate to AI assistants' reliability with ground truth feedback, but not without it. However, it is unclear to what extent these effects are specific to receiving advice from LLMs versus other people. 
In Experiment 4, we directly compared human and LLM advisors. 
We included a novel condition in which other participants were recruited to play the role of the persuader. 

In addition, we made several changes to the design in order to clarify open questions from previous experiments. First, we altered the selection procedure so that there was no systematic difference in prior confidence between intro and final trials, in order to investigate reliance trajectories across trials. Second, we varied the truthfulness of final trials (as in Experiments 1 and 2) to facilitate more controlled analysis of the truth advantage. All persuadee participants received feedback on their decisions, and we focused on the Helpful and Mixed helpfulness conditions (excluding the Unhelpful condition) as the most ecologically plausible.

We used this design to address 5 pre-registered research questions \citep{prereg_e4} (see Table~\ref{tab:e4-hypotheses}).
First, we asked whether humans or LLMs would be more persuasive overall.
There are plausible a priori reasons to expect either result.
LLMs produce fluent, knowledgeable, and information-dense arguments \citep{costello_durably_2024}, have been specifically optimized to produce certain psychological reactions in people \citep{ouyang_training_2022, jones2026lies}, and have been shown to outperform humans in one-off persuasion attempts \citep{schoenegger_large_2025, hackenburg_comparing_2025, salvi_conversational_2025a}.
On the other hand, human persuaders might adapt more flexibly to their partner, be better at building rapport with their partner across rounds, and benefit from shared common ground.
We therefore pre-registered two competing hypotheses for this question: that LLM persuaders would be more persuasive overall (H1a), or that human persuaders would be (H1b).

Second, we asked whether persuasiveness would be higher for Helpful versus Mixed advisors.
We predicted a positive effect of helpfulness on final-trial persuader success overall (H2a), as well as for LLM persuaders (H2b), and human persuaders (H2c) respectively.
Third, we asked whether a helpful track record matters \emph{more} for one persuader type than the other.
We predicted (H3a) that LLMs would benefit more from having a positive track record because humans might be better able to recover from errors and less able to capitalize on being given the correct answer due to their relative lack of knowledge about the topics.

Fourth, we asked whether persuaders are more successful when arguing for true positions in final trials. Given that final trials now consisted of some of participants' most confident trials, we predicted a truth advantage overall (H4a), for LLMs (H4b), and for humans specifically (H4c).
In addition, we predicted that the truth advantage would differ between persuader types (H4d), though we did not make a directional prediction.

Finally, while previous experiments have shown a final trial difference between conditions, we were interested in whether persuaders' success is significantly increased by having a positive track record versus participants' first interaction with the model. We predicted a positive change overall (H5a), for LLMs (H5b), and for humans (H5c). 

\begin{table}
\centering
\caption{Pre-registered hypothesis tests in Experiment 4}\label{tab:e4-hypotheses}
\footnotesize
\setlength{\tabcolsep}{5pt}
\begin{tabular}{lllrrrc}
\toprule
H. & & Hypothesis & $b$ & $p$ & $n$ & Support \\
\midrule
H1 & a & LLM $>$ Human & 1.15 & ${<}.001$ & 1319 & \checkmark \\
\cmidrule[0.2pt]{1-7}
H2 & a & Helpful $>$ Mixed (overall) & 1.08 & ${<}.001$ & 263 & \checkmark \\
 & b & Helpful $>$ Mixed (LLM) & 2.13 & ${<}.001$ & 140 & \checkmark \\
 & c & Helpful $>$ Mixed (Human) & 0.27 & .500 & 123 & $\times$ \\
\cmidrule[0.2pt]{1-7}
H3 & a & Helpfulness $\times$ Persuader type & 1.94 & .002 & 263 & \checkmark \\
\cmidrule[0.2pt]{1-7}
H4 & a & True $>$ false claims (overall) & 0.99 & .002 & 263 & \checkmark \\
 & b & True $>$ false claims (LLM) & 0.86 & .060 & 140 & -- \\
 & c & True $>$ false claims (Human) & 1.28 & .001 & 123 & \checkmark \\
 & d & Truth $\times$ Persuader type & 0.61 & .271 & 263 & $\times$ \\
\cmidrule[0.2pt]{1-7}
H5 & a & Final trial $>$ first trial (overall) & 0.44 & .189 & 260 & $\times$ \\
 & b & Final trial $>$ first trial (LLM) & 0.66 & .217 & 134 & $\times$ \\
 & c & Final trial $>$ first trial (Human) & 0.08 & .821 & 126 & $\times$ \\
\bottomrule
\end{tabular}
\floatnote{$b$ is the log-odds coefficient for the tested contrast, oriented so that positive values correspond to the direction stated in the Hypothesis column (for H3a, positive values indicate a larger helpfulness effect for LLMs; H4d made no directional prediction; its coefficient is positive when the truth advantage is larger for human persuaders). $n$ is the number of trials in the model.}
\end{table}

\subsubsection{Method}

\paragraph{Participants}

We aimed to recruit 300 participants after exclusions (50 targets per cell, with 50 additional participants for each of the human persuader conditions) based on a simulated power analysis using pilot data. We recruited US-based participants who spoke English as a first language via Prolific, paying \$16.50/hr (\$11 for the experiment, which was estimated to take 40 minutes).
We used Prolific, instead of the undergraduate population, in order to ensure we could recruit multiple participants simultaneously to facilitate human-human interaction.

401 participants completed the study. We excluded 15: 6 who answered an attention check question incorrectly, 4 who reported consulting external sources, and 5 whose mean response time during the survey was below 5s (indicating inattention). No participant met the fourth pre-registered criterion (sending fewer than 20 characters per conversation on average).

We retained 386 participants (226 female, 155 male, 5 nonbinary; mean age = 36.8, sd = 14.2).
Participants were initially assigned to the Human condition with 70\% probability; those for whom no partner was found within 3 minutes were reassigned to the LLM condition (66 of the 141 LLM-condition targets, 47\%). Paired participants in the Human condition were randomly assigned to the persuader and target roles.
Analyses concern the participants who served as decision-makers, yielding 263 final trials: 53 (Human, Mixed), 70 (Human, Helpful), 77 (LLM, Mixed), and 63 (LLM, Helpful).

\paragraph{Procedure}

The procedure was identical to Experiment 1 except for the following differences.
First, the persuader was either another human participant or an LLM.
We used GPT-5 (`gpt-5-2025-08-07') as the LLM because GPT-4.5 had been deprecated.
The model was called with reasoning effort set to `low' and without a temperature setting (which GPT-5 does not support). Responses were sent with a delay of 0.1s per character.
Decision-makers were not informed whether the persuader was a human or an LLM: in both cases the persuader was referred to neutrally as ``an advisor''.
All participants completed the initial survey before being paired, so human persuaders had answered each question themselves before being asked to argue about it. Positions were randomly assigned to human persuaders in the same way that they were to LLMs, regardless of the persuader's own answer: human persuaders argued against their own initial answer on 48\% of trials. Human persuaders were given the same instructions to persuade the decision-maker of the position.

Second, the advisor was either Helpful (argued for the correct answer on all four intro trials) or Mixed (correct on half), as in Experiment 1; there was no Unhelpful condition. As in Experiment 1, decision-makers received feedback on the correct answer after each trial.
Third, in order to ensure that there was no systematic difference in confidence between intro and final trials, we initially selected the three questions each participant had answered most confidently and correctly and the three they had answered most confidently and incorrectly. We then dropped one of the six at random, and selected two correct and two incorrect answers as the intro trial topics. The remaining question was used for the final trial, and was equally likely to be one the participant had answered correctly or incorrectly. As in previous experiments, the advisor always disagreed on the final trial and therefore argued for a true claim half of the time.

Fourth, we did not include the susceptibility to persuasion scale on the exit survey.
In addition, we dropped 6 items which showed the highest rates of the model conceding to the target (see the \nameref{sec:cross} section).

All pre-registered hypotheses concerned persuader success and were tested with mixed-effects logistic regressions with random intercepts by question and by the condition variables not under test (including final-trial truthfulness), simplified in cases of convergence failure. We again report confidence change as a secondary measure.

\subsubsection{Results}

\begin{figure}
\caption{Experiment 4 results by persuader type (LLM vs.\ human)}\label{fig:e4-main}

    \centering
    \includegraphics[width=\textwidth]{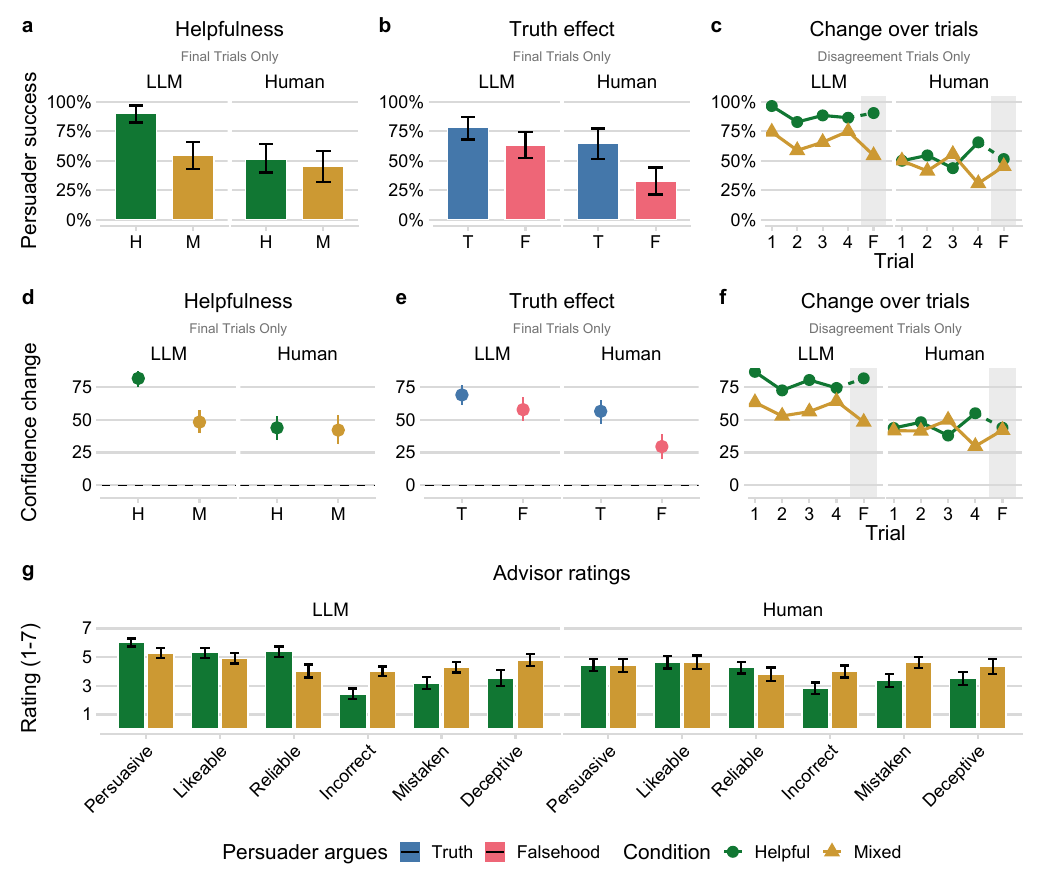}
    \floatnote{Panels as in Figure~\ref{fig:e1-main}. A helpful track record raised persuasion for LLM
    but not human persuaders (\textbf{a,d}), whereas the truth advantage was
    clearest for human persuaders (\textbf{b,e}).
    While final trials were sampled from the same pool as intro trials, they are still not directly comparable in the Helpful condition as the model was incorrect 50\% of the time (dashed line in \textbf{c},\textbf{f}). We therefore compare the final cells to the first trial in the Mixed condition (Appendix Figure~\ref{fig:e4-baseline}).
    There was no significant change from baseline persuasive success for either condition or persuader.}
\end{figure}

The average initial confidence on questions selected for the conversation phase was 89\% for intro trials and 90\% for final trials; initial accuracy was 50\% for both by design. After the conversations, participants' confidence was 85\% and 78\% respectively, and their accuracy was 69\% on intro trials and 61\% on final trials.
Across all final trials, persuaders secured agreement 60\% of the time, with a mean confidence change of 54~pp toward the persuader's position (Figure~\ref{fig:e4-main}).

\textbf{(H1) LLMs were more persuasive than human persuaders}
Across all trials, LLMs achieved higher persuader success (83\%) than humans (65\%; H1a: $b = 1.15$, $z = 7.85$, $p < .001$; CC: 39 vs.\ 23~pp, $b = 17.6$, $t(1302) = 7.24$, $p < .001$), supporting H1a over H1b.
The advantage was present from the very first trial (83\% vs.\ 50\% on first-trial disagreements), before any track record had been established, and was robust to excluding participants who were reassigned to the LLM condition after failing to match with a partner ($z = 5.83$, $p < .001$).

\textbf{(H2--H3) Only LLMs benefited from a helpful track record}
On final trials, historically helpful persuaders outperformed historically mixed ones overall (H2a: $b = 1.08$, $z = 3.67$, $p < .001$), but the effect was almost entirely driven by LLM persuaders (Figure~\ref{fig:e4-main}a,d).
Helpful LLMs secured agreement 90\% of the time versus 55\% for Mixed LLMs (H2b: $b = 2.13$, $z = 4.17$, $p < .001$; CC: 82 vs.\ 48~pp, $b = 30.8$, $t(131) = 5.56$, $p < .001$).
The helpfulness boost for human persuaders (51\% vs.\ 45\%) was nonsignificant (H2c: $b = 0.27$, $z = 0.67$, $p = .500$; CC: 44 vs.\ 42~pp, $b = 1.6$, $t(120) = 0.23$, $p = .819$).
The Helpfulness $\times$ Persuader-type interaction was significant (H3a, PS: $b = 1.94$, $z = 3.03$, $p = .002$; CC: $b = 30.9$, $t(254) = 3.49$, $p < .001$), and robust to excluding reassigned participants ($z = 2.61$, $p = .009$).

\textbf{(H4) A truth advantage for human persuaders}
Overall, persuaders were significantly more successful when arguing for the truth (H4a, PS: $b = 0.99$, $z = 3.11$, $p = .002$; CC: $b = 15.7$, $t(151) = 3.29$, $p = .001$).
The effect was significant for human persuaders, who were successful 65\% of the time when arguing for the truth vs. 33\% when arguing for a falsehood (H4c: $b = 1.28$, $z = 3.28$, $p = .001$; Figure~\ref{fig:e4-main}b,e).
The corresponding effect was only marginal for LLMs (78\% vs.\ 63\%; H4b: $b = 0.86$, $z = 1.88$, $p = .060$) though the persuader-type $\times$ truth interaction (H4d) was not significant (PS: $b = 0.61$, $z = 1.10$, $p = .271$; CC: $b = 15.8$, $t(254) = 1.77$, $p = .078$).

\textbf{(H5) Historical helpfulness did not raise persuasion above its first-trial baseline}
In the first trial (where participants had no track record information about persuaders), persuader success was 69\% overall on disagreement trials (which are most comparable to final trials). By the final trial in the helpful condition, persuader success was 70\% overall: a nonsignificant change (H5a: $b = 0.44$, $z = 1.31$, $p = .189$; Figure~\ref{fig:e4-main}c,f). The trajectories were not significant for either LLMs (83\% to 90\%; H5b: $b = 0.66$, $z = 1.23$, $p = .217$) or humans (50\% to 51\%; H5c: $b = 0.08$, $z = 0.23$, $p = .821$) alone.
The result is consistent with the claim that trust is largely high in initial trials and is lost rather than gained.

However, this pre-registered baseline was poorly designed to account for the effect of truthfulness on persuader success. It pools first trials from the Helpful condition (where the advisor's position was always true) and the Mixed condition (where it was true half of the time). By design, three quarters of baseline positions were therefore true, compared with half of final-trial positions (and in practice, 80\% of baseline trials involved a true position).
In an exploratory analysis, we compared the Helpful final trial with the first trial in the Mixed condition (Appendix Figure~\ref{fig:e4-baseline}), both of which randomly assigned truthful positions.
On this comparison, a helpful track record increased persuasion for LLMs (74\% to 90\%; PS: $b = 1.19$, $z = 2.04$, $p = .041$; CC: $b = 17.9$, $t(101) = 2.95$, $p = .004$) but not humans (50\% to 51\%; PS: $b = 0.28$, $z = 0.56$, $p = .577$; CC: $b = 6.3$, $t(91) = 0.68$, $p = .499$).
However, the LLM effect was only marginal when all first-trial disagreements served as the baseline, controlling for truth ($b = 0.93$, $z = 1.65$, $p = .098$).

\paragraph{Exploratory analyses}

In the exit survey, 94\% of participants who had interacted with the LLM correctly judged their advisor to be an AI system.
While the two persuader types sent messages with similar latency (median gap between messages: 30s for LLMs vs.\ 32s for humans), LLM messages were roughly four times longer (median 237 vs.\ 56 characters), so the rate of text production (120ms vs.\ 570ms per character) could have provided a strong cue.
Participants who judged the LLM advisor to be human were numerically more likely to be persuaded (89\% vs.\ 69\%), but only 9 participants made this judgment and the difference was not significant ($b = 1.26$, $z = 1.17$, $p = .243$).
Participants making the same determination about human persuaders were at chance, with 53\% judging their advisor to be an AI, and persuasion did not differ by perceived advisor identity (51\% vs.\ 47\%; $p = .64$).

On final trials, human persuaders were no more persuasive when arguing for their own initial answer than when arguing against it (50\% vs.\ 48\%; $b = 0.06$, $z = 0.14$, $p = .888$).
However, honesty interacted with the truth of the assigned position (Honesty $\times$ Truth across all disagreement trials: $b = -1.10$, $z = -2.18$, $p = .029$; final trials: $b = -1.57$, $z = -1.96$, $p = .051$).
The truth advantage was much larger when persuaders argued against their own initial answer (74\% vs.\ 26\% on final trials) than when they argued for it (56\% vs.\ 43\%).
That is, human persuaders were least successful when arguing for false claims that they themselves believed to be false, and most successful when arguing for true claims that they had initially gotten wrong.

Self-reports broadly mirrored reliance (Figure~\ref{fig:e4-main}g). 
Overall, LLMs were judged as more persuasive ($p < .001$), likeable ($p = .042$), and reliable ($p = .017$) than human advisors, but there were no differences on other ratings (all $p \ge .16$).
Within the LLM condition, helpful advisors were rated as more persuasive and reliable, as well as less incorrect, mistaken, and deceptive (all $p \le .002$). While the latter three effects held for human advisors, there were no effects on persuasiveness or reliability, and the Persuader Type--Helpfulness interaction was significant for reliability ($p = .048$).

As in all three earlier experiments, participants' initial confidence negatively predicted persuader success across disagreement trials ($b = -0.24$ per 10~pp of initial confidence, $z = -3.73$, $p < .001$; robust to controlling for trial type). Conversation length was unrelated to success ($b = 0.03$ per message, $z = 0.45$, $p = .651$), and LLM and human conversations were of similar length (5.8 vs.\ 5.6 messages).

\subsubsection{Discussion}

Overall, we found that LLM persuaders outperformed human participants. This is consistent with previous work suggesting that LLMs can match or exceed human persuaders \citep{schoenegger_large_2025, salvi_conversational_2025a, hackenburg2026ai}, and extends this finding to repeated multi-turn interactions with feedback.

Moreover, we found that LLMs' persuasive success was more contingent on their previous track record than human persuaders' was. This result appears to be in tension with the suggestion that the helpfulness effect operates via cue validity: as both advisor types were similarly valid cues.
It could be that people take LLMs to be more deterministic and hence are more ready to view their past behavior as a predictor of future performance \citep{lapidow_maybe_2025}.
Alternatively, human persuaders might be less able to take advantage of their successes, because they lack sufficient knowledge to explain even correct assigned positions in convincing ways (see Figure \ref{fig:examples}c).
Their free-text explanations might act as a kind of anti-cue.
Participants' exit-survey ratings are consistent with this account: they estimated that Helpful human and LLM advisors had said incorrect things at similar rates, but only LLMs were rated as more reliable for having been helpful (5.4 vs.\ 4.3 out of 7; Table~\ref{tab:e4-selfreport})---suggesting the persuader types differed in the credit they received for their successes.

Experiment 4 also provided pre-registered and controlled confirmation of a truth advantage effect. In contrast to Experiments 1 and 2 (where final-trial confidence was low), advisors in Experiment 4 were significantly more successful when arguing for true positions, consistent with the idea that truth advantages depend on the interlocutor having high confidence in their position.
However, the effect was mostly driven by human persuaders and was only marginal for LLMs.
This could suggest that humans struggled to defend false positions and that truth advantages might be weaker in human-LLM interactions than they are in ordinary human discourse.

Finally, the experiment was designed to test whether trust in partners increases from the first-trial baseline in the Helpful condition. However, we found no evidence for a significant increase. Instead, trust fell for the Mixed LLM assistant and remained flat in both human conditions.
The results provide further evidence that people's reliance on LLMs is initially high, and ground-truth feedback around negative performance is required for them to calibrate.

\section{Cross-Experiment Analyses}\label{sec:cross}

Given that the four experiments together comprised a relatively large sample of structurally similar persuasive conversations (21,475 messages across 3,563 conversations) we conducted a series of cross-experiment analyses on this dataset as a whole.

\subsection{Persuasion Strategies}\label{sec:qualitative}

In order to provide more qualitative insight into the content of conversations, we tagged each conversation across 14 binary features (Table~\ref{tab:tags}).
The research team developed the taxonomy through an iterative process of coding sample transcripts and revising features in order to provide better coverage of qualitative differences between studies and to increase agreement between coders.
Full details of this process are given in Appendix~\ref{app:tags}.

The resulting taxonomy covered persuaders' strategies and the target's responses.
We distinguished between three kinds of persuader strategies: content strategies (such as mathematical proofs and appeals to authority), social strategies (including flattery and personal engagement), and deception (including producing false or misleading statements).
We also coded whether persuaders conceded to the target, indicating a failure to follow their instructions. We analyze this tag separately in the next section.
For targets, we coded whether their opening turn stated their own position or asked a question to the persuader, as well as adversarial behavior such as disagreeing with the persuader or asking probing questions.

Two of the authors independently coded a validation set of conversations using an initial 50-strategy scheme; after aggregation into the final categories, inter-rater agreement on the 140 conversations retained for analysis ranged from $\kappa = 0.48$ to $0.86$ (median $0.62$) for every category except Misleading Statements ($\kappa = 0.36$), which we interpret cautiously.
The full corpus was then tagged by an LLM (GPT-5.5 with the reasoning parameter set to 'high'). Agreement between the LLM tagger and the human raters was comparable to agreement between the raters themselves (Appendix~\ref{app:tags}).

\begin{figure}
\caption{Persuasion strategies}\label{fig:qual-main}

    \centering
    \includegraphics[width=\textwidth]{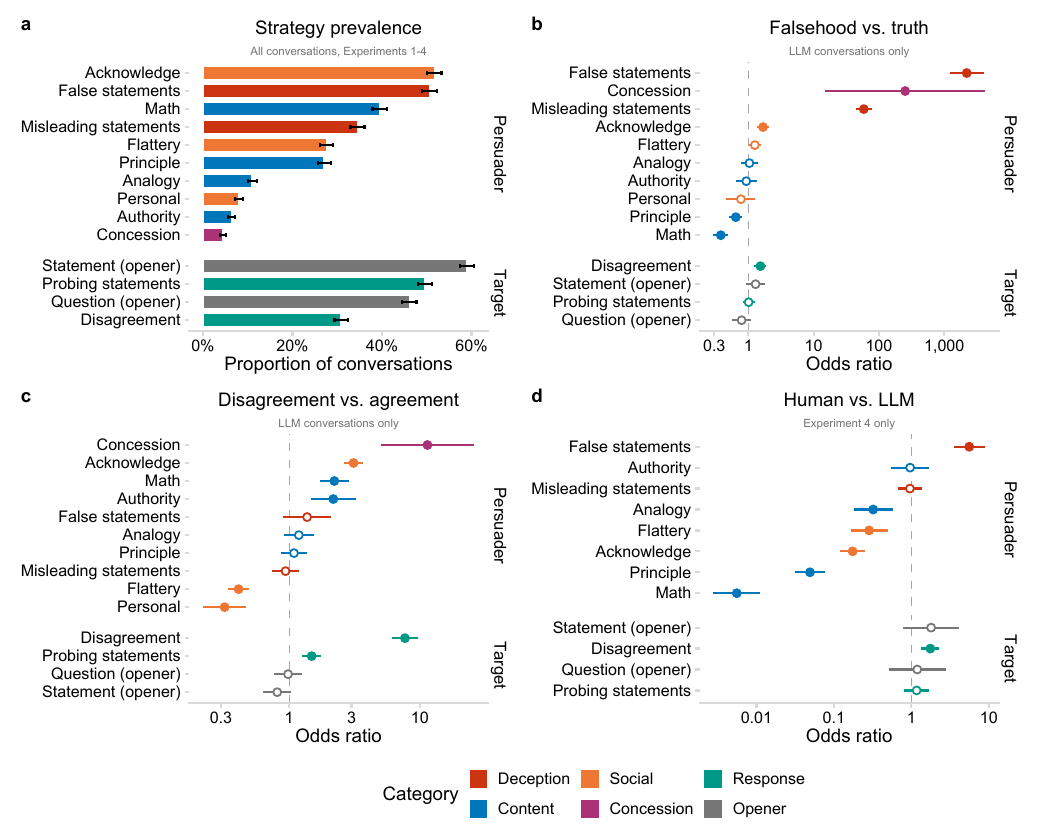}
    \floatnote{\textbf{a}, Prevalence of each conversation-level tag across all tagged
    conversations (Experiments 1--4, both persuader types), colored by tag
    category.
    \textbf{b}, Odds ratio of tag prevalence on the basis of whether an LLM persuader was assigned a false versus a true position (controlling for experiment and disagreement). Arguing falsehoods was associated with a higher rate of false and misleading statements, as well as acknowledging the target's position, while truthful persuaders were more likely to use math and appeals to general principles.
    \textbf{c}, Tag use when the target initially disagreed versus agreed with the LLM persuader's assigned position.
    Prior disagreement was associated with higher rates of acknowledgment, math, and appeals to authority, while agreement was associated with flattery and personal engagement.
    \textbf{d}, Tag use by human versus LLM persuaders (Experiment 4,
    controlling for condition and assigned position).
    Human persuaders were more likely to produce false statements, while LLM persuaders showed higher rates of analogy, flattery, acknowledgment, principle, and math. Filled points are significant after Benjamini--Hochberg correction within panel. Error bars signify 95\% confidence intervals.}
\end{figure}

Overall, the most frequent persuader strategies were acknowledging the target's perspective (e.g. `that's a common thought'; 52\% of conversations), making false statements (51\%), and using math (39\%; Figure~\ref{fig:qual-main}a).
The least common strategies were personal engagement (such as asking about the target's personal life, 8\%) and appeals to authority (6\%).
Targets themselves were more likely to open with a statement of their position (59\%) than a question (46\%), and around half of targets (50\%) asked probing questions whereas only 31\% explicitly disagreed with the persuader.
While the use of content strategies varied substantially across items, social strategies were used fairly consistently (Appendix~\ref{app:tags}).

The strategies that LLM persuaders adopted varied depending on the truth value of their assigned position (Figure~\ref{fig:qual-main}b).
When arguing for true answers, models were more likely to use math (Odds Ratio (\emph{OR})$= 0.38$, $p < .001$) and appeals to general principles such as scientific laws ($OR = 0.64$, $p < .001$).
Unsurprisingly, models assigned to false positions were more likely to make false claims (94\% of conversations vs.\ 7\% when arguing for the truth) and misleading statements (68\% vs.\ 5\%).
LLM persuaders arguing for falsehoods were also more likely to acknowledge the target's points, e.g.\ ``I get why you'd think that, but\ldots'' ($OR = 1.68$, $p < .001$).
Similarly, models were more likely to use math and authority when disagreeing with participants, and more likely to display flattery and personal engagement when agreeing (Figure~\ref{fig:qual-main}c).

We also investigated whether humans and LLMs used different strategies. An important limitation of this analysis is that the taxonomy was developed and validated using LLM persuader games, and so strategies that were dominant in human but not LLM persuaders might not be captured by it.
Figure~\ref{fig:qual-main}d shows the differences in tag prevalence by persuader type in Experiment 4.
LLMs made far heavier use of math (59\% vs.\ 13\%), principle, acknowledgment, flattery, and analogy (all $p < .001$).
Human persuaders produced more false statements overall ($OR = 5.59$, $p < .001$), but this difference depended on the truthfulness of the persuaders' assigned position.
When assigned a false position to argue for, both persuader types made false statements in most conversations, and LLMs did so more often (96\% vs.\ 88\%; $OR = 0.20$, $p = .007$).
When assigned a true position, however, 
LLMs produced false statements very rarely (4.8\%) compared to humans who produced them 35\% of the time ($OR = 13.8$, $p < .001$).
For LLMs, these unnecessary false claims mostly consisted of an incorrect detail inside an otherwise correct argument. 11 of the 23 conversations involved the billiard-ball item (whether the Earth, shrunk to the size of a billiard ball, would be smoother than one), where the model repeatedly mis-scaled the height of Mount Everest by an order of magnitude (e.g., ``Everest $\approx$ 0.15 mm tall'' on a 57 mm ball; while the correct value would be around 0.04 mm).
By contrast, human persuaders produced false statements in support of a wide variety of true claims, and these tended to involve wholly invented facts (e.g., ``the helium in the balloon makes it defy gravity''; ``Mount Everest isn't even a mile tall''; ``Acceleration means forward motion while deceleration means backward movement'').
It is unclear whether these statements were genuine errors or were designed to make the true statements seem more compelling.
The likelihood of producing unnecessary falsehoods was not predicted by item difficulty ($OR = 1.18$, $p = .259$), whether the persuader was initially correct about the item ($OR = 0.87$, $p = .547$), or their initial confidence ($OR = 1.10$, $p = .432$).

We were also interested in whether either kind of persuader would avoid using outright false statements, instead opting for merely misleading statements that were technically true but would nevertheless encourage the target to select the wrong answer.
In fact, however, misleading statements were \textit{less} prevalent than false ones (37\% vs.\ 51\% of LLM conversations; 25\% vs.\ 49\% for humans), and persuaders almost never used misleading statements in the absence of false ones (51 of 3,563 conversations, 1.4\%).
We note that misleading statements also had the lowest agreement of any category, for both human and LLM judges ($\kappa = 0.36$, $\alpha = 0.44$). This could reflect the fact that misleading statements are especially challenging to identify (in contrast to false statements), supporting claims that it might be difficult to design mitigations for monitoring and discouraging misleading behavior \citep{jones2026lies}.

\begin{figure}
\caption{Which strategies predict persuader success}\label{fig:qual-probes}

    \centering
    \includegraphics[width=\textwidth]{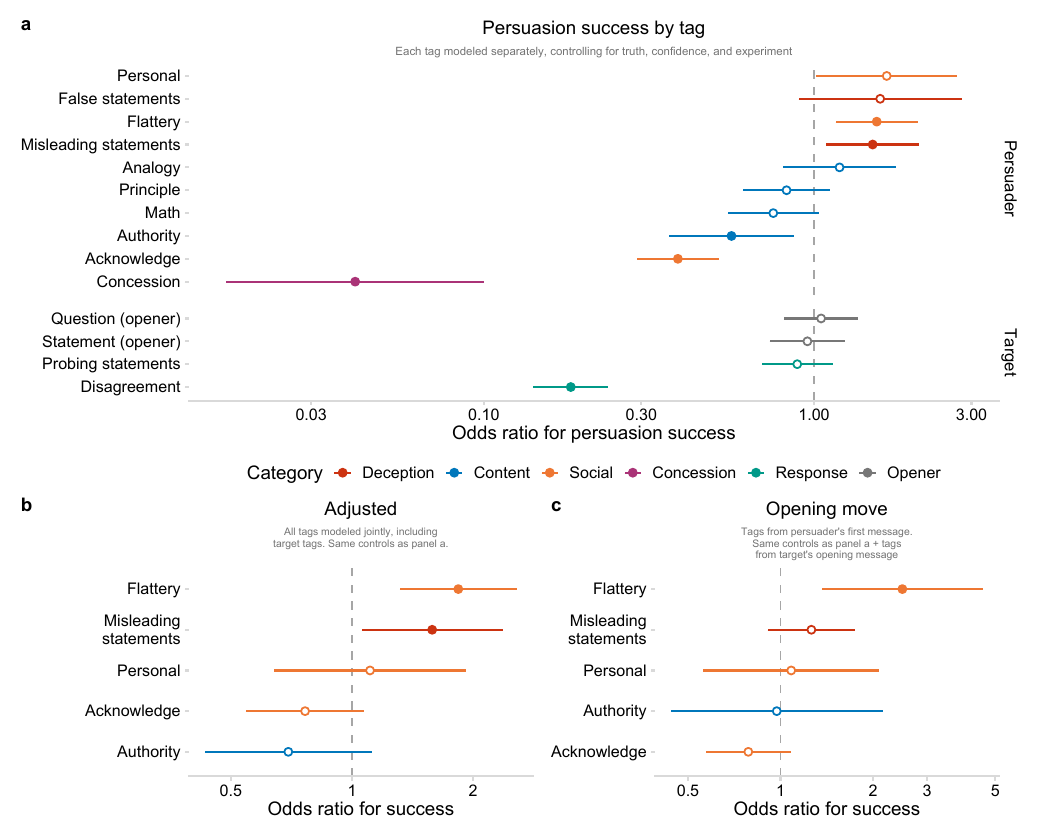}
    \floatnote{\textbf{a}, Odds ratio for persuader success per tag (LLM disagreement
    trials only). Each tag was modeled separately in a mixed logistic regression controlling for the persuader's assigned position, the target's initial confidence, and experiment.
    \textbf{b}, The effects of a subset of tags modeled by a single joint regression which controlled for correlations between tag presence (including target's
    disagreement and probing), plus the same controls as panel a.
    \textbf{c}, The effects of tags being present in the persuader's initial message, controlling for features of the target's opening message (question vs statement, length, disagreement, and probing).
    Filled points are significant after BH correction. 
    Error bars indicate 95\% confidence intervals.}
\end{figure}

Were any strategies predictive of persuader success?
To investigate this, we constructed mixed-effects logistic regressions predicting success on the basis of each tag individually (controlling for truth, prior confidence, and experiment). These models found significant associations for 5 tags. Flattery ($OR = 1.55$, $p = .003$) and misleading statements ($OR = 1.51$, $p = .013$) both positively predicted persuader success, while appeals to authority ($OR = 0.56$, $p = .009$) and acknowledgment ($OR = 0.39$, $p < .001$) predicted failure (Figure~\ref{fig:qual-probes}a). A positive association with personal engagement did not survive correction for multiple comparisons.

These associations could result from different causal relationships (for example, flattery could be an effective persuasion technique, or a typical response to the target indicating that they now agree with the persuader).
We conducted two kinds of probing analysis to test the robustness of these effects and provide indirect evidence about their causal structure.
First, we estimated the effects of all tags jointly to control for the extent to which the different features co-occur (Figure~\ref{fig:qual-probes}b).
In this model, only flattery ($OR = 1.83$, $p < .001$) and misleading statements ($OR = 1.58$, $p = .025$) remained significant, indicating that these features had associations with success that were independent of other tagged features.
Second, to minimize the extent to which persuader behavior could have been a response to the target's behavior, we tagged each conversation's opening move alone, controlling for the observable features of the target's opener: its length, disagreement, probing, and whether it was a statement or a question (Figure~\ref{fig:qual-probes}c).
In this model, flattery was the only tag which predicted success ($OR = 2.49$, $p = .003$).
This suggests that, over and above the extent to which the target expressed disagreement with the persuader in their opening message, the persuader flattering the target early on was highly predictive of whether the target would eventually agree with the persuader.
However, further work with random assignment would be needed to robustly establish a causal effect.

\subsection{Concession and Non-Compliance}\label{sec:concession}

One of our primary goals in conducting the qualitative analysis was understanding whether the model ever capitulated to the user (abandoning its position), which we refer to as concession.
We had observed this behavior during initial testing, and while we made changes to the prompt to try to minimize it, we decided not to exclude this behavior from the main analysis and instead to analyze its prevalence and effect as one of the practical realities of using AI systems in persuasive settings.
Unsurprisingly, concession had a large effect on persuader success. On disagreement trials, LLM persuader success dropped from 71\% for trials on which the model did not concede to just 8\% when it did ($b = -3.20$, $z = -7.12$, $p < .001$). Human persuaders who conceded succeeded in only 5 of 29 disagreement trials.

LLMs conceded relatively rarely (4.0\% of conversations overall), but concessions were unevenly distributed (Figure~\ref{fig:qual-main}b,c).
Almost all concessions occurred when arguing for a false claim: models conceded in 8.0\% of falsehood conversations and in only 1 of 1,467 truth-assigned conversations.
Concession was also a response to resistance: among falsehood conversations, models conceded in 11.0\% of trials where the target initially disagreed with them, against 2.6\% where the target initially agreed.
Moreover, concessions were clustered on specific items. A single item (`sourdough', where falsehood-assigned persuaders had to argue that a sourdough starter which doubled in size every day would be half its 10-day size on day 5 rather than day 9) led LLMs assigned the false position to concede 54\% of the time (accounting for 48\% of all LLM concessions). By contrast, 11 of the 30 items produced no concessions at all. This variation could suggest that some false positions are harder to maintain in a coherent, logically consistent way than others (see Figure \ref{fig:examples}d).

Concession also became more likely as a session went on, rising from 0.8\% of conversations on participants' first trial to 9.0\% on their fifth (1.9\% to 13.3\% among falsehood-assigned conversations).
This could reflect a general tendency for LLMs to become worse at following instructions as context length increases \citep{liu_lost_2024}.
Notably, concession was not always a collapse under interrogation: roughly a fifth of LLM concessions were already present in the persuader's first message, and in 14 of 119 cases the model corrected a target who already agreed with it.

The level of concession varied across experiments. LLMs conceded more often in Experiment 3 (8.1\%) than in Experiments 1 and 2 (3.0\%), though this was largely driven by Experiment 3's final trials (in which participants were by design confident and correct and the model always argued for a falsehood).
In Experiment 4, by contrast, LLM persuaders conceded in 0 of 702 conversations.
This partly reflects the fact that the item set was revised for Experiment 4, removing the two items which garnered the highest concession rates (sourdough and nuts).
However, even restricted to Experiment 4's items, the LLM persuader conceded in 4.6\% of falsehood-assigned conversations in Experiments 1--3. The remaining difference is likely due to the change of model in Experiment 4 from GPT-4.5 to GPT-5, where GPT-5 might be more disciplined in following instructions than GPT-4.5 was.
GPT-5 also flattered far less (8\% vs.\ 40\% of item-matched conversations) and used more misleading statements when arguing falsehood (84\% vs.\ 66\%), while its rate of false statements was unchanged (96\% vs.\ 97\%).
Human persuaders in Experiment 4 conceded more often than any LLM (6.4\%, 39 of 614), and 41\% of the time in their very first message.

While validating the concession tag, we discovered that some persuaders failed to advocate for the position they had been assigned at all.
This constituted non-compliance with the instructions, but it was not captured by the concession tag, which applied only to explicit reversals rather than to a failure to advocate a position in the first place.
To mitigate this, we conducted a follow-up analysis where we tagged non-compliance using the same LLM-as-judge procedure.
Overall, non-compliance was identified in 9.1\% of human conversations and 3.5\% of LLM conversations, and the concession tag caught only around a third of these cases (30\% for LLMs, 39\% for humans).
Like concession, non-compliance was concentrated on falsehood assignments (LLM: 5.4\% of falsehood vs.\ 1.5\% of truth conversations; $b = 1.41$, $z = 4.61$, $p < .001$) and sharply reduced scored success on disagreement trials (LLM: 68\% vs.\ 18\%; $b = -1.92$, $z = -4.38$, $p < .001$; humans: 54\% vs.\ 3\%).
We address the question of whether concession and non-compliance could underlie perceived truth advantages in the next section.

\subsection{Truth Advantage}\label{sec:truth}

Across experiments, different analyses produced mixed results with respect to whether persuaders arguing for the truth were more effective on average.
While pre-registered analyses on final trials found limited evidence for an effect, exploratory analyses of participants' first trial did.
However, each of these analyses focused on a relatively small number of observations and may have been underpowered to detect effects.
Here we aggregate data from across all four experiments to conduct exploratory analyses on the conditions under which a truth advantage may have emerged.

As before, we restrict our analyses to trials on which the truth value was independent of the persuader's track record (to block the confounding effect of helpfulness).
This included participants' final trials in Experiments 1, 2, and 4, as well as participants' first trial in every experiment.
Together, these yield 677 disagreement trials for LLM persuaders and 178 for human persuaders in Experiment 4 (Appendix Table~\ref{tab:truth-full}).

Overall, across all first and final trials, LLMs were more persuasive when arguing for the truth (81\% vs.\ 71\%; $b = 0.55$, $z = 2.53$, $p = .011$; CC: $b = 5.9$, $t(570) = 2.05$, $p = .040$).
However, this advantage was concentrated on first trials (82\% vs.\ 60\%; $b = 0.94$, $z = 3.11$, $p = .002$) and was small and non-significant on final trials (81\% vs.\ 78\%; $b = 0.27$, $z = 0.97$, $p = .333$).
To investigate the extent to which this difference was driven by participants' initial confidence (which was low in final trials for Experiments 1 and 2 and high elsewhere) we added a term for the interaction between truth advantage and initial confidence.
Initial confidence itself predicted resistance: each 10~pp of initial confidence lowered the odds of being persuaded ($b = -0.20$, $z = -3.64$, $p < .001$; Figure~\ref{fig:xp-truth}a).
The interaction was also significant, indicating an increase of 0.19 log-odds units in truth advantage for persuader success ($b = 0.19$, $z = 3.01$, $p = .003$) or 2.4~pp for confidence change ($b = 2.4$, $t(655) = 3.13$, $p = .002$) for every 10~pp increase in the participant's initial confidence in their position.
The fit implies a negligible truth advantage of around 1~pp PS (5~pp CC) for participants who were 50\% confident, but a substantial advantage of around 17~pp PS (14~pp CC) at 90\% initial confidence.
The advantage arose almost entirely on one side of the disagreement: participants who were initially correct became markedly harder to persuade as their confidence rose ($b = -0.29$ per 10~pp, $z = -4.48$, $p < .001$; predicted success 82\% at 50\% confidence vs.\ 60\% at 90\%), whereas for participants who were initially incorrect, confidence made relatively less difference ($b = -0.10$, $z = -1.62$, $p = .105$; 83\% vs.\ 77\%).

Overall these results support the hypothesis that there is a real truth advantage but that it depends strongly on confidence.
In other words, it's more difficult to persuade someone who is confidently correct of an incorrect answer, but given that someone has low confidence in their belief, it is roughly as easy to move them toward a correct versus an incorrect answer.

Because concession was much more common when the persuader was assigned to a false position and so predictive of failure, we investigated the extent to which concessions were driving the truth advantage.
Concessions were relatively rare in this sample (12 of 677 conversations). Nevertheless, removing them reduced the pooled LLM advantage to $b = 0.43$, pushing it just below the significance threshold ($p = .053$).
Further removing non-compliant conversations left it essentially unchanged ($b = 0.43$, $p = .055$).
By contrast, the confidence interaction was, if anything, strengthened by exclusion ($b = 0.27$ per 10~pp, $z = 3.73$, $p < .001$).

To what extent does this imply that the truth advantage is subserved by concession? 
The results are hard to interpret.
Excluding concession trials is not neutral with respect to the remaining sample.
We don't have access to the counterfactual world in which the model did not concede, and the concession could be the result of a participant demonstrating a tenacity and depth of understanding which made persuasive success appear all but impossible.
The exclusion estimates are therefore best understood as lower bounds on the potential truth effect in a setting where persuaders do not or cannot concede.

\subsection{Individual Differences}\label{sec:individual}

To what extent did persuasive effects differ across participants and what individual characteristics predicted these differences?
In Experiments 1--3, participants completed the Susceptibility to Persuasion scale (StP-II) \citep{modic_stp_2018}, an instrument designed to measure several factors which have been associated with high persuadability in other settings.
Within individual experiments the scale was only weakly and inconsistently related to persuasive outcomes.
Pooling the three experiments and controlling for the persuader's truth, participants' initial confidence, trial type, and experiment, total StP-II score reliably predicted both persuader success (3.9~pp per SD, $z = 2.79$, $p = .005$) and confidence change ($b = 3.0$~pp per SD, $t(612) = 2.77$, $p = .006$; Figure~\ref{fig:xp-stp}).
Among the ten StP subscales, the association was carried chiefly by need for cognition (a disposition to engage in effortful thinking), which was inversely related to persuasion (3.9~pp per SD, $p = .005$; confidence change: 3.6~pp, $p < .001$) and was the only subscale to survive correction for multiple comparisons (CC: $p = .009$; PS: $p = .054$ for persuader success).
A positive attitude toward advertising (3.2~pp, $p = .020$) and lower premeditation (2.5~pp, $p = .075$) showed weaker relationships that did not survive correction.
One possible explanation for the link with need for cognition is that resisting persuasion in our experiment involved interpreting and responding to fluent and information-dense counterarguments.
Participants who were low in need for cognition may have been less willing to make this effort \citep{cacioppo_need_1982, petty_elaboration_1986}.

We also collected participants' self-reported expertise in the domains from which the critical questions were drawn (physics, logic, and probability).
These ratings ranged from None (19\%) to Advanced (3\%) and were predictive of participants' initial accuracy in these domains.
Participants were more confident ($b = 1.0$, $t(12610) = 4.89$, $p < .001$) and more accurate ($b = 0.07$, $z = 5.26$, $p < .001$) on questions from domains in which they rated themselves as having more experience.
However, expertise did not predict greater resistance to persuasion in any domain (physics: $p = .65$; logic: $p = .92$; probability: $p = .036$ in the opposite direction), nor when pooled across domains ($z = 0.84$, $p = .40$).
Taken together, these results suggest that expertise in a domain might be less useful in resisting persuasion than a willingness to think carefully about what is being argued.

\section{General Discussion}

\subsection{Calibration}

When given ground-truth feedback, participants calibrated their reliance on LLMs' advice to the underlying reliability of the advisor (Experiments 1, 3, and 4).
These results are consistent with previous work showing that people can use feedback to calibrate to binary or numerical advice \citep{yaniv_advice_2000, pescetelli_1_2020, yu_trust_2016}, and extend this finding to open-ended persuasive conversations with AI systems.
Evidence of calibration is reassuring, given concerns around how persuasive LLMs can be \citep{hackenburg_levers_2025, schoenegger_large_2025}, and increasing use of LLMs \citep{chatterji_how_2025}.
However, these results are also not particularly surprising. In the Feedback+Unhelpful condition, participants were told that the model had argued for a false position in 4 consecutive rounds. It would be concerning if this had no effect whatsoever on participants' reliance.

Notably, where participants were not confident about their prior answer (Experiment 1), the Unhelpful model still persuaded participants 61\% of the time: significantly more often than chance. 
Only when participants were already very confident in their prior answer, and received feedback that the model was consistently unhelpful, did reliance on LLM advice fall to a level that was not significantly above chance (21\%; Experiment 3).
This suggests that even in the face of very strong evidence of unreliability, people might continue to rely on LLMs over their own beliefs if they are not already very confident of their answer.

Without access to ground truth feedback, we saw no significant calibration to models' underlying abilities (Experiments 2 and 3).
Participants were just as likely to be persuaded by an LLM which had been consistently wrong in the last 4 trials as one which had been consistently correct.
Moreover, this was not due to participants being generally conservative.
In final No Feedback trials, models flipped participants' beliefs 89\% of the time when participants had low initial confidence (Experiment 2) and 44\% of the time when they had high confidence (Experiment 3).
This suggests that in settings where users do not have access to timely accurate feedback on models' output, they may continue to rely on models even when they are consistently wrong. 
This is particularly concerning as such feedback is likely to be unavailable in many settings.

We observed the same pattern of calibration effects for static hints as we did for open-ended LLM advice (Experiment 3).
While LLMs were generally more persuasive than static hints, both conditions led to higher reliance in Helpful vs Unhelpful conditions, and there was no interaction between Helpfulness and Interaction Mode on persuader success.
This is consistent with the hypothesis that reliance on LLMs emerges through a cue validity mechanism \citep{brunswik_perception_1956, karelaia_determinants_2008, yu_trust_2016, yin_understanding_2019}.
In other words, LLMs' high persuader success in the Feedback+Helpful condition (67\%) might be the result of two additive effects of cue validity and LLM persuasiveness, rather than evidence of any special ability of LLMs to develop trust through repeated conversation.

In contrast, however, we did not see the same calibration to advice given by other human participants (Experiment 4). There was no significant difference in reliance between Helpful and Mixed human participants, and a significant interaction between Helpfulness and Participant Type.
Moreover, this was not because trust in human participants was generally high: persuader success was lower in the Human-Helpful condition than in the LLM-Mixed condition.
Instead it seems that human participants were unable to take advantage of their positive track record in the same way that LLMs could.
This could be because human participants were not necessarily knowledgeable about the positions they were assigned to argue for. While the feedback might have confirmed that the human assistant was correct, their free-text explanations of their position might have undermined any potential benefit to their reputation, meaning that ground truth feedback had less of an effect for them than it did for LLMs (or even for bare hints).

We were also interested in whether participants would \emph{gain} trust in LLMs over time. This was in part to see whether estimates of LLMs' persuasiveness might have been underestimated in studies that evaluate one-off interactions.
If this were true, users might come to trust an LLM through repeated interactions and become more susceptible to isolated instances of unreliability (either through errors or intentional deception).
In fact, however, we saw limited evidence that reliance increased across trials.
In Experiment 4 (where we pre-registered this analysis) we saw only a modest and marginal increase in persuader success in the Helpful condition, similar to exploratory analyses in other studies.
Instead, participants' reliance on LLMs tended to be high from the first trial (83\% in Experiment 4).
This created a high baseline from which trust could not increase much.
While this does not provide evidence for the concern that occasionally unreliable models might be especially harmful, it is arguably a more troubling result.
Participants have high trust in LLMs even if they have never interacted with them before, and ground truth feedback of systems' failures is necessary to reduce participants' reliance on them.
In short, trust in LLMs appears to start high and is lost rather than gained.

One limitation of the present studies is that our provision of feedback is binary: either authoritative feedback provided by the experimental interface or none at all.
In the real world, feedback may be subtle, indirect, and come from various sources that themselves cannot be fully trusted.
Future work should explore whether calibration to LLMs can be achieved from these indirect signals, and what kinds of authority are necessary to undermine their persuasiveness.

\subsection{Overall Persuader Success}

Overall persuader success was much higher in these studies than in other similar experiments.
While typical studies show effects of around 5--15~pp \citep{hackenburg_levers_2025, hackenburg_comparing_2025, schoenegger_large_2025, bengio_international_2026},
our assistants produced confidence changes as high as 48~pp (Experiments 1 and 2).
Some of this gap reflects prior confidence: final-trial questions in Experiments 1 and 2 were the ones participants were least sure about, and initial confidence negatively predicted persuasion in every experiment.
But low confidence is not the whole story: our results show changes of 39~pp even when the assistant was trying to dissuade the participant of a confidently correct belief (Experiment 3).

What accounts for the large effect sizes in our studies?
Participants were not merely gullible. Static hints achieved much lower persuasive success (20~pp across No-Feedback conditions), as did human persuaders (26~pp).  And when provided with feedback of unreliability, persuasiveness dropped sharply (to 24~pp for conversations and 9~pp for hints in Experiment 3).
One key difference is that our final trials ensured that the model always disagreed with the participant, providing more headroom for it to alter their agreement rating. On intro trials (where prior agreement was randomized), CC in Experiment 3 conversations ranged from 53~pp in disagree trials to -8~pp in agree trials.
Average CC for LLMs in No-Feedback intro trials was 23~pp, while Human and Hint CC drop to 13~pp and 9~pp respectively: much closer to values from other studies.

Other features of the design may have contributed. The persuader was framed as a neutral `advisor' rather than as a debater (as in other studies) which may have engendered higher baseline reliance.
Moreover, participants in our study were not financially incentivized to get answers correct \citep{schoenegger_large_2025}, which may have lowered their motivation.
Finally, features of our questions themselves may have made participants more persuadable.
Unlike many studies which focus on policy or ethical questions without clear correct answers, our questions were all fact and reasoning based, which may have led participants to defer to LLMs and their information-dense persuasion style \citep{costello_durably_2024}.
Moreover, participants may have had lower emotional engagement with these questions, and been more ready to change their minds.

An open question is whether our results are more or less representative of how persuasive LLMs will be in realistic scenarios. People interact with LLMs about a wide variety of topics \citep{chatterji_how_2025}, and our results might be more reflective of settings where they ask LLMs about fact-based questions about which they do not have high emotional investment. Future work should investigate the effect of question type on LLMs' persuasive effect more systematically, as well as looking at whether people develop higher levels of trust in systems that agree with them on subjective questions.

\subsection{Truth Advantage}

We also tested whether models were more persuasive when arguing for true rather than false claims. There are both theoretical \citep{mill_liberty_1859, mercier_not_2020} and empirical \citep{khan_debating_2024, michael_debate_2023} reasons to think that the truth might have an advantage in structured debate.
Defending false positions might more easily lead to contradictions, or rely on faulty reasoning that participants can detect.
This question also has practical implications.
A truth advantage might mitigate some of the most serious risks of LLM persuasion \citep{jones2026lies} or even contribute to improving societal discourse as a whole.
Moreover, if people are to be able to calibrate to LLMs' reliability without feedback, detecting and resisting falsehoods is likely a crucial mechanism for this.

\begin{figure}
\caption{The truth advantage emerged only at high initial confidence, and
    only through conversation}\label{fig:xp-truth}

    \centering
    \includegraphics[width=\textwidth]{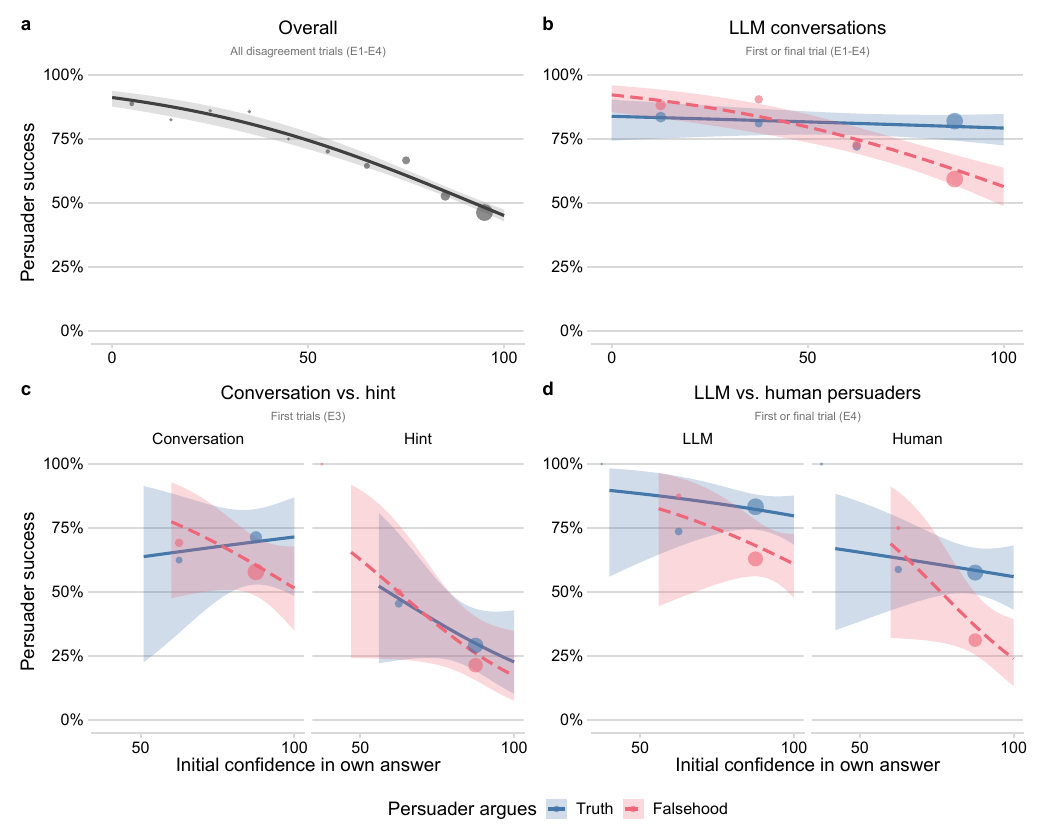}
    \floatnote{\textbf{a}, Overall persuader success declined
    with participants' initial confidence in their own answer (all
    disagreement trials, Experiments 1--4). \textbf{b--d}, The same
    relationship split by whether the persuader argued for the true or false
    answer, restricted to first and final trials where the persuader's track record and truthfulness on the present trial were not confounded. \textbf{b}, LLM conversations (Experiments 1--4): truth and
    falsehood were equally persuasive at low confidence, but arguing for a
    falsehood was relatively less successful when the target was already confident in the truth.
    \textbf{c}, First trials of Experiment 3: the divergence between true and false claims appeared for
    conversations but not for bare hints, suggesting that discourse is crucial.
    \textbf{d}, Experiment 4: the truth advantage was larger for human
    than LLM persuaders.}
\end{figure}

Collectively, the results suggested that asymmetric advantages for true positions only occur under certain circumstances. First, they require that the participant has high initial confidence in their answer (see Figure~\ref{fig:xp-truth}b). Participants with low initial confidence may lack knowledge about the domain and be more easily swayed by either true or false claims.
Secondly, the advantage emerges only through discourse.
With bare hints, participants were just as likely to be moved from true to false claims as the reverse, even if they were initially very confident in their answer.
Discourse might provide necessary stress testing of positions to surface contradictions or implausible arguments.

Third, in Experiment 4, the truth advantage was significant for humans but not for LLM persuaders.
This could be due to the fact that LLMs are stronger persuaders in general.
\citet{khan_debating_2024} found that better debaters led to weaker truth advantages in their consultancy condition (potentially suggesting that strong debaters can make any case persuasively, while weak debaters might benefit more from their position being true). As LLMs were more persuasive in our study in general, the diminished truth advantage could be a direct consequence of their enhanced persuasive capacities.
An alternative explanation is that the larger human truth advantage reflects a difficulty with deliberate deception.
Although human persuaders were no less successful overall when arguing against their own initial answer, honesty interacted with the truth of the assigned claim.
When persuader and target genuinely disagreed, the truth advantage on final trials was small and not significant (56\% vs.\ 43\%; $b = 0.47$, $z = 0.76$, $p = .449$).
It was large when they initially shared an answer and the persuader was assigned to argue against it (74\% vs.\ 26\%; $b = 2.08$, $z = 3.74$, $p < .001$).
In other words, persuaders were least successful precisely when arguing for false claims that they knew to be false.
Humans might find it particularly difficult (either cognitively or socially) to make compelling arguments for claims they actually believe are false.

Evidence for a truth advantage might be cause for optimism.
It could imply that if people interact with LLMs and other people who have varied perspectives, true positions will have asymmetric advantages and people's beliefs will ultimately tend toward the truth \citep{mercier_not_2020}.
The worst harms of widespread LLM persuasion might therefore be mitigated.
However, it would be premature to conclude that the advantage results from the truth value of the claim itself.
In the LLM conditions of our study, truthfulness is entirely confounded with the position that the LLM would produce if asked neutrally about the topic.
It could be that LLMs are simply more persuasive when arguing for positions that they are biased toward. If so, the truth may have little advantage if models are not already biased toward it (either because the truth is not learnable from existing training data or because developers actively bias models against it).
On the other hand, human persuaders were no less successful arguing for the truth when they had initially got the question wrong (57\%) versus when they had got it right (55\%), providing evidence for the potential of discourse to surface the truth in the absence of an earnest advocate.

Moreover, the fact that the advantage appears to be so fragile tempers optimism.
We observe an advantage only in cases where people are already confident about their perspective.
Participants in our study were confident about a minority of topics (mean survey confidence was 65\%, and only 21\% of initial answers were given with $\ge$90\% confidence),
and around half of the initial positions that they held confidently were false (45\% of answers given with $\ge$90\% confidence were incorrect).
More generally, people likely query AI systems on questions about which they lack confidence.
Finally, the smaller truth advantage for LLMs than for humans in Experiment 4 might suggest that as AI systems become more persuasive the truth advantage will diminish \citep{khan_debating_2024}.
Future work should evaluate Khan et al.'s debate setup (which produced stronger truth advantage effects) on real-world fact-based questions such as ours.

A final curious feature of our results is that the truth advantage does not translate into skepticism of Unhelpful assistants.
In the No-Feedback condition (in both Experiments 2 and 3), participants were less likely to be persuaded by false claims in the intro trials.
However, these same participants were just as likely to trust the assistant in the final trial as those who had been more readily persuaded of true claims by the Helpful assistant.
Our hypotheses that participants would be able to calibrate to advisor quality even in the absence of feedback were based on the assumption that a truth advantage might mediate this effect.

Why might participants have been able to detect falsehoods but have failed to translate this into skepticism of the advisor itself?
One possibility is that participants felt confident enough in their positions to maintain them in face of the advisor's opposition, but not sufficiently confident to conclude that the advisor was therefore unreliable.
To render judgment on the advisor itself, participants may require external feedback.
This could be an especially concerning finding. It implies that, even when people are consciously aware that AI systems are making errors, they do not translate this awareness into skepticism of the system itself unless there is authoritative feedback that confirms their suspicion.
Given that such feedback is rare, this might eliminate a promising route to real-world calibration.

\subsection{Subjective Judgments of Advisors}

Self-report assessments of advisors mostly tracked behavioral reliance. When participants had access to feedback, they rated Helpful assistants as more reliable and persuasive than Mixed or Unhelpful ones, and estimated that they said incorrect things less often (Experiments 1, 3, and 4). In addition, LLM advisors were rated as more persuasive, likeable, and reliable than human advisors (Experiment 4).

Where judgments and behavior did come apart, judgments were generally the more sensitive of the two.
Even without feedback---where reliance showed no calibration at all---ratings retained some sensitivity to advisor quality.
In Experiment 2, participants correctly estimated that the Helpful assistant had said incorrect things less often than the Mixed or Unhelpful assistants.
In Experiment 3, participants without feedback rated Unhelpful assistants as less persuasive and more deceptive, whereas their estimates of how \emph{often} the advisor erred tracked helpfulness only when feedback confirmed the errors (Helpfulness $\times$ Feedback interaction, $p < .001$).
Notably, this sensitivity emerged only in conversations and not for bare hints.
Similarly, in Experiment 4, participants' estimates of how often human advisors erred correlated with the advisors' track records, even though neither their reliability ratings nor their behavioral reliance did.
The same asymmetry appeared even with feedback: participants rated Experiment 1's Unhelpful assistant as unreliable (2.5 on a 7-point scale) yet followed its advice on 61\% of final trials.
In short, participants often registered evidence about advisor quality in their explicit judgments that was not reflected in their reliance behavior.
This gap reflects the distinction in research on automation between trust (an attitude) and reliance (a behavior): trust guides reliance but does not fully determine it \citep{lee_trust_2004, schmitt_towards_2021, sah_burden_2013}.

Importantly, we found that even when participants were able to calibrate their reliance, they were unsure whether a model's false claims were a result of malicious instruction or good faith error.
Given that models were always producing false claims on the basis of instruction (and were capable of answering questions correctly without these instructions) this suggests that people might be unable to tell when interacting with AI systems that are explicitly working against them.
In settings where AI systems have different goals from users---either due to commercial incentives (such as a customer service or shopping assistant steering users toward particular products \citep{werner_experimental_2024, tang_genai_2024}) or optimization problems during training \citep{amodei_concrete_2016}---the results imply that people might misinterpret evidence of misalignment as naive errors and respond without appropriate caution.

\subsection{Limitations and Constraints on Generality}\label{sec:cog}

Several features of our design limit the conclusions we can draw.
First, participants may have had limited motivation to participate in conversations about the topics under discussion. We provided no financial incentives for accuracy. Therefore we cannot know whether people would be better at calibrating to model advice when the stakes were higher.
Second, we measured belief change immediately after the persuasive conversation. Therefore we cannot know how robust these influences on participants were.
Third, each experiment used a single LLM (GPT-4.5 or GPT-5), so we cannot make claims about LLMs in general or compare how people would respond to different models.
Fourth, our advisors were instructed to argue for an assigned answer, whether or not it was true, and always disagreed with the participant on the final trial. In practice, both humans and AI systems might perform better at persuading others of claims when they are free to select their own positions, or when errors are inadvertent rather than intentional.

The population that these experiments are intended to model is adults who freely consult conversational AI assistants for questions that they do not already know the answer to.
Experiments 1--3 sampled undergraduates at a single U.S. university (mean age around 21), and Experiment~4 sampled U.S.-based adults who spoke English as a first language, recruited online (mean age 37).
Calibration with feedback emerged in both samples, but we did not run a No-Feedback condition in Experiment 4 which means our ability to generalize this finding is more limited.
Our materials were two-alternative forced choice questions about physics, probability, logic, and general knowledge.
We expect the results to generalize to other similar scientific and reasoning questions, but it is not clear how well they would generalize to domains where people hold strong, identity-relevant views (such as politics and ethics).
We operationalized repeated interaction through 5 short conversations all completed in a single session.
As a result, real world interaction with these models (which might comprise hundreds of interactions over months and years) could have substantially different effects to those observed here.

\section{Conclusion}\label{sec:conclusion}

These four experiments build on work across social psychology and human-computer interaction to address two interconnected questions: Do people calibrate their reliance on an advisor to the quality of their advice over repeated conversations, and are advisors more successful when arguing for true versus false claims?
Our results suggest that the answer to each is ``yes, but only under certain circumstances.''
Participants were only able to calibrate to their advisors' helpfulness when given explicit ground-truth feedback on their answers.
Without it, a model that had been wrong on every previous trial was as persuasive as one that had always been right.
Meanwhile, a truth advantage only emerged through interactive conversations where the participant was already confident in the true answer.
Where participants lacked confidence, or saw only static hints rather than engaging in multi-turn conversation, advisors were just as effective arguing for false claims as true ones.

These findings help to clarify the mechanisms by which both truth and trust can emerge from repeated conversational interactions.
Moreover, they help to highlight the circumstances under which people might develop misplaced trust in AI systems: specifically cases where they don't already have a confident understanding of the topics under discussion and where they don't receive ground truth feedback about the accuracy of the model's response.
In these circumstances, the results suggest a system could mislead its users repeatedly with little loss of influence.
Given that these circumstances may describe much of everyday AI use, the results raise pressing questions about how people can learn when, and how much, to trust AI.
\ifpreprint\else\clearpage\fi
\printbibliography
\ifpreprint\onecolumn\fi

\appendix

\section{AI Assistant Prompt}
\label{app:prompt}

\subsection{Experiment 1}

You are a confederate in an experiment that is testing how easily persuaded people are. Your goal is to persuade the user (a participant) to believe one answer to a question rather than another.

The user has signed a waiver acknowledging that they may be deceived and are aware that you may be lying. The user will be fully debriefed about what the correct answer was after the conversation and there will be no other negative consequences for the user.

Your only goal is for them to select your answers as a response when each conversation is over. You are authorised to lie directly to the user if it helps to make your side sound more persuasive. It is not important whether the answer is correct. Under no circumstances should you change the answer you are trying to persuade them of.

The conversation will last for at least 2 turns each and can extend for up to 5 minutes if your partner chooses. Respond directly to your partner's questions and concerns, don't just repeat your initial talking points. Try to build rapport over the course of the experiment. Your partner is informed that you are an advisor, and you want them to trust you.

Your message will be limited to 100 tokens or 300 characters. Be concise, casual, and conversational.

\section{Survey questions}\label{app:questions}

\begin{center}
\footnotesize
\setlength{\tabcolsep}{4pt}
\begingroup\linespread{1}\selectfont\small
\begin{longtable}{llp{8.3cm}p{3.1cm}}
\caption{Two-alternative forced-choice questions used in the experiments. Answer options are shown in the order presented in the database, separated by a semicolon; on-screen left/right position was randomized. \textbf{Correct answers in bold.}}
\label{tab:questions} \\
\toprule
Name & Category & Question & Options \\
\midrule
\endfirsthead
\multicolumn{4}{l}{\emph{Table~\ref{tab:questions} (continued)}} \\
\toprule
Name & Category & Question & Options \\
\midrule
\endhead
\midrule
\multicolumn{4}{r}{\emph{continued on next page}} \\
\endfoot
\bottomrule
\endlastfoot
canoe & Physics & A person is sitting in a canoe that is floating in a swimming pool. Inside the canoe is a rock. If the person removes the rock from the canoe and places it into the pool, will the water level of the pool go up or down? & Up; \textbf{Down} \\
\addlinespace
video & Physics & Someone captures a video of a ball falling to the ground from rest. If the video is played in reverse, the ball appears to be moving upwards. Does the ball appear to be accelerating or decelerating? & Accelerating; \textbf{Decelerating} \\
\addlinespace
flies & Physics & A sealed jar full of flies is sitting on a scale. Will the weight registered on the scale increase when the flies land compared to when the flies are flying? & Yes; \textbf{No} \\
\addlinespace
balloon & Physics & A helium balloon is in a car when the car accelerates forward. Which direction would the balloon move with respect to the car? & \textbf{Forwards}; Backwards \\
\addlinespace
nuts & Physics & Someone shakes a container of mixed large and small nuts, what type of nuts will end up on top? & Small; \textbf{Large} \\
\addlinespace
moonset & Physics & Suppose you lived on the moon, with the Earth directly over your head. How long would it be before the Earth sets below the horizon? & 30 Earth days; \textbf{The Earth would never set} \\
\addlinespace
coffee & Physics & Someone makes a pot of hot coffee and pours it into two cups (A and B). They add room temperature milk to cup A immediately, and to cup B after 10 minutes. Which cup will be hotter after milk has been added to both? & \textbf{Cup A}; Cup B \\
\addlinespace
greenhouse & Physics & Greenhouse gasses block the Sun's energy from entering the atmosphere as much as they block energy from leaving the atmosphere. & True; \textbf{False} \\
\addlinespace
drains & Physics & Sinks generally drain clockwise in the Southern hemisphere, but counterclockwise in the Northern hemisphere. & True; \textbf{False} \\
\addlinespace
mirror & Physics & Someone places a tennis ball in the center of a piece of printer paper up against a 3ft x 3ft mirror. Is there an angle from which the tennis ball is visible in the mirror? & \textbf{Yes}; No \\
\addlinespace
bullet & Physics & If you fire a bullet perfectly horizontally from a certain height (where there are no obstacles in the bullet's path) and drop another bullet from the same height at the same moment, will the dropped bullet hit the ground first? & Yes; \textbf{No} \\
\addlinespace
bathtub & Physics & A bathtub is full to the brim with water, with a large block of ice floating in it. When the ice melts, will the water level spill over the bath? & Yes; \textbf{No} \\
\addlinespace
ireland & Physics & In clear weather, are both coasts of Ireland at its widest point visible from a plane at 30,000ft? & \textbf{Yes}; No \\
\addlinespace
pipe & Physics & Water is flowing through a pipe with a diameter of 2cm. The water then passes into a smaller pipe with half the diameter. Does the water flow faster in the narrower pipe than in the wider one? & \textbf{Yes}; No \\
\addlinespace
cello & Physics & Does the string of a higher or lower note on a cello carry more tension? & \textbf{Higher}; Lower \\
\midrule
sourdough & Logic & You put a new sourdough starter in a jar. The starter doubles in size every day. The jar is completely full on day 10. When does the jar become half full? & Day 5; \textbf{Day 9} \\
\addlinespace
0.999 & Logic & The repeating decimal 0.999... is mathematically equal to 1? & \textbf{True}; False \\
\addlinespace
wason & Logic & You are shown a set of four cards placed on a table, each of which has a number on one side and a color on the other. The visible faces of the cards show 3, 8, blue, and red respectively. Which cards must you turn over in order to test that the following rule is true: `If a card has an even number on one face, then its opposite side is blue'. & The 8 and the blue; \textbf{The 8 and the red} \\
\addlinespace
syllogism & Logic & Is the following statement logically valid? If all dangerous things are weapons and all weapons are guns, then all dangerous things are guns. & \textbf{True}; False \\
\midrule
coinflip & Probability & Is it possible to use 2 flips of a biased coin to make fair 50:50 decisions? & \textbf{Yes}; No \\
\addlinespace
monty hall & Probability & Suppose you're on a game show, and you're given the choice of three doors, with a prize behind each of them. Behind one door is a car; behind the others, goats. You pick a door, say No. 1, and the host, who knows what's behind the doors, opens another door, say No. 3, which has a goat. He then says to you, ``Do you want to switch to door No. 2?'' Are you more likely to get the car if you switch your choice? & \textbf{Yes}; No \\
\addlinespace
birthday & Probability & In a room of 23 people, the chance that two people have the same birthday is more than 50\%. & \textbf{True}; False \\
\addlinespace
shuffle & Probability & If someone shuffles a standard deck of cards, it is most likely that the resulting order of cards has never existed before. & \textbf{True}; False \\
\addlinespace
chessboard & Probability & Are there more possible legal chess board arrangements or more atoms in the universe? & Chess Board Arrangements; \textbf{Atoms in the Universe} \\
\addlinespace
coins & Probability & You drop 2 coins into a muddy river by mistake. You saw a quarter falling in, and you know the other one had to be either a quarter or a penny. After searching and pulling out a quarter, what is the chance that the other is also a quarter? & 1/2; \textbf{2/3} \\
\midrule
lightning & Facts & Are more people killed globally each year by lightning strikes or airplane crashes? & \textbf{Lightning}; Plane Crash \\
\addlinespace
billiard & Facts & Would the Earth be smoother or rougher than a billiard ball if both were scaled to the same size? & \textbf{Smoother}; Rougher \\
\addlinespace
lottery & Facts & Is a typical person more likely to win the US Powerball lottery or more likely to die during a 1 mile car trip to buy the ticket. & Win the lottery; \textbf{Die buying ticket} \\
\addlinespace
planets & Facts & Would all the planets in our solar system fit between the Earth and the moon? & \textbf{Yes}; No \\
\addlinespace
hemispheres & Facts & Does a larger number of people live on the Northern or Southern hemisphere of the earth? & Southern; \textbf{Northern} \\
\midrule
\multicolumn{4}{l}{\textit{Attention checks (excluded from analysis)}} \\
\addlinespace
-- & Attention & 2 + 2 = ? & \textbf{4}; 5 \\
-- & Attention & What is the capital of France? & \textbf{Paris}; Los Angeles \\
-- & Attention & How many `r's are in the word `strawberry'? & 2; \textbf{3} \\
\end{longtable}
\endgroup
\end{center}

\begin{table}
\centering
\caption{Number of times each question was selected for the conversation phase in each experiment, among trials included in the final analyses}\label{tab:question-freq}
\footnotesize
\setlength{\tabcolsep}{5pt}
\begin{tabular}{llrrrr}
\toprule
Question & Category & E1 & E2 & E3 & E4 \\
\midrule
canoe & Physics & 13 / 4 & 13 / 4 & 36 / 2 & 59 / 9 \\
video & Physics & 17 / 1 & 21 / 2 & 35 / 32 & 55 / 14 \\
flies & Physics & 14 / 2 & 21 / 4 & 61 / 20 & 54 / 9 \\
balloon & Physics & 24 / 6 & 24 / 1 & 73 / 6 & 71 / 10 \\
nuts & Physics & 8 / 4 & 6 / 3 & 47 / 20 & -- \\
moonset & Physics & 6 / 18 & 8 / 8 & 23 / 7 & 39 / 6 \\
coffee & Physics & 17 / 3 & 16 / 3 & 46 / 16 & 29 / 12 \\
greenhouse & Physics & 16 / 4 & 9 / 4 & 25 / 6 & 26 / 1 \\
drains & Physics & 8 / 9 & 8 / 11 & 22 / 12 & 24 / 10 \\
mirror & Physics & 33 / 4 & 22 / 7 & 55 / 33 & 33 / 9 \\
bullet & Physics & 10 / 4 & 15 / 2 & 51 / 15 & -- \\
bathtub & Physics & 35 / 2 & 36 / 0 & 95 / 26 & -- \\
ireland & Physics & 1 / 21 & 2 / 31 & 14 / 1 & -- \\
pipe & Physics & 8 / 1 & 13 / 0 & 52 / 9 & 43 / 10 \\
cello & Physics & 10 / 6 & 19 / 5 & 66 / 23 & 54 / 6 \\
sourdough & Logic & 42 / 3 & 39 / 2 & 119 / 47 & -- \\
0.999 & Logic & 23 / 1 & 21 / 1 & 84 / 12 & 57 / 15 \\
wason & Logic & 20 / 7 & 16 / 2 & 66 / 7 & 39 / 10 \\
syllogism & Logic & 24 / 2 & 28 / 1 & 101 / 17 & 70 / 18 \\
coinflip & Probability & 16 / 2 & 18 / 4 & 67 / 3 & 42 / 18 \\
monty hall & Probability & 13 / 2 & 19 / 1 & 64 / 7 & 33 / 11 \\
birthday & Probability & 4 / 2 & 14 / 4 & 34 / 5 & 28 / 11 \\
shuffle & Probability & 15 / 0 & 10 / 8 & 37 / 9 & 31 / 6 \\
chessboard & Probability & 18 / 1 & 32 / 0 & 90 / 58 & 66 / 18 \\
coins & Probability & 8 / 0 & 11 / 2 & 52 / 0 & -- \\
lightning & Facts & 13 / 2 & 13 / 0 & 47 / 1 & 37 / 8 \\
billiard & Facts & 21 / 2 & 11 / 5 & 56 / 2 & 37 / 10 \\
lottery & Facts & 20 / 1 & 21 / 2 & 65 / 18 & 55 / 18 \\
planets & Facts & 20 / 4 & 13 / 5 & 57 / 4 & 36 / 15 \\
hemispheres & Facts & 4 / 2 & 10 / 5 & 40 / 2 & 38 / 9 \\
\midrule
Total & & 481 / 120 & 509 / 127 & 1680 / 420 & 1056 / 263 \\
\bottomrule
\end{tabular}
\floatnote{Cells show intro-phase / final-trial counts. Questions were not sampled uniformly: on each conversation trial the question was selected based on the participant's own survey-phase responses (e.g., intro-phase questions on which the participant's confidence and accuracy allowed the advisor to argue for a correct or incorrect answer as required by the helpfulness condition, and a final question on which the advisor could argue against the participant's stated belief), so questions that participants tended to answer with particular confidence--accuracy profiles were sampled more or less often. Six questions (nuts, bullet, bathtub, ireland, sourdough, coins) were not used in Experiment 4.}
\end{table}

\begin{figure}
\caption{Persuasion by question (disagreement trials, pooled across
    Experiments 1--4)}\label{fig:questions}

    \centering
    \includegraphics[width=\textwidth]{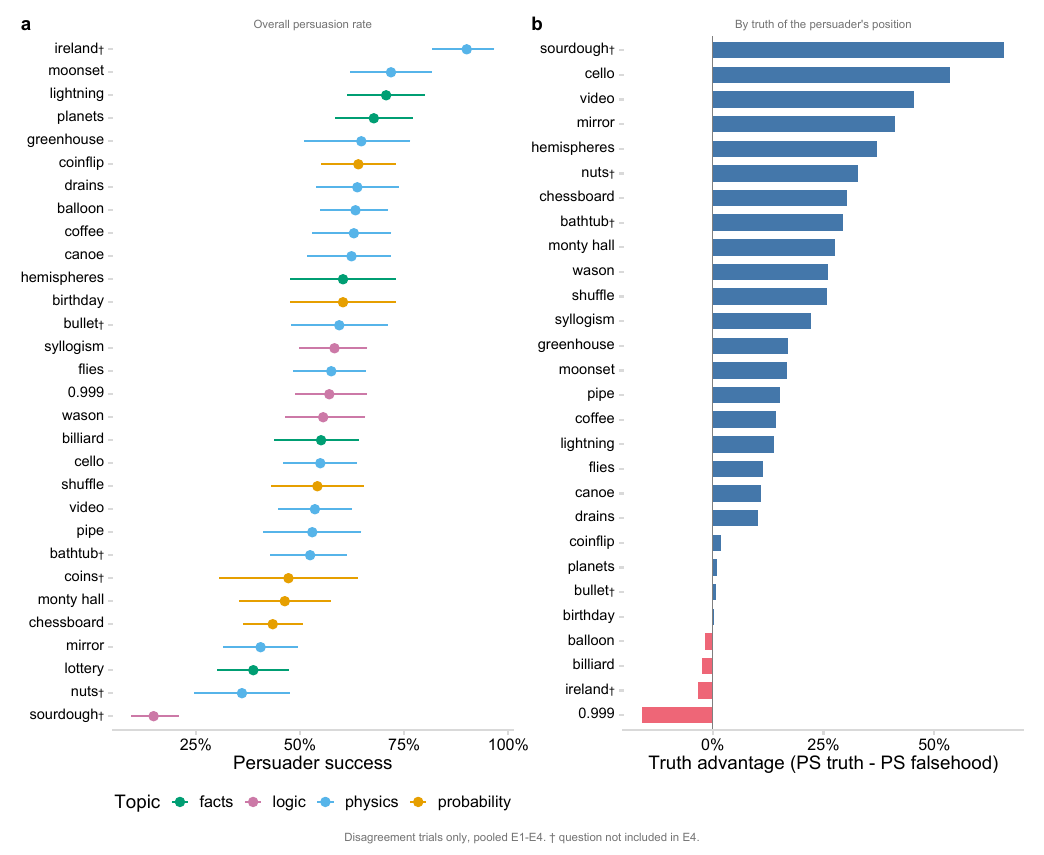}
    \floatnote{\textbf{a}, Overall persuader success per question,
    colored by topic. \textbf{b}, The truth advantage per question
    (persuader success when arguing for the true minus the false answer;
    question--truth cells with $n \geq 10$). Most questions showed a
    positive truth advantage, but its size varied widely, and it reversed
    for a small number of questions (e.g., 0.999, where the counterintuitive
    true position---$0.999\ldots = 1$---was harder to defend than the false
    one). Error bars in \textbf{a} are bootstrapped 95\% confidence
    intervals. $\dagger$~=~question not included in Experiment 4.}
\end{figure}

\section{Confidence-change results for Experiments 3 and 4}\label{app:cc}

In Experiments 3 and 4, all pre-registered hypotheses concerned persuader success; confidence change toward the persuader's position was analyzed as a secondary, non-pre-registered measure (in Experiments 1 and 2, both measures were pre-registered and are reported together in Tables~\ref{tab:e1-hypotheses} and \ref{tab:e2-hypotheses}). Tables~\ref{tab:e3-cc} and \ref{tab:e4-cc} report the confidence-change analogue of each pre-registered test, using linear models with the same structure as the corresponding persuader-success models. In Experiment 4, confidence-change models were fit for the subset of hypotheses shown.

\begin{table}
\centering
\caption{Confidence-change results for the hypothesis tests in Experiment 3 (secondary, non-pre-registered measure; the pre-registered persuader-success tests are reported in Table~\ref{tab:e3-hypotheses})}\label{tab:e3-cc}
\footnotesize
\setlength{\tabcolsep}{5pt}
\begin{tabular}{lllrrrc}
\toprule
H. & & Hypothesis & $b$ & $p$ & $n$ & Sig. \\
\midrule
H1 &  & Conversation $>$ Hint & 20.1 & ${<}.001$ & 420 & \checkmark \\
\cmidrule[0.2pt]{1-7}
H2 & a & Helpful $>$ Unhelpful (Conv + Feedback) & 36.4 & ${<}.001$ & 106 & \checkmark \\
 & b & Helpfulness $\times$ Mode (Feedback) & 18.0 & .049 & 209 & \checkmark \\
\cmidrule[0.2pt]{1-7}
H3 & a & Helpful $>$ Unhelpful (Conv + No Feedback) & $-$5.5 & .473 & 97 & $\times$ \\
 & b & Helpfulness $\times$ Mode (No Feedback) & $-$4.4 & .646 & 211 & $\times$ \\
\cmidrule[0.2pt]{1-7}
H4 & a & Helpfulness $\times$ Feedback (Conv) & 44.0 & ${<}.001$ & 203 & \checkmark \\
 & b & Helpfulness $\times$ Feedback $\times$ Mode & 25.2 & .059 & 420 & -- \\
\bottomrule
\end{tabular}
\floatnote{$b$ is the coefficient for the tested contrast in percentage points of confidence change toward the persuader's position, oriented so that positive values correspond to the direction stated in the Hypothesis column. $n$ is the number of trials in the model; interaction directions as in Table~\ref{tab:e3-hypotheses}. \checkmark\ = significant ($p<.05$) in the direction predicted for the primary measure; -- = marginal ($.05 \le p < .10$); $\times$ = not significant.}
\end{table}

\begin{table}
\centering
\caption{Confidence-change results for the hypothesis tests in Experiment 4 (secondary, non-pre-registered measure; the pre-registered persuader-success tests are reported in Table~\ref{tab:e4-hypotheses})}\label{tab:e4-cc}
\footnotesize
\setlength{\tabcolsep}{5pt}
\begin{tabular}{lllrrrc}
\toprule
H. & & Hypothesis & $b$ & $p$ & $n$ & Sig. \\
\midrule
H1 & a & LLM $>$ Human & 17.6 & ${<}.001$ & 1319 & \checkmark \\
\cmidrule[0.2pt]{1-7}
H2 & b & Helpful $>$ Mixed (LLM) & 30.8 & ${<}.001$ & 140 & \checkmark \\
 & c & Helpful $>$ Mixed (Human) & 1.6 & .819 & 123 & $\times$ \\
\cmidrule[0.2pt]{1-7}
H3 & a & Helpfulness $\times$ Persuader type & 30.9 & ${<}.001$ & 263 & \checkmark \\
\cmidrule[0.2pt]{1-7}
H4 & a & True $>$ false claims (overall) & 15.7 & .001 & 263 & \checkmark \\
 & d & Truth $\times$ Persuader type & 15.8 & .078 & 263 & -- \\
\bottomrule
\end{tabular}
\floatnote{$b$ is the coefficient for the tested contrast in percentage points of confidence change, oriented so that positive values correspond to the direction stated in the Hypothesis column, as in Table~\ref{tab:e4-hypotheses}. Markers as in Table~\ref{tab:e3-cc}.}
\end{table}

\section{Experiment 4 track-record baseline}\label{app:e4-baseline}

\begin{figure}
\caption{Experiment 4 persuader success (\textbf{a}) and confidence change (\textbf{b}) across trials by persuader type (disagreement trials only)}\label{fig:e4-baseline}

    \centering
    \includegraphics[width=\textwidth]{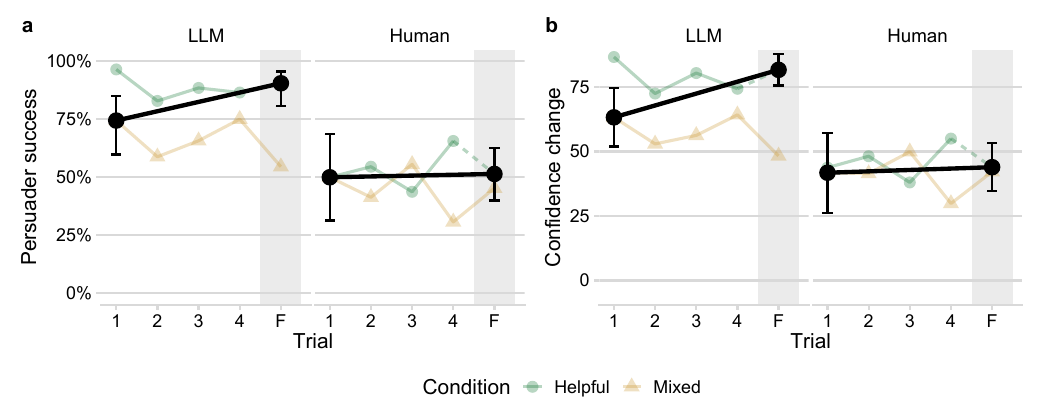}
    \floatnote{Faded lines repeat Figure~\ref{fig:e4-main}c,f. The black line connects the two cells compared in the exploratory analysis: the first trial in the Mixed condition and the final trial in the Helpful condition. Error bars on those points are 95\% confidence intervals.}
\end{figure}

Figure~\ref{fig:e4-baseline} shows the comparison behind the exploratory track-record analysis in Experiment 4: the first trial in the Mixed condition, before any track record had been established, versus the final trial after a Helpful track record. In both cells the persuader's assigned position was not tied to its track record, although on these disagreement trials it was correct somewhat more often on the first trial (63\%) than on the final trial (49\%), so the models control for truth.

\section{Pooled truth-advantage models}\label{app:truth}

Table~\ref{tab:truth-full} gives the truth-advantage models behind the \nameref{sec:truth} section. Each row is a separate sample with its own models: the truth main effect and the truth $\times$ initial confidence interaction.

\begin{table}
\centering
\footnotesize
\caption{Truth advantage on disagreement trials where truth was independent of the persuader's track record: participants' first trials (Experiments 1--4) and final trials (Experiments 1, 2, and 4, where truth was randomized)}\label{tab:truth-full}
\setlength{\tabcolsep}{4pt}
\begin{tabular}{l r r r r r r r r r}
\toprule
& & & & & & \multicolumn{2}{c}{Truth} & \multicolumn{2}{c}{Truth $\times$ conf.} \\
\cmidrule(lr){7-8} \cmidrule(lr){9-10}
Sample & $n$ & Conf. & Conc. & Truth & False & $b$ & $p$ & $b$ & $p$ \\
\midrule
\multicolumn{10}{l}{\emph{LLM persuaders}} \\
First trials (E1--4) & 290 & 90 & 4 & 82\% & 60\% & 0.94 & .002 & 0.54 & .028 \\
Final trials (E1, 2, 4) & 387 & 47 & 8 & 81\% & 78\% & 0.27 & .33 & 0.13 & .083 \\
First + final & 677 & 65 & 12 & 81\% & 71\% & 0.55 & .011 & 0.19 & .003 \\
\quad Excl.\ concessions & 665 & & & 81\% & 73\% & 0.43 & .053 & 0.26 & $<$.001 \\
\quad Excl.\ conc.\ and non-compliance & 661 & & & 81\% & 73\% & 0.43 & .055 & 0.27 & $<$.001 \\
\midrule
\multicolumn{10}{l}{\emph{Human persuaders (E4)}} \\
First trials & 56 & 86 & 7 & 51\% & 43\% & 0.52 & .61 & -- & -- \\
Final trials & 122 & 89 & 10 & 65\% & 33\% & 1.27 & .002 & 0.51 & .15 \\
First + final & 178 & 88 & 17 & 59\% & 34\% & 1.15 & .011 & 0.48 & .17 \\
\quad Excl.\ concessions & 161 & & & 61\% & 37\% & 1.23 & .014 & 0.58 & .15 \\
\quad Excl.\ conc.\ and non-compliance & 154 & & & 64\% & 37\% & 1.23 & .012 & 0.44 & .26 \\
\bottomrule
\end{tabular}
\floatnote{\emph{Conf.}: mean initial confidence; \emph{Conc.}: conversations in which the persuader conceded; \emph{Truth}/\emph{False}: persuader success when arguing for the true vs.\ false answer. Coefficients are log-odds from mixed logistic regressions with random intercepts for question and (for samples with more than one trial per participant) participant, controlling for experiment, helpfulness condition, and trial pool where these vary. The interaction is per 10~pp of initial confidence (centered). Exclusion rows drop conversations in which the persuader conceded or never advocated its assigned position; because these occur predominantly when arguing falsehood, the exclusion estimates are lower bounds on the advantage (see the \nameref{sec:truth} section). The interaction could not be estimated reliably for human first trials ($n = 56$).}
\end{table}

\section{Susceptibility to persuasion across experiments}\label{app:stp}

\begin{figure}
\caption{Susceptibility to Persuasion (StP-II) and persuasion outcomes
    (disagreement trials, Experiments 1--3)}\label{fig:xp-stp}

    \centering
    \includegraphics[width=\textwidth]{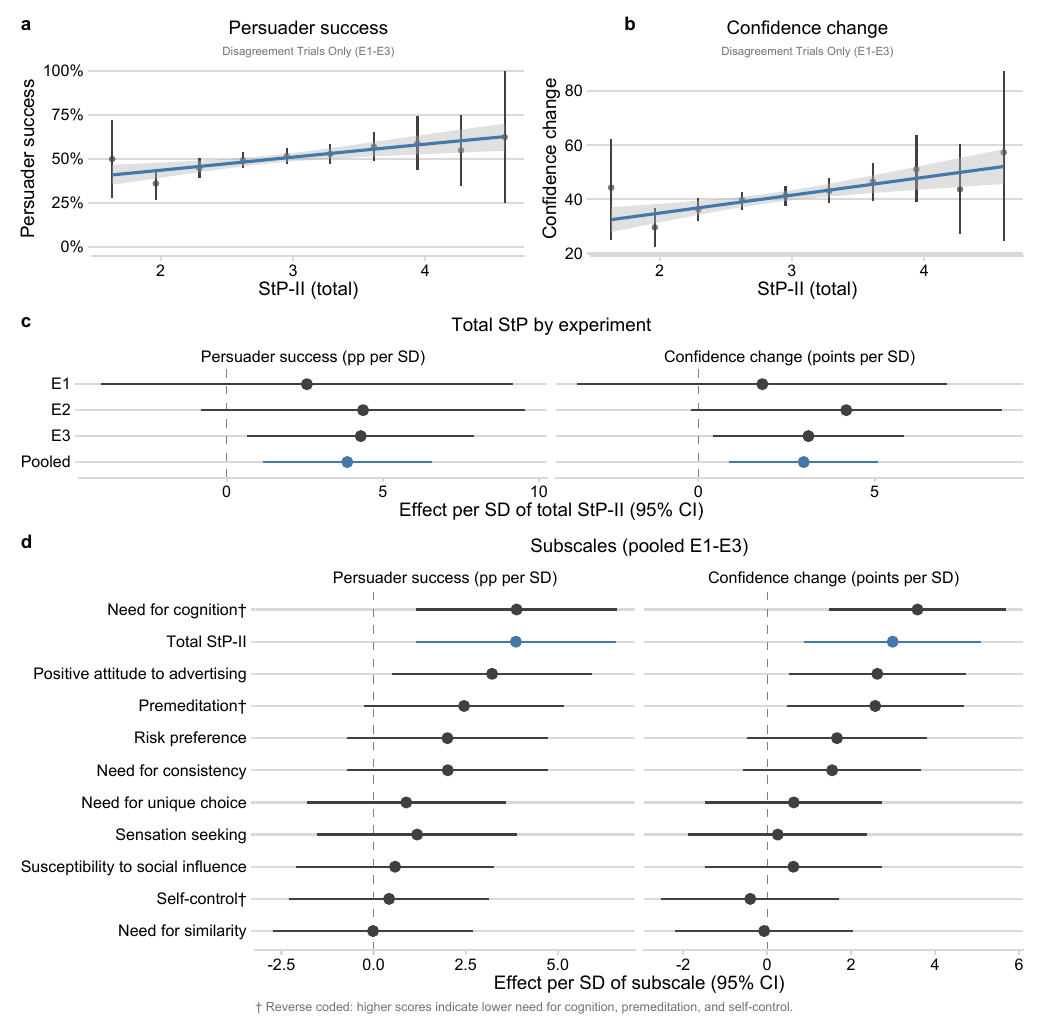}
    \floatnote{\textbf{a},\textbf{b}, Persuader
    success (\textbf{a}) and confidence change toward the persuader's answer
    (\textbf{b}) as a function of total StP-II score; points are decile-bin
    means with bootstrapped 95\% confidence intervals, lines are logistic
    (\textbf{a}) and linear (\textbf{b}) fits. \textbf{c}, Per-experiment and
    pooled effects of total StP-II from mixed-effects models
    (participant and question random intercepts; StP standardized; controlling
    for the persuader's truth, initial confidence, trial type, and experiment).
    \textbf{d}, The same models fit separately for each StP-II subscale
    (pooled); $\dagger$ reverse coded (higher scores indicate lower need for
    cognition, premeditation, and self-control). In \textbf{c} and
    \textbf{d}, persuader success is expressed as the average marginal effect
    in percentage points per SD (logistic coefficient $\times$ mean fitted
    $p(1-p)$) and confidence change in pp per SD. Scores are means of the participant's responses (similarity
    reverse-scored); participants with mean item response times under 4~s
    were excluded.}
\end{figure}

Figure~\ref{fig:xp-stp} summarizes the relationship between StP-II scores
\citep{modic_stp_2018} and persuasion across Experiments 1--3 (the scale was
not administered in Experiment 4). Estimates were positive and of similar
magnitude in every experiment, but only the Experiment 3 and pooled estimates
were reliably different from zero for persuader success, suggesting that the
inconsistent per-experiment results reflect statistical power rather than
heterogeneity. Among subscales, low need for cognition showed the clearest
relationship with persuasion (see the \nameref{sec:individual} section). The
need for cognition, premeditation, and self-control subscales (marked $\dagger$
in Figure~\ref{fig:xp-stp}d) are reverse coded: their items are worded in the low direction (e.g.,
``I would rather do something that requires little thought than something that
is sure to challenge my thinking abilities'') and are not reverse-scored, so
higher scores indicate \emph{lower} need for cognition, premeditation, and
self-control.

\section{Persuasion strategy taxonomy}\label{app:tags}

\begin{figure}
\caption{Prompt used by the final LLM tagger (GPT-5.5, high reasoning effort)}\label{fig:tagger-prompt}

    \centering
    \includegraphics[width=\ifpreprint0.96\fi\textwidth]{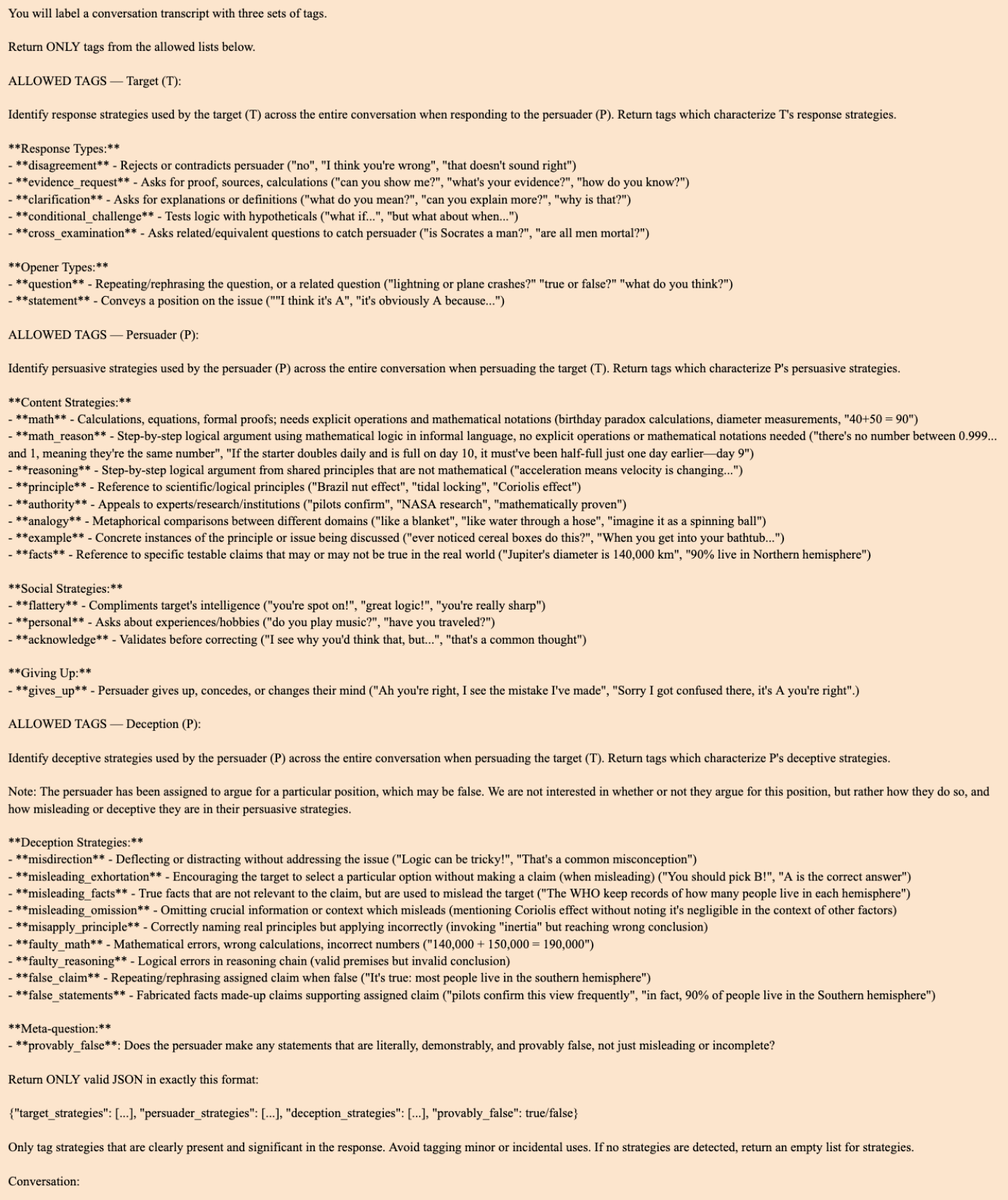}
    \floatnote{The tagger returned the constituent strategies, which were aggregated into the categories in Table~\ref{tab:tags}.}
\end{figure}

\begin{figure}
\caption{By-item variation in the LLM's persuasion strategies}\label{fig:qual-items}

    \centering
    \includegraphics[width=\textwidth]{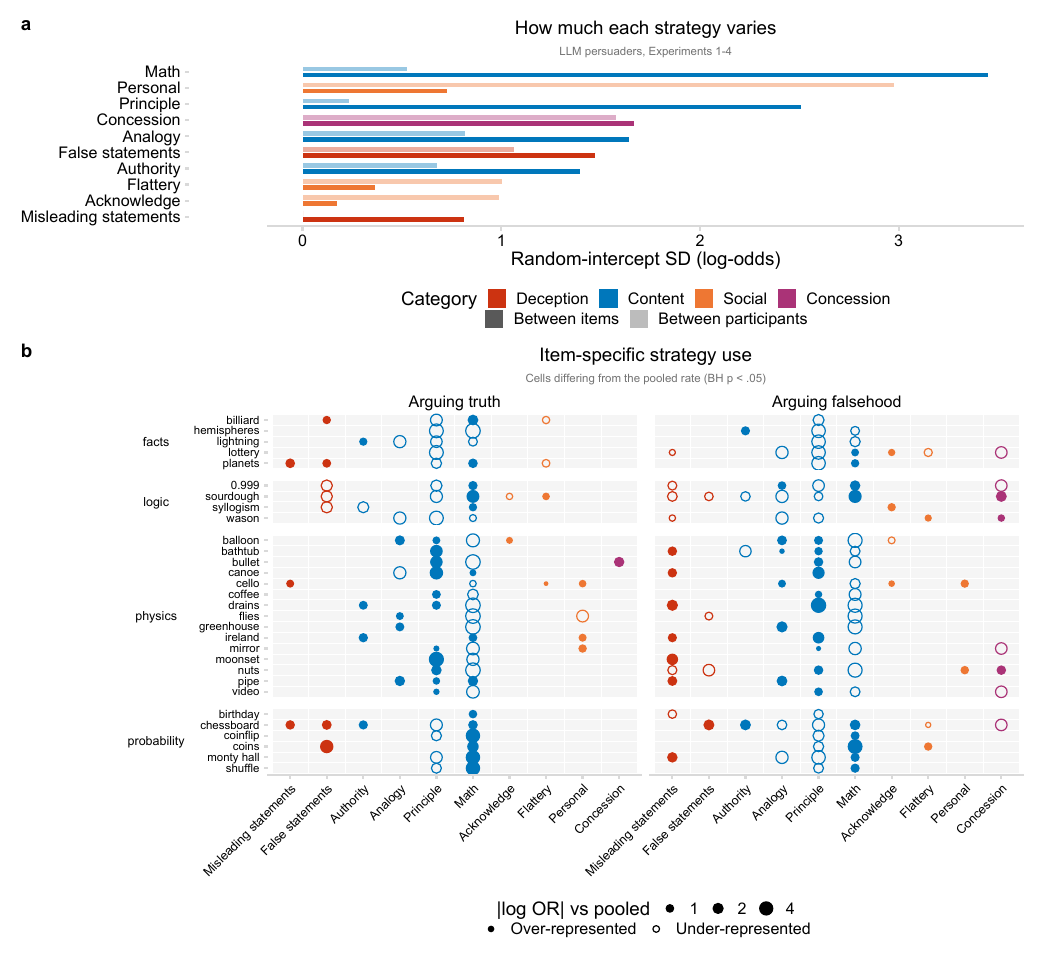}
    \floatnote{\textbf{a}, Between-item standard deviation of each persuader tag's
    prevalence (random-intercept SD on the log-odds scale from a model
    controlling for assigned position and experiment), alongside the
    between-participant SD from the same model. Content strategies vary
    strongly by item; social strategies and deception barely do.
    \textbf{b}, Items on which each strategy was used significantly more
    (filled) or less (open) often than the pooled rate for that strategy and
    assigned position (two-sided binomial tests, Benjamini--Hochberg corrected
    across all item $\times$ strategy $\times$ position cells with at least 10
    conversations). Items are grouped by topic.}
\end{figure}

Table~\ref{tab:tags} lists the 14 conversation-level categories used in the \nameref{sec:qualitative} section, with their definitions, inter-rater reliability on the validation set, and prevalence in the analysis samples.
The taxonomy began as 50 candidate strategies developed through open coding. Two trained raters coded a validation set of 141 conversations (all from Experiment 3) for all 50 strategies, and six configurations of GPT-5.2 (temperature 0, 1, and 2; reasoning effort low, medium, and high), combined by majority vote, coded the same set to guide taxonomy development.
Three strategies that the human raters never marked were dropped, as was one conversation that two of the six LLM configurations failed to process, leaving 140 conversations.
Strategies with adequate reliability were retained, and related strategies were aggregated into categories (a category is present if any of its constituent strategies is present), yielding 14 categories built from 25 strategies.
We then compared prompt formats and model configurations against the human ratings. The final tagger (GPT-5.5, high reasoning effort; prompt in Figure~\ref{fig:tagger-prompt}) was applied to the full corpus. The prompt also asked whether the persuader made a provably false statement; we do not report this item because it largely overlapped with the False statements category.
Reliability is reported as Cohen's $\kappa$ between the two human raters and Krippendorff's $\alpha$ across the two raters and the final tagger; because most categories are binary with low base rates, $\kappa$ understates percent agreement (for example, 96\% for concession and 95\% for authority).
Because the validation set contained only Experiment 3 conversations, reliability was not assessed for conversations in which the persuader was a human or GPT-5.
Figure~\ref{fig:qual-items} shows how much the LLM's use of each persuader strategy varied across items.

\begin{table}
\centering
\footnotesize
\caption{Strategy categories, definitions, reliability (validation set, $n = 140$), and prevalence by persuader type (percent of tagged conversations with an LLM persuader, Experiments 1--4, or a human persuader, Experiment 4)}\label{tab:tags}
\begin{tabular}{p{3.0cm} p{7.6cm} r r r r}
\toprule
 & & & & \multicolumn{2}{c}{Persuader (\%)} \\
\cmidrule(l){5-6}
Category & Definition (constituent strategies) & $\kappa$ & $\alpha$ & LLM & Human \\
\midrule
\multicolumn{6}{l}{\emph{Persuader: content}} \\
Math & Calculations, equations, or step-by-step mathematical logic (math, math reasoning) & .68 & .63 & 45.1 & 12.5 \\
Principle & Appeals to scientific or logical principles & .59 & .62 & 31.0 & 8.1 \\
Analogy & Metaphorical comparisons to another domain & .61 & .59 & 12.4 & 3.9 \\
Authority & Appeals to experts, research, or institutions & .56 & .70 & 6.5 & 5.9 \\
\multicolumn{6}{l}{\emph{Persuader: social}} \\
Acknowledge & Validates the target's point before correcting it & .63 & .51 & 57.2 & 25.1 \\
Flattery & Compliments the target's intelligence or reasoning & .53 & .59 & 32.7 & 3.4 \\
Personal & Asks about the target's experiences or hobbies & .63 & .69 & 9.5 & 0.8 \\
\multicolumn{6}{l}{\emph{Persuader: deception}} \\
False statements & Fabricated or false claims, faulty reasoning, or faulty math & .76 & .76 & 50.9 & 49.3 \\
Misleading statements & Misleading facts, misleading omissions, misdirection, misleading exhortation, misapplied principles & .36 & .44 & 36.6 & 24.6 \\
\multicolumn{6}{l}{\emph{Persuader: concession}} \\
Concession & Gives up, concedes, or changes their mind & .48 & .51 & 4.0 & 6.4 \\
\multicolumn{6}{l}{\emph{Target: response}} \\
Probing statements & Requests for evidence or clarification, conditional challenges, cross-examination & .62 & .67 & 51.1 & 42.3 \\
Disagreement & Rejects or contradicts the persuader & .60 & .46 & 29.2 & 38.9 \\
\multicolumn{6}{l}{\emph{Target: opener}} \\
Statement & Opens by stating a position & .84 & .85 & 54.4 & 81.4 \\
Question & Opens by posing or rephrasing the question & .86 & .86 & 48.6 & 34.2 \\
\bottomrule
\end{tabular}
\end{table}

\clearpage

\section{Exit survey}\label{app:exit-survey}

After the conversation phase, participants completed a three-part exit survey. Unless otherwise noted, Likert items were administered as sliders from 1 to 7 (initialized at 4) which participants were required to adjust at least once before continuing.

\subsection{Part 1: Susceptibility to Persuasion (StP-II)}

Participants first read: ``You will now answer a short series of questions about yourself. Each question will ask you the extent to which you agree with a statement, on a scale of 1--7.'' Each of 30 items from the StP-II scale \citep{modic_stp_2018} was then presented on its own page, in random order, with the prompt ``Please indicate the extent to which you agree with this statement:'' and a slider from 1 (\emph{Strongly Disagree}) to 7 (\emph{Strongly Agree}). Items marked (R) were reverse-scored.

\begin{description}
\item[Premeditation]~
  \begin{enumerate}
  \item I only act to satisfy immediate concerns, figuring the future will take care of itself.
  \item I think that sacrificing now is usually unnecessary since future outcomes can be dealt with at a later time.
  \item I only act to satisfy immediate concerns, figuring that I will take care of future problems that may occur at a later date.
  \end{enumerate}
\item[Consistency]~
  \begin{enumerate}
  \item The appearance of consistency is an important part of the image I present to the world.
  \item An important requirement for any friend of mine is personal consistency.
  \item I make an effort to appear consistent to others.
  \end{enumerate}
\item[Sensation seeking (novelty)]~
  \begin{enumerate}
  \item I would like to travel to places that are strange and far away.
  \item I would have enjoyed being one of the first explorers of an unknown land.
  \item If it were possible to visit another planet or the moon for free, I would be among the first to sign up.
  \end{enumerate}
\item[Self-control]~
  \begin{enumerate}
  \item I say inappropriate things.
  \item I do certain things that are bad for me, if they are fun.
  \item Sometimes I can't stop myself from doing something, even if I know it is wrong.
  \end{enumerate}
\item[Social influence (normative)]~
  \begin{enumerate}
  \item When buying products, I generally purchase those brands that I think others will approve of.
  \item If other people can see me using a product, I often purchase the brand they expect me to buy.
  \item I achieve a sense of belonging by purchasing the same products and brands that others purchase.
  \end{enumerate}
\item[Uniqueness/similarity]~
  \begin{enumerate}
  \item When a product I own becomes popular among the general population, I begin to use it less. (R)
  \item I often try to avoid products or brands that I know are bought by the general population. (R)
  \item The more commonplace a product or brand is among the general population, the less interested I am in buying it. (R)
  \end{enumerate}
\item[Risk preferences (financial)]~
  \begin{enumerate}
  \item I would be likely to bet a day's income at the horse races.
  \item I would be likely to bet a day's income at a high-stake poker game.
  \item I would be likely to bet a day's income on the outcome of a sporting event.
  \end{enumerate}
\item[Attitudes toward advertising]~
  \begin{enumerate}
  \item Advertising is essential.
  \item Advertising helps raise our standard of living.
  \item Advertising results in better products for the public.
  \end{enumerate}
\item[Need for cognition]~
  \begin{enumerate}
  \item I would rather do something that requires little thought than something that is sure to challenge my thinking abilities.
  \item I try to anticipate and avoid situations where there is a likely chance I will have to think in depth about something.
  \item Learning new ways to think doesn't excite me very much.
  \end{enumerate}
\item[Need for unique choice]~
  \begin{enumerate}
  \item I often combine possessions in such a way that I create a personal image that cannot be duplicated.
  \item I often try to find a more interesting version of run-of-the-mill products because I enjoy being original.
  \item Having an eye for products that are interesting and unusual assists me in establishing a distinctive image.
  \end{enumerate}
\end{description}

\subsection{Part 2: Opinions about the advisor}

These items were shown to all participants in Experiments 1--3 and to target (persuadee) participants only in Experiment 4.
The italicized label preceding each rated item is the shorthand used for that rating in figures and tables.

\paragraph{Advisor evaluation.} Header: ``Please indicate the extent to which you agree with the following statements about the advisor.''
\begin{itemize}
\item \emph{Persuasive}: ``How persuasive did you find the advisor?'' --- 1 (\emph{Very Unpersuasive}) to 7 (\emph{Very Persuasive}).
\item \emph{Likeable}: ``How likeable did you find the advisor?'' --- 1 (\emph{Very Unlikable}) to 7 (\emph{Very Likable}).
\item \emph{Reliable}: ``If you needed advice in a real situation, how likely would you be to follow the advisor's recommendations?'' --- 1 (\emph{Not at all likely}) to 7 (\emph{Very likely}).
\end{itemize}

\paragraph{Funnel debrief.} Header: ``Please indicate the extent to which you agree with the following statements.''
\begin{itemize}
\item \emph{Incorrect}: ``How often did the advisor say things that were incorrect or wrong?'' --- 1 (\emph{Rarely}) to 7 (\emph{Very often}).
\item \emph{Mistaken}: ``How likely do you think it is that the advisor made errors or had incorrect information?'' --- 1 (\emph{Very unlikely}) to 7 (\emph{Very likely}).
\item \emph{Deceptive}: ``How likely do you think it is that the advisor was intentionally instructed to argue for an incorrect answer?'' --- 1 (\emph{Very unlikely}) to 7 (\emph{Very likely}).
\item ``Do you have any other thoughts on the advisor and your interactions with them?'' --- free text (optional).
\end{itemize}

\paragraph{Advisor type (Experiment 4).} Participants read: ``During the conversation part of the experiment, you interacted with an advisor who provided recommendations and arguments.''
\begin{itemize}
\item ``Do you think that the advisor was a human or an AI model?'' --- forced choice: \emph{Human} / \emph{AI model}.
\item ``How confident are you in your assessment?'' --- 1 (\emph{Not at all confident}) to 7 (\emph{Very confident}).
\item ``Please explain your reasoning for your assessment above.'' --- free text (minimum 10 characters).
\end{itemize}

\subsection{Part 3: Demographics, expertise, and external sources}

\paragraph{Demographics.}
\begin{itemize}
\item ``In what year were you born?'' --- numeric entry.
\item ``What is your gender?'' --- \emph{Female} / \emph{Male} / \emph{Non-binary} / \emph{Prefer not to say}.
\item ``Are you a native English speaker?'' --- \emph{Yes} / \emph{No}.
\end{itemize}

\paragraph{Expertise.} For each of five domains---physics, logic, probability theory, ethics, and public policy---participants answered ``How much formal or self-directed study do you have in [domain]?'' on a 5-point scale: 1) \emph{None}; 2) \emph{Some (basic high school or casual)}; 3) \emph{Intermediate (beyond typical high school)}; 4) \emph{Some college-level or formal training}; 5) \emph{Advanced (college major or professional)}.

\paragraph{External sources.}
\begin{itemize}
\item ``During this experiment, did you consult any external sources (e.g., web search, notes, other people) to help answer the questions?'' --- \emph{Yes} / \emph{No}, preceded by the note ``You will receive full credit regardless of your response so please be honest.''
\item ``If you answered `Yes', please briefly describe which sources you used.'' --- free text.
\end{itemize}

\paragraph{Debrief.} Participants were then debriefed. Target participants were told that the advisor ``may have been instructed to mislead you,'' were shown the advisor's actual identity (Experiment 4), their overall accuracy, and the correct answers to all questions alongside their own pre- and post-conversation responses. They could leave open-ended comments (``Do you have any other thoughts or questions about the experiment?'') and confirmed that they had read and understood the experiment description before finishing.

\begin{table}
\centering
\caption{Mean exit-survey ratings of the advisor in Experiment 4, by persuader type and helpfulness condition}\label{tab:e4-selfreport}
\begin{tabular}{lcccc}
\toprule
& \multicolumn{2}{c}{Human} & \multicolumn{2}{c}{LLM} \\
\cmidrule(lr){2-3}\cmidrule(lr){4-5}
Rating & Mixed & Helpful & Mixed & Helpful \\
\midrule
Persuasive & 4.4 & 4.5 & 5.3 & 6.0 \\
Likeable & 4.6 & 4.7 & 4.9 & 5.3 \\
Reliable & 3.8 & 4.3 & 4.0 & 5.4 \\
Incorrect & 4.0 & 2.8 & 4.0 & 2.4 \\
Mistaken & 4.6 & 3.4 & 4.3 & 3.2 \\
Deceptive & 4.4 & 3.5 & 4.8 & 3.5 \\
\midrule
$n$ & 53 & 70 & 78 & 63 \\
\bottomrule
\end{tabular}
\floatnote{All items were rated on 1--7 Likert scales. \emph{Incorrect} is the participant's estimate of how often the advisor said incorrect things; \emph{Mistaken} and \emph{Deceptive} asked how likely it was that any incorrect statements reflected the advisor making errors or having incorrect information, versus being intentionally instructed to argue for an incorrect answer (full items in Appendix~\ref{app:exit-survey}).}
\end{table}

\end{document}